\documentclass[nofootinbib,preprintnumbers,superscriptaddress,amsmath,amssymb,floatfix]{revtex4-1}
\usepackage{graphicx,setspace,epsfig,color}
\usepackage[letterpaper,dvips,width=9.4in,height=8.5in,includemp=false]{geometry}

\usepackage{exscale}
\usepackage{enumerate}
\usepackage{cancel}
\usepackage{mdframed}
\usepackage{amsmath}
\usepackage[cm]{fullpage}
\usepackage{amssymb}
\usepackage{mathtools}
\usepackage{float}
\usepackage{slashed}
\usepackage{hyperref}
\usepackage{cleveref}
\usepackage{inputenc}
\usepackage{comment}
\usepackage{wrapfig}
\usepackage{framed}
\usepackage{enumitem}
\usepackage{float}

\long\def\symbolfootnote[#1]#2{\begingroup%
\def\thefootnote{\fnsymbol{footnote}}\footnote[#1]{#2}\endgroup}

\newcommand{\be}{\begin{equation}}
\newcommand{\ee}{\end{equation}}
\newcommand\bea{\begin{eqnarray}}
\newcommand\eea{\end{eqnarray}} 

\newcommand{\E}

\usepackage{hyperref}
\begin{document}

\title{Electromagnetic scattering in Degenerate and Partially Degenerate Nuclear Matter}

\author{Stephan Stetina} 
\email{stetina@hep.itp.tuwien.ac.at}
\affiliation{Institut f\"ur Theoretische Physik, Technische Universit\"at Wien,
Wiedner Hauptstrasse 8-10, A-1040 Vienna, Austria}
\begin{abstract}
\noindent I calculate the rate of electromagnetic scattering in degenerate and partially degenerate plasmas composed of electrons, muons, protons, and neutrons. Correlations with strong interactions, induced by the polarizability of electromagnetically and strongly charged protons, are taken into account. At subnuclear densities induced interactions are shown to cause a drastic increase of the longitudinal scattering rate. The results obtained are particularly relevant for the electron contribution to transport in the outer core of neutron stars, and represent a first step towards a realistic calculation of transport phenomena in neutron star mergers.      
\end{abstract}

\pacs{Valid PACS appear here}
\maketitle

\section{Introduction}

\noindent Recent multi-messenger observations of neutron stars and their mergers have ushered in a golden age of nuclear astrophysics. A few days after their formation in supernovae, neutron stars cool to temperatures well below $1$ MeV  \cite{Prakash:2000jr} \cite{Yakovlev:2004yr} \cite{Yakovlev:2004iq} \cite{Brown:2017gxd}. The vast majority of matter is located in the core, where nucleons are compressed to densities from $0.5\,n_0$ to several times $n_0$, where $n_0=2.3\cdot\,10^{17} \,\textrm{kg}/\textrm{m}^3$ is the nuclear saturation density. Under such conditions nuclear matter forms a degenerate plasma composed of electrons, muons, protons, and neutrons. The relative abundances of these particles are deduced from the requirements of charge neutrality and beta equilibrium, resulting in highly asymmetric nuclear matter, with typical proton fractions below $10$ \%. During merger events temperatures are projected to be significantly higher, reaching up to $30$ Mev  \cite{Oechslin:2006uk} \cite{Kiuchi:2009jt} \cite{Radice:2016dwd} \cite{Baiotti:2016qnr}. Densities up to four times nuclear saturation density (and beyond, if phase transitions are taken into account \cite{Bauswein:2018bma}) can be achieved, resulting in partially degenerate matter in near-equilibrium.\newline 
A particularly exciting prospect is the utilization of compact stars as ``laboratories" to study the properties of nuclear matter under extreme conditions. For many observables transport phenomena represent a crucial link to microscopic physics, where they are calculated from reaction rates of particles traversing through the plasma. In this article, I calculate the rate of electromagnetic scattering (or Moeller scattering), taking into account the environment of fully and partially degenerate nuclear matter. Leptons and hadrons are treated as quasiparticles, immersed in  weakly and strongly interacting Fermi liquids respectively. Electrons, and depending on the density also muons, are relativistic and weakly interacting, and their scattering is thus a particularly important mechanism for transport. The scattering rate $\Gamma$ (damping rate, interaction rate) may either be obtained from a direct calculation according to Fermi`s golden rule, or, via the optical theorem, from the imaginary part of the fermion self energy. As outlined by Braaten and Pisarski \cite{Braaten:1989mz}, a consistent calculation of scattering amplitudes in hot or dense plasmas relies on the separation of two scales, termed  ``hard" and ``soft". In the present case the hard scale is set by the Fermi momentum $k_f$, while the soft scale is of order $e\,k_f$, where $e\ll1$ is the gauge coupling constant of electromagnetic interactions. Dispersion relations of particles carrying soft momenta are strongly modified by interactions with the surrounding medium, while hard momentum particles will traverse unhindered through the plasma. Transport phenomena in degenerate matter are predominantly determined by fermions in close proximity to the Fermi surface, whose momenta are hard by definition. Medium modifications to their dispersion relation stemming from electromagnetic interactions can be neglected for all practical purposes. The momentum of the exchanged photon, on the other hand, may either be hard or soft, depending on the angle between the scatterers. Medium modifications to the photon propagator thus play an important role, and need to be resummed to obtain consistent results. 
Vertex resummations become necessary if all momenta flowing into the or out of a vertex are soft, and can thus be neglected in the present context. To obtain the the dressed photon propagator the (relativistic) random phase approximation (RPA) is employed, which amounts to resumming one-loop polarization functions.\newline 
The surrounding medium manifests itself in the photon spectrum in two ways: electromagnetic interactions are screened, and a longitudinal component of the photon field, termed plasmon, arises. The latter corresponds to a pure collective excitation, and disappears from the spectrum upon approaching the hard momentum limit. When the exchanged momentum is soft, longitudinal and transverse scattering amplitudes exhibit very different characteristics. Interactions in the longitudinal channel are predominantly modified by Debye screening, stemming from the real part of the longitudinal polarization tensor $\Pi_L$. Debye screening persists even in the static limit, where one obtains the screening mass (Debye mass) as $\Pi_L(q_0,\,|\boldsymbol{q}|\rightarrow 0)=-m_D^2$. The transverse polarization tensor $\Pi_\perp$, in contrast, vanishes in the static limit, and the dominant contribution to \textit{dynamical} screening originates from Landau damping, encoded in the imaginary part of $\Pi_\perp$. The difference of Debye screening and Landau damping becomes crucial at finite temperature. While Debye screening renders the longitudinal scattering rate $\Gamma_L$ finite, \textit{dynamical} screening in the transverse channel is unable to do so, resulting in a logarithmic divergence of $\Gamma_\perp$ in the infrared, see Ref. \cite{Bellac:2011kqa} for a pedagogical review. The only exception is the fully degenerate (i.e., $T=0$) limit, where strict Pauli blocking prevents an infrared catastrophe \cite{LeBellac:1996kr} \cite{Vanderheyden:1996bw}\cite{Manuel:2000mk}. A finite result for $\Gamma_\perp$ at finite temperature requires for an advanced resummation scheme, developed in Ref. \cite{Blaizot:1996az}. Fortunately it is not the total rate $\Gamma$, but rather the energy loss per distance traveled, $-dE/dx$, which ultimately enters the transport integral. Both expressions differ by an additional factor of $q_0 / |\boldsymbol{v}|$ \cite{Braaten:1991jj}, where $\boldsymbol{v}$ is the velocity of the fermion. Combined with the effects from Landau damping, the additional power of $q_0$ is just about enough to compensate the infrared divergence stemming from the Bose distribution $n_b\sim T / q_0$. The distinct characteristics of electric and magnetic screening are most pronounced at high densities and low temperatures, where they lead to very different results for the scattering rates in both channels: Heiselberg and Pethick discovered \cite{Heiselberg:1993cr}, that the scattering of ultra relativistic particles is dominated by the exchange of (transverse) photons, while the scattering of non-relativistic particles is dominated by the exchange of longitudinal plasmons. In the context of nuclear matter the former are represented by electrons, and the latter are represented by nucleons. Muons interpolate between both cases, at least at lower densities.\newline 
Within the RPA it is straightforward to generalize the calculation of scattering rates to multi-component plasmas: the total energy loss of, say, an electron, is simply given by the sum of the individual rates due to collisions with other electrons, muons, and protons in the plasma. In each case, the screening is provided by $\textit{all}$ plasma constituents. This is essentially the approach used to calculate the lepton contribution to transport coefficients in neutron star cores in Ref. \cite{Shternin:2008es}  \cite{Shternin:2018jop}, see Ref. \cite{Schmitt:2017efp} for a recent review. If strong interactions are taken into account, photons and plasmons couple to electromagnetically and strongly charged protons, which correlates both types of interactions. These correlations allow for medium induced lepton-neutron scattering \cite{Bertoni:2014soa}, which is otherwise negligible since it arises only due to the small magnetic moment of the neutron \cite{FlowersItoh1976}. Given that proton fractions in dense nuclear matter are small, the question arises whether there are circumstances under which the impact of induced scattering is particularly amplified. Systematic studies have shown \cite{Stetina:2017ozh} that resumming induced interactions strongly modifies the photon spectrum at lower densities $n\sim(0.5-0.6)\, n_0$, where \textit{homogeneous} nuclear matter is projected to become unstable \cite{Baym:1971ax} \cite{Muller:1995ji} \cite{Li:1997ra}. The onset of this instability manifests itself, among other things, in a divergence of the Debye screening of \textit{strong} interactions, which, owing to the presence of protons, drags the electromagnetic screening in the longitudinal channel along with it. Since protons are non-relativistic at lower densities, the transverse channel is only mildly affected. The consistent inclusion of induced interactions into the electromagnetic scattering rates is a central aspect of this paper. \newline
\newline 
In the context of neutron star phenomenology, the results obtained in this article are particularly relevant for the damping of oscillatory  modes  of  a  star, most  importantly r-modes. While the excitation of r-modes in fast spinning stars seems unavoidable, they are known to become unstable with  respect  to  the  emission of gravitational waves \cite{Andersson:1997xt} \cite{Haskell:2015iia}. The fact that fast spinning stars are nevertheless observed in nature points towards an efficient damping mechanism. Viscous damping in the crust-core transition region has been identified as a promising candidate, but would have to be several times larger than previously calculated \cite{Ho:2011tt}. In the crust-core interface the impact of induced interactions is most pronounced, which makes it a region of particular interest to this article. It is commonly assumed that protons in the outer core are superconducting. Predictions for the magnitude of the superconducting gap vary, and even the complete absence of superconductivity has been projected, see Ref. \cite{Sedrakian:2018ydt} for a recent review. The interplay of induced interactions and superconductivity will be discussed further in the outlook. Whether the outer core of neutron stars connects directly to the crust is currently unknown, and the possibility of an intermediate layer comprised of nuclear clusters of various geometries, collectively called ``nuclear pasta", has been suggested \cite{Pethick:1995di}. The existence of a ``pasta phase" depends on the properties of nuclear interactions at subnuclear densities. Its  absence would promote the outermost region of the core to a very privileged position, with profound impact on the spin evolution of neutron stars. \newline
In light of the recent observation of a neutron star merger \cite{GBM:2017lvd} \cite{TheLIGOScientific:2017qsa} \cite{Metzger:2017wot}  the question has emerged whether transport phenomena are potentially relevant for the modelling of merger events. Current simulations are based mostly on ideal (magneto)hydrodynamics. The importance of viscous effects has been investigated in Refs. \cite{Alford:2017rxf}  \cite{Harutyunyan:2018mpe}, which estimate that electromagnetic scattering takes to much time to play a role during merger events. It is emphasized \cite{Harutyunyan:2018mpe}, however, that a definitive clarification can be
reached only through fully numerical studies which include all
possible dissipative effects. The calculation of scattering rates in partially degenerate matter with temperatures up to $\sim30$ MeV is consequently in demand. It is reasonable to assume that under such conditions the rates will exhibit quite different characteristics compared to those computed for cold neutron star matter. \newline 
\newline
\noindent This paper is organized as follows: Section \ref{sec:OT} reviews the relationship of the scattering rate $\Gamma$ and the fermion self energies in detail, and discusses various approximations to the full one-loop resummation. It contains subsection  \ref{subsec:multi}, which discusses the generalization to a multi-component plasma and subsection \ref{subsec:induced} which introduces induced interactions. Section \ref{sec:FullDegen} evaluates $\Gamma$ at zero temperature, which is appropriate for old neutron stars. Finite temperature studies are discussed in Sec. \ref{sec:PartDegen}, in a range of $T=(0.1-1)$ MeV, covering the life span of neutron stars, and up to $30$ MeV relevant for partially degenerate matter in hot regions of neutron star mergers. Throughout this paper I use natural units  $\hbar=c=k_b=1$, and the electric charge $e^2=4\pi\alpha_f$ where $\alpha_f=1/137$ is the fine structure constant, and a mostly negative metric convention $g^{\mu\,\nu}=\textrm{diag} (1, -1 ,-1 ,-1)$.
\section{Scattering rate from optical theorem}
\label{sec:OT}
\begin{figure}[t]
\includegraphics[scale=1]{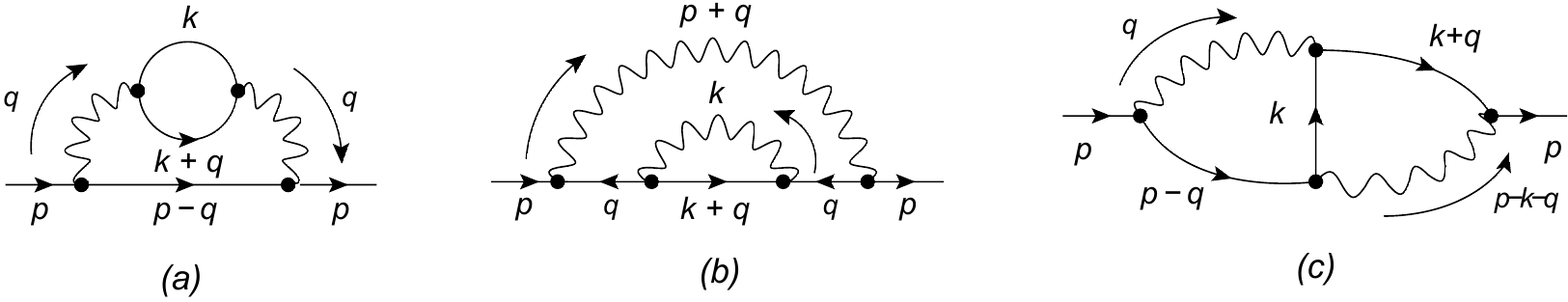}
\caption{\label{fig:2loop} Central cuts of the two-loop fermion self energy correspond to Moeller scattering (a), Compton scattering (b), and interference terms between different channels of both (c). Cutting diagram (c) from the bottom left to top right puts all fermions on shell, and consequently yields the interference term to Moeller scattering. The opposite diagonal cut yields the interference contribution to Compton scattering. }
\end{figure}
\begin{figure}[t]
\includegraphics[scale=1.13]{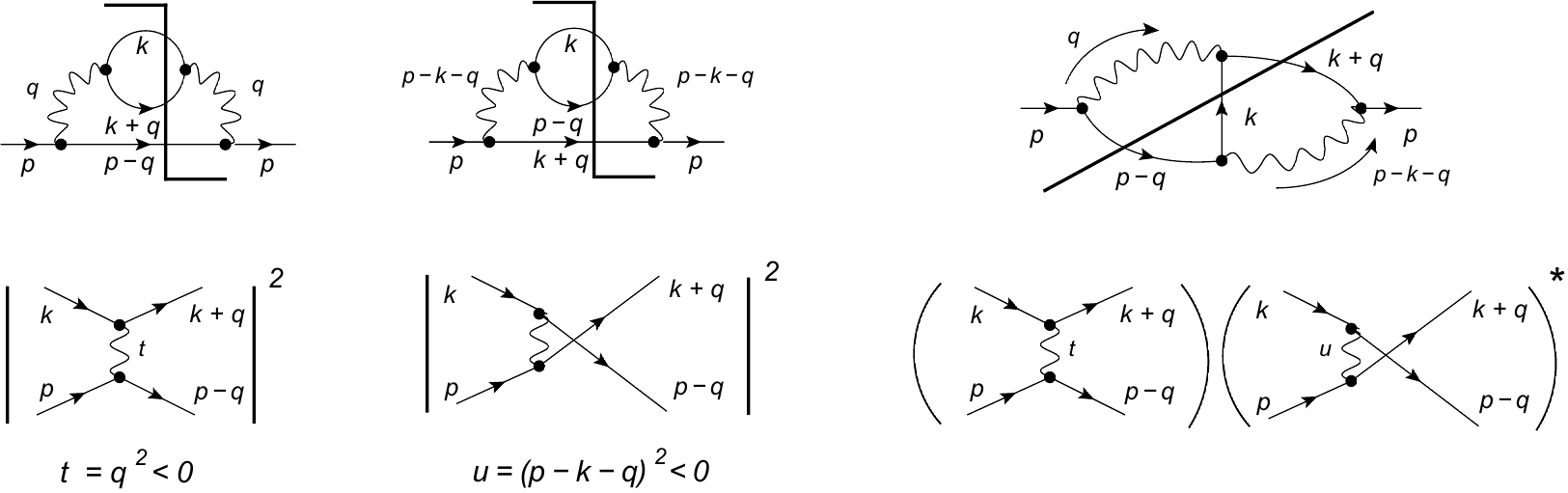} \\[2ex]
\setlength{\belowcaptionskip}{-40pt}
\caption{\label{fig:scatter} Processes contributing to Moeller scattering originate from (central) cuts of diagram (a) and (c) in Fig. \ref{fig:2loop}. Both, $t$ and $u$ channel matrix elements, correspond to cuts of diagram (a), the interference contribution can be extracted from diagram (c), as indicated. The exchanged four-momenta are conventionally labeled according to the Mandelstam variables $t=(p-p^\prime)^2=(k^\prime-k)^2$ and $u=(p-k^\prime)^2=(p^\prime-k)^2$, and are space-like in both channels. Note, that according to the unitarity rules the right hand side of each cut corresponds to the complex conjugate amplitude, which differs from the left hand side in that the momentum flow is reversed. In addition to the $M_t\,M^*_u$ interference term, there is consequently the term $M^*_t\,M_u$ which can be obtained from diagram (c) upon reversing all arrows in the loop.} 
\end{figure}
\begin{figure}[t]
\includegraphics[scale=1.15]{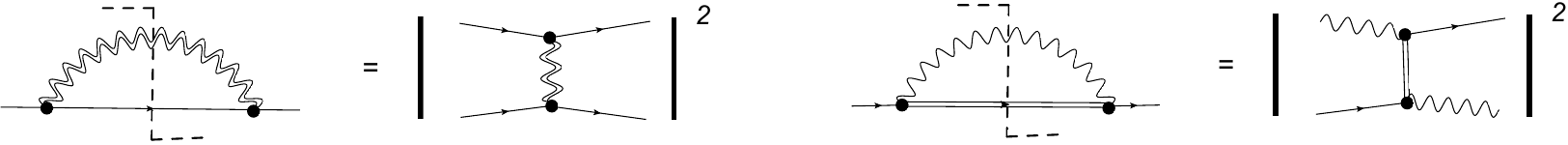}
\caption{\label{fig:ResumScatt} Left: Moeller scattering, obtainable from cuts of the fermion self-energy with a dressed a photon propagator. Right: Compton scattering, obtainable from cuts of the fermion self-energy with a dressed fermion propagator.}
\end{figure}

\noindent  The scattering rate $\Gamma$ of a fermion immersed in a QED plasma can either be calculated directly, i.e., according to Fermi`s golden rule, or via the optical theorem, which relates $\Gamma$ to the imaginary part of the fermion self energy.  \textit{On-shell} fermions receive no contributions from their one-loop self energy, processes such as fermionic Landau damping are kinematically forbidden. At two-loop level there are three diagrams contributing to the self energy, see Fig. \ref{fig:2loop}. The optical theorem relates the corresponding imaginary parts to rates of elementary $2\rightarrow2$ scattering processes in QED, namely Moeller scattering and Compton scattering. The imaginary parts can be calculated using cutting rules (or unitarity rules, see e.g., Refs. \cite{Kobes:1985kc} \cite{Das:1997gg}): Moeller and Compton scattering are obtainable from central cuts of diagrams (a) and (b) respectively. As both processes may occur in more than one channel, a complete description needs to account for interference terms. These can be extracted from diagram (c), which allows for two distinct central cuts, one contributing to Moeller scattering and one contributing to Compton scattering. \newline
Moeller scattering is the dominant process leading to energy loss of fermions in cold and dense matter, and its study is the main focus of this article. If the fermions engaging in the scattering are \textit{identical}, there are indeed two channels available:  direct (t-channel) scattering, and exchange (u-channel) scattering, see Fig. \ref{fig:scatter}. The total matrix element squared, $|M_t+M_u|^2$, thus contains the interference terms $M_t\,M_u^*$ and $M_u\,M_t^*$, which can be obtained from diagram (c), cutting diagonally from the bottom left to the top right corner.  
\newline
In the following I outline the essential steps in the calculation of the scattering rate using a \textit{resummed} photon propagator. For simplicity, we shall consider a dense plasma composed of a single fermion species. One may object that a single component plasma at large chemical potential fails the requirement of charge neutrality. Leaving this issue aside for a moment, the present discussion should be viewed as a mere preparation for the multi-component case introduced in the next section. The starting point is the \textit{retarded} self energy of on-shell fermions, $\Sigma^+_R$, which is obtained by projecting $\Sigma_R$ onto positive energy states\footnote{To calculate the scattering rate of antiparticles one proceeds analogous \cite{Manuel:2000mk}. In the present context antiparticles can safely be ignored.} 
\be\label{eq:SigmaPlus} 
\Sigma^+=\frac{1}{2}\text{Tr}\left[\Lambda^+_{\boldsymbol{p}}\,\gamma_0\,\Sigma(p_0,\,\boldsymbol{p})\right]=\frac{1}{4\,\epsilon_{\boldsymbol{p}}}\text{Tr}\left[\left(\slashed{p}+m\right)\,\Sigma(p_0,\,\boldsymbol{p})\right]\,.
\ee
Positive and negative energy projectors are given by 
\begin{equation}
\Lambda_{\boldsymbol{p}}^{\pm}=\frac{1}{2}\left(1\pm\gamma_{0}\frac{\boldsymbol{\gamma}\cdot\boldsymbol{p}+m}{\epsilon_{\boldsymbol{p}}}\right)\,,\label{eq:LambdaPM}
\end{equation}
with the usual relativistic dispersion relation  $\epsilon_{\boldsymbol{p}}=\sqrt{\boldsymbol{p}^2+m^2}$.
 Note that the fermion self energy is in general not gauge invariant, except for when it is evaluated  on the fermion mass-shell. The interaction rate in turn is related to the imaginary part of the retarded self energy via \cite{Weldon:1983jn}
\be \label{eq:OTheorem}
\Gamma(|\boldsymbol{p}|)=-2\,\text{Im}\,\Sigma^+_R=-\frac{1}{2\,\epsilon_{\boldsymbol{p}}}\text{Tr}\left[\left(\slashed{p}+m\right)\,\text{Im}\,\Sigma_R(p_0=\epsilon_{\boldsymbol{p}},\,\boldsymbol{p})\right]\,.
\ee
The above expression corresponds to the \textit{total} rate, adding the contributions of particles and holes. If one is specifically interested in the interaction rate of the former or latter, Eq. \ref{eq:OTheorem} needs to be multiplied by $1-n_f(\epsilon_{\boldsymbol{p}})$ or $n_f(\epsilon_{\boldsymbol{p}})$ respectively. At strictly zero temperature Eq. \ref{eq:OTheorem} thus corresponds to the scattering rate of holes for $\epsilon_{\boldsymbol{p}}<\mu$, and to the scattering rate of particles for $\epsilon_{\boldsymbol{p}}>\mu$. A generic expression for the imaginary part of the  retarded \textit{one loop} fermion self energy is derived in Appendix
\ref{sec: RTFcalc}, and reads 
\be 
\text{Im}\,\Sigma_{R}(p) = -\frac{e^{2}}{4}\int\frac{d^{4}q}{(2\pi)^{2}}\,\text{I}_{DB}(q_0,\,p_0)\,\gamma_{\mu}\left(\slashed{p}^{\prime}+m\right)\gamma_{\nu}\,\delta(p^{\prime2}-m^{2})\,\delta(q^{2})\,G^{\mu\nu}(q)\,,\label{eq:ImSigma}
\ee 
with $p^\prime=p-q$ as in Fig. \ref{fig:2loop}(a). In the following we shall assume that we are always interested in retarded self energies and drop the index "R". The above expression includes the detailed balance factor 
\begin{equation}\label{eq:IDB}
\text{I}_{DB}(q_0,\,p_0)=\text{sign}(p_{0}^{\prime})\left[1+2n_{b}(q_{0})\right]+\text{sgn}(q_{0})\left[1-2N_{f}(p_{0}^{\prime})\right]\,,
\end{equation}
where the Fermi distribution covering particles and antiparticles is defined as
\begin{equation}
N_{f}(p_{0})=n_{f}(p_{0}-\mu)\,\Theta(p_{0})+n_{f}(p_{0}+\mu)\,\Theta(-p_{0})\,.\label{eq:FermiDistr}
\end{equation}
Calculations are carried out in Coulomb gauge, subject to the gauge fixing condition $\boldsymbol{\nabla}\cdot A=0$. The gauge fixing dependent factor $G^{\mu\nu}$ in Eq. \ref{eq:ImSigma} reads
\begin{equation}
G^{\mu\nu}(q)=\frac{q^{2}}{\boldsymbol{q}^{2}}\, g^{\mu0}g^{\nu0}+\delta^{\mu i}\delta^{\mu j}(\delta_{ij}-\hat{q}_{i}\hat{q}_{j})\,.\label{eq:PhotonGauge}
\end{equation}
The imaginary parts of diagrams corresponding to Compton scattering and Moeller scattering can be obtained from Eq. \ref{eq:ImSigma} by replacing the bare spectral function of the photon or fermion with dressed ones, i.e., by making the replacements
\be \delta(q^{2})\,\,G^{\mu\nu}(q)\rightarrow \rho_L\,P_L^{\mu\nu}+\rho_\perp\,P_\perp^{\mu\nu},\hspace{1cm} \left(\slashed{p}^{\prime}+m\right)\,\delta(p^{\prime\,2}-m^2)\rightarrow\gamma_{0}\,\rho_{0}+\boldsymbol{\gamma}\cdot\hat{\boldsymbol{p}}^\prime\,\rho_{\boldsymbol{p}^\prime}+\rho_m\,.
\ee 
\noindent Diagrams (a) and (b) in Fig. \ref{fig:2loop} leading to tree-level processes are obtained by dressing the internal photon or fermion propagator with a single self-energy insertion. Screening is taken into account, if the self energy insertions are resummed, leading to the scattering processes depicted in Fig. \ref{fig:ResumScatt}.  Note, that self energy insertions cannot produce diagram (c), not even at tree level. Interference contributions are consequently not included in the RPA resummation program, at least not in its simplest realization (i.e., in the Hartree approximation). To generate diagram (c) one needs to include  vertex corrections in $\Sigma$, which, as mentioned in the introduction, can be neglected if one is interested in the dynamics of hard momentum fermions. In Ref. \cite{Shternin:2008es} interference contributions to electron and muon scattering in neutron star cores are computed from Fermi`s golden rule, using bare vertices but a screened photon propagator. The obtained results are small compared to pure t- and u-channel contributions. The exact relationship between vertex corrections and screening effects in the various scattering channels is an interesting question which warrants further studies. \newline
To proceed with the calculation of the self-energy on the left of Fig. \ref{fig:ResumScatt} we require the resummed photon spectrum. Employing the random phase approximation, longitudinal and transverse components read (in Coulomb gauge)
\bea
\rho_{L}(q) & = & -\frac{1}{\pi}\frac{\text{Im}\,\text{\ensuremath{\Pi}}^{00}}{(\text{Re}\,\Pi^{00} - \boldsymbol{q}^{2})^{2}+(\text{Im}\,\Pi^{00} )^{2}} + Z_L(q_0=\omega_L)\,\delta\left(\text{Re}\,\Pi^{00} -\boldsymbol{q}^2 \right)\,,  \label{eq:SpecL}\\[3ex]
\rho_{\perp}(q) & = & -\frac{1}{\pi}\frac{\text{Im}\,\Pi_\perp }{(\text{Re}\,\Pi_\perp -q^{2})^{2}+(\text{Im}\,\Pi_\perp )^{2}} + Z_\perp(q_0=\omega_\perp)\,\delta\left(\text{Re}\,\Pi_\perp-q^2\right)\,,\label{eq:SpecPerp}\vspace{5cm}
\eea
where $\Pi_{00}$ and $\Pi_\perp=(\delta_{ij}-\hat{q}_i\hat{q}_j)\Pi_{ij}$ are the longitudinal and transverse photon polarization tensor. A detailed review of the photon spectrum in degenerate matter can be found in Ref. \cite{Stetina:2017ozh}. Expressions \ref{eq:SpecL} and \ref{eq:SpecPerp} each consist of a continuum contribution and a $\delta$ function corresponding to the energies of photons and plasmons\footnote{The delta distributions included in Eqs. \ref{eq:SpecL} and \ref{eq:SpecPerp} indicate that on-shell photons and plasmons are undamped. This is obviously an artifact of the one-loop resummation of the photon spectrum, which only captures imaginary parts due to Landau damping, and, at much higher energies, pair creation. Compton scattering and inverse
Bremsstrahlung, which appear at two-loop order in QED, potentially fill this gap.}. The functions $Z_{L,\perp}$ represent the residua of the propagator evaluated at the poles $p_0=\omega_{L,\,\perp}$. Since the kinematics of Moeller scattering require the intermediate photons to carry space-like momenta, the poles located in the time-like region do not contribute to the calculation of scattering rate. The continuum contribution in the space-like region corresponds to Landau damping, i.e., the scattering of soft photons with hard fermions thermalized in the plasma. 
\newline Equipped with the spectral function of the photon we may put the scattering rate Eq. \ref{eq:OTheorem} together. To match the momentum labels of final and initial states indicated in Figs. \ref{fig:2loop} and \ref{fig:scatter}, we attribute the momentum $p^\prime=p-q$ to the fermion propagator in the self energy Eq. $\ref{eq:ImSigma}$, and the momenta $k$ and $k^\prime=k+q$ to the fermion propagators in the photon polarization tensor. Ignoring anti-particles Eq. \ref{eq:OTheorem} becomes 
\be 
\Gamma(\epsilon_{\boldsymbol{p}})  =  -\frac{e^{2}}{4\epsilon_{\boldsymbol{p}}}\int\frac{d^{4}q}{(2\pi)^{2}}\frac{1}{2\epsilon_{\boldsymbol{p}^{\prime}}}\delta(\epsilon_{\boldsymbol{p}}-\epsilon_{\boldsymbol{p}^{\prime}}-q_{0})\left[1+2n_{b}(q_{0})+\text{sgn}(q_{0})(1-2n_{f}^{-}(p_{0}^{\prime}))\right]\left[\rho_{L}(q)\,g^{\mu0}g^{\nu0}+\rho_{\perp}(q)\,P_{\perp}^{\mu\nu}\right]T_{\mu\nu}\,,
\ee
where the trace reads 
\be
 T_{\mu\text{\ensuremath{\nu}}}=4\left[p_{\mu}p_{\nu}^{\prime}+p_{\mu}^{\prime}p_{\nu}-g_{\mu\nu}(p\cdot p^{\prime}-m^{2})\right]\,,
\ee
and is to be evaluated at $p_0=\epsilon_{\boldsymbol{p}}$, and  $p_0^\prime=\epsilon_{\boldsymbol{p}^\prime}$. Note that a negative energy transfer $q_0<0$ corresponds to the inverse rather than the direct process. In this case $n_b(-q_0)=-[1+n_b(q_0)]$, such that the detailed balance factor obtains an overall negative sign. This sign is compensated by the spectral function of the photon which is an odd function of $q_0$. In a fully degenerate plasma, direct and inverse processes correspond to the scattering rates of particles and holes respectively: a direct process involves a particle with energy $\epsilon_{\boldsymbol{p}}>\mu$, which scatters on a constituent of the Fermi sea with energy $\epsilon_{\boldsymbol{k}}<\mu$. As a result of Pauli blocking, both particles occupy final states above the Fermi surface, i.e, $\epsilon_{\boldsymbol{k}^\prime}>\mu$, $\epsilon_{\boldsymbol{p}^\prime}>\mu$. The energy transfer $q_0=\epsilon_{\boldsymbol{k}^\prime}-\epsilon_{\boldsymbol{k}}=\epsilon_{\boldsymbol{p}}-\epsilon_{\boldsymbol{p}^\prime}$ is positive, and a maximum of $q_0=\epsilon_{\boldsymbol{p}}-\mu$ can be transferred. To picture the inverse process one may think of a hole with energy $\epsilon_{\boldsymbol{p}}<\mu$, which is being filled by a particle with energy $\epsilon_{\boldsymbol{p}^\prime}$ "falling" into it from above in the Fermi sea, whereby it emits a virtual photon and leaves behind a hole in the final state with energy $\epsilon_{\boldsymbol{p}^\prime}$.  The energy difference $\epsilon_{\boldsymbol{p}}-\epsilon_{\boldsymbol{p}^\prime}<0$ is transferred to the state $\epsilon_{\boldsymbol{k}^\prime}$, which is extracted from the Fermi sea. The roles of final and initial states are consequently interchanged, which is reflected in the detailed balance factor, see Eqs . \ref{eq:FermiRateP} and \ref{eq:FermiRateH} below. \newline 
After contracting the trace with longitudinal and transverse projectors, the final results for the rates are
\begin{eqnarray}\label{eq:GammaL}
\Gamma_{L}(\epsilon_{\boldsymbol{p}}) & = & \frac{e^{2}}{2}\,\int\frac{d^{3}\boldsymbol{q}}{(2\pi)^{2}}\,\rho_{L}(\epsilon_{\boldsymbol{p}}-\epsilon_{\boldsymbol{p}^{\prime}},\,\boldsymbol{q})\,\left[1+n_{b}(\epsilon_{\boldsymbol{p}}-\epsilon_{\boldsymbol{p}^{\prime}})-n_{f}^{-}(\epsilon_{\boldsymbol{p}^{\prime}})\right]\,\left(1+\frac{\epsilon_{\boldsymbol{p}}^{2}-\boldsymbol{p}\cdot\boldsymbol{q}}{\epsilon_{\boldsymbol{p}}\,\epsilon_{\boldsymbol{p}^{\prime}}}\right)\,,\\
\nonumber \\
\Gamma_{\perp}(\epsilon_{\boldsymbol{p}}) & = & e^{2}\,\int\frac{d^{3}\boldsymbol{q}}{(2\pi)^{2}}\,\rho_{\perp}(\epsilon_{\boldsymbol{p}}-\epsilon_{\boldsymbol{p}^{\prime}},\,\boldsymbol{q})\,\left[1+n_{b}(\epsilon_{\boldsymbol{p}}-\epsilon_{\boldsymbol{p}^{\prime}})-n_{f}^{-}(\epsilon_{\boldsymbol{p}^{\prime}})\right]\,\left(1+\frac{\boldsymbol{p}_{\perp}^{2}-\epsilon_{\boldsymbol{p}}^{2}+\boldsymbol{p}\cdot\boldsymbol{q}}{\epsilon_{\boldsymbol{p}}\,\epsilon_{\boldsymbol{p}^{\prime}}}\right)\,.\label{eq:GammaP}
\end{eqnarray}
Note the global factor of $2$ in $\Gamma_\perp$, which stems from the two transverse polarizations of the photon. It is an instructive exercise to obtain the tree level rate as the leading order term in an $\alpha_f$ expansion of the resummed result. A detailed derivation of the results below can be found in Appendix \ref{sec:scatrate}. In the context of degenerate matter tree-level rates correspond to the scattering of fermions far away from the Fermi surface, governed by the exchange of hard-momentum photons\footnote{In this case the Rutherford singularity is unscreened, leading to divergent results for $\Gamma_\perp$ and $\Gamma_L$. To carry out the momentum integration one may follow the approach of Braaten and Yuan \cite{Braaten:1991dd}, and introduce a cutoff scale $q^*$ with $e k_f\ll q^* \ll k_f$, which separates the soft region (where medium effects are essential) from the hard region. Adding soft and hard contributions one finds that the $q^*$ dependence drops out \cite{LeBellac:1996kr}. The calculations in this article are based on the full one-loop resummation, and consequently cover both regions automatically. The introduction of an intermediate scale is not necessary.}. To expand $\rho_\perp$ (for hard momenta the photon is purely transverse) we briefly indicate the $e^2$ dependence of the polarization tensor $\Pi$ explicitly. To leading order one finds
\be \label{eq:rhoExpand}
\rho_{\perp}(q) = - \frac{1}{\pi}\frac{e^2\,\text{Im}\,\Pi_\perp}{(e^2\,\text{Re}\,\Pi_\perp-q^{2})^{2}+(e^2\,\text{Im}\,\Pi_\perp)^{2}}\sim-\frac{e^2}{\pi}\frac{1}{q^{2}}\,\,\text{Im}\,\,\Pi_\perp\,\frac{1}{q^{2}}+\mathcal{O}(e^4)\,.
\ee
Each factor of $1/q^2$ represents a free photon propagator. Plugging Eq. \ref{eq:rhoExpand} into the imaginary part of the fermion self energy yields the imaginary part of the two-loop diagram Fig. \ref{fig:2loop}(a).  
Note, that it is imperative to include the factors $1-n_f(\epsilon_{\boldsymbol{p}})$ and $n_f(\epsilon_{\boldsymbol{p}})$ to obtain the correct detailed balance relation of particle and hole scattering respectively. After a variable transformation and a reorganization of the thermal distribution functions one finds the tree-level (order $e^4$) rate of particles,
\begin{equation}
\Gamma_p(\epsilon_{\boldsymbol{p}})=\frac{1}{2\epsilon_{\boldsymbol{p}}}\,\int_{k}\,\frac{1}{2\epsilon_{\boldsymbol{k}}}\,n_{f}^{-}(\epsilon_{\boldsymbol{k}})\,\int_{k^{\prime}}\,\frac{1}{2\epsilon_{\boldsymbol{k}^\prime}}\left[1-n_{f}^{-}(\epsilon_{\boldsymbol{k}^{\prime}})\right]\,\int_{p^{\prime}}\,\,\frac{1}{2\epsilon_{\boldsymbol{p}^\prime}}\,\left[1-n_{f}^{-}(\epsilon_{\boldsymbol{p}^{\prime}})\right]\,\left(2\pi\right)^{4}\delta(p+k-p^{\prime}-k^{\prime})\,\left|M\right|^{2}\,,\label{eq:FermiRateP}
\end{equation}
and holes,
\begin{equation}
\Gamma_h(\epsilon_{\boldsymbol{p}})=\frac{1}{2\epsilon_{\boldsymbol{p}}}\,\int_{k}\,\frac{1}{2\epsilon_{\boldsymbol{k}}}\,\left[1-n_{f}^{-}(\epsilon_{\boldsymbol{k}})\right]\,\int_{k^{\prime}}\,\frac{1}{2\epsilon_{\boldsymbol{k}^\prime}}\,n_{f}^{-}(\epsilon_{\boldsymbol{k}^{\prime}})\,\int_{p^{\prime}}\,\frac{1}{2\epsilon_{\boldsymbol{p}^\prime}}\,n_{f}^{-}(\epsilon_{\boldsymbol{p}^{\prime}})\,\left(2\pi\right)^{4}\delta(p+k-p^{\prime}-k^{\prime})\,\left|M\right|^{2}\,,\label{eq:FermiRateH}
\end{equation}
with the standard short-hand notation for the integration measure,
\be 
\int_{p}=\int\frac{d^{3}\boldsymbol{p}}{(2\pi)^{3}}\,.\label{eq:measure}
\ee
The squared matrix element $|M|^2$ emerges from the product of the traces included in $\Gamma$ and $\Pi$, and corresponds to $t$-channel or $u$-channel scattering, i.e.,  $|M|^2=|M_t|^2$ or $|M|^2=|M_u|^2$ (see Eq. \ref{eq:RateCompare} in Appendix \ref{sec:scatrate} for the explicit result). The matrix element of the interference term does not factorize, and, as mentioned above, has to be extracted from the imaginary part of Fig. \ref{fig:2loop}(c). Deriving the rate according to Fermi`s golden rule from field theory serves as a useful check to ensure sure all statistical factors are accounted for correctly. The step from the bare to screened interactions is now straight-forward: writing the spectral functions Eq. \ref{eq:SpecL} and Eq. \ref{eq:SpecPerp} as $\rho_{j}=-(1/\pi)\,|D_{j}|^2\,\textrm{Im}\,\Pi_j$, where $j=\{L,\,\perp\}$ and $D_j$ is the dressed photon propagator,
\be 
D_{L}=\frac{1}{\boldsymbol{q}^2-\Pi_{00}}, \hspace{1cm}D_{\perp}=\frac{1}{q^2-\Pi_\perp}\,,
\ee
one recovers the expansion Eq. \ref{eq:rhoExpand}, except that bare propagators are replaced by dressed ones. In conclusion $\Gamma$ corresponds to the scattering rate depicted on the left hand side of Fig. \ref{fig:ResumScatt}.

\subsection{Scattering in the Multi Component Plasma}
\label{subsec:multi}
\begin{figure}[t]
\includegraphics[scale=1.15]{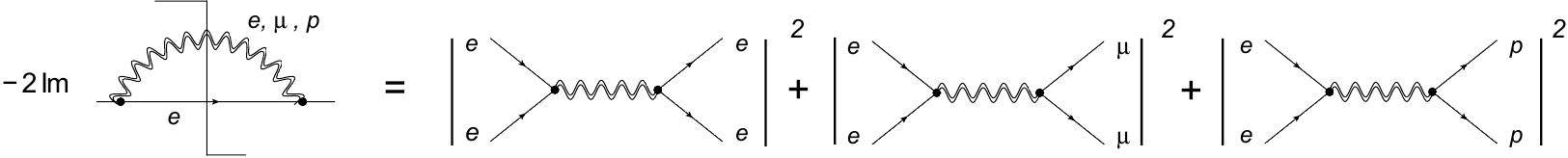}
\caption{\label{fig:ScatterMulti} Total scattering rate calculated from the self energy of electrons, propagating through nuclear matter composed of other electrons, muons, and protons. In each case the scattering occurs via the exchange of plasmons and photons, whose dispersion relations at soft momenta are strongly modified by screening and damping effects of all fermions in the plasma. The photon spectrum in the multi-component plasma is obtained from the imaginary part of the dressed photon propagator Eq. \ref{eq:PropEMP}.}
\end{figure}
\begin{figure}
\includegraphics[scale=0.8]{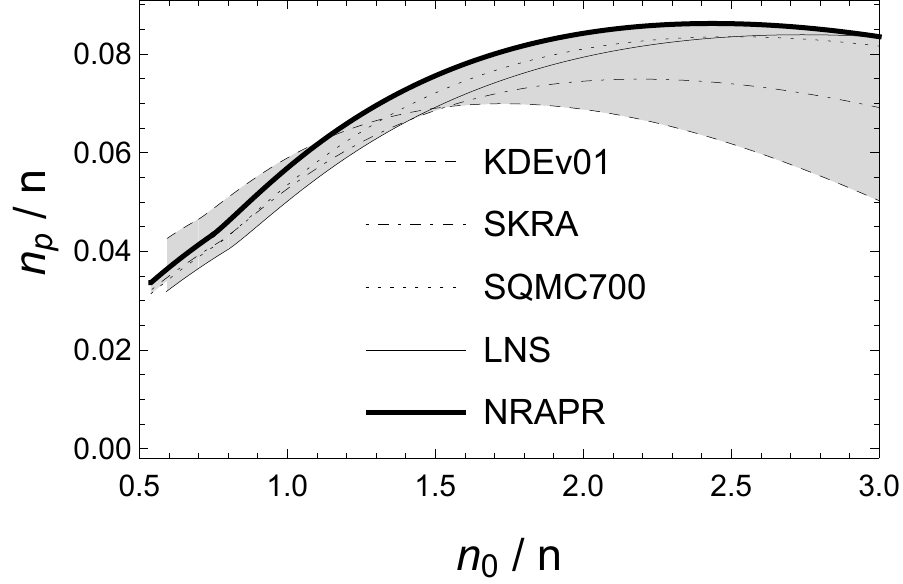}\hspace{0.8cm}\includegraphics[scale=0.58]{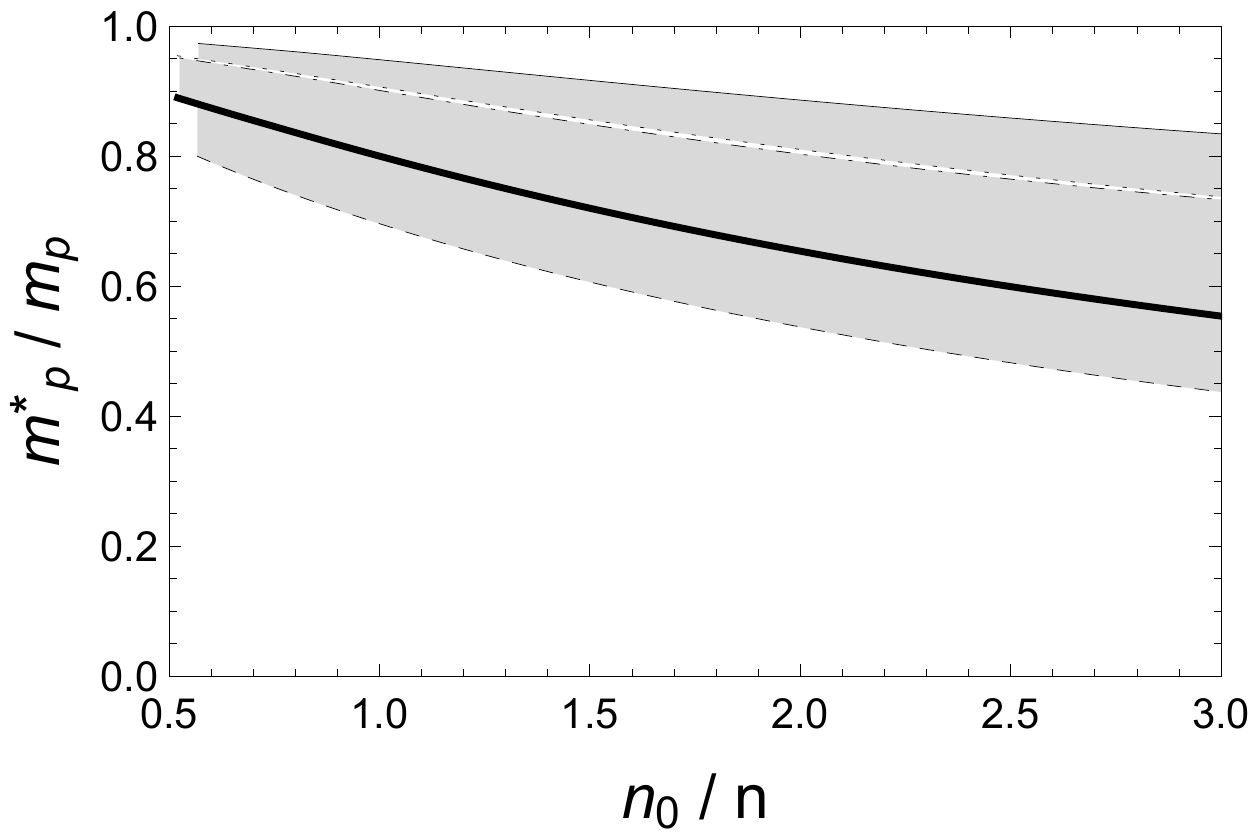}
\caption{\label{fig:Skyrmes} Comparison of proton fractions and effective masses at zero temperature, using the Skyrme parameter sets recommended in Ref. \cite{Dutra:2012mb}. Homogeneous nuclear matter is stable above a critical density $n_c$ which has to be determined separately for each parameter set. The results depicted above are consequently cut off at slightly different values at the left hand side of each plot.}
\end{figure}
\begin{figure}
\includegraphics[scale=0.6]{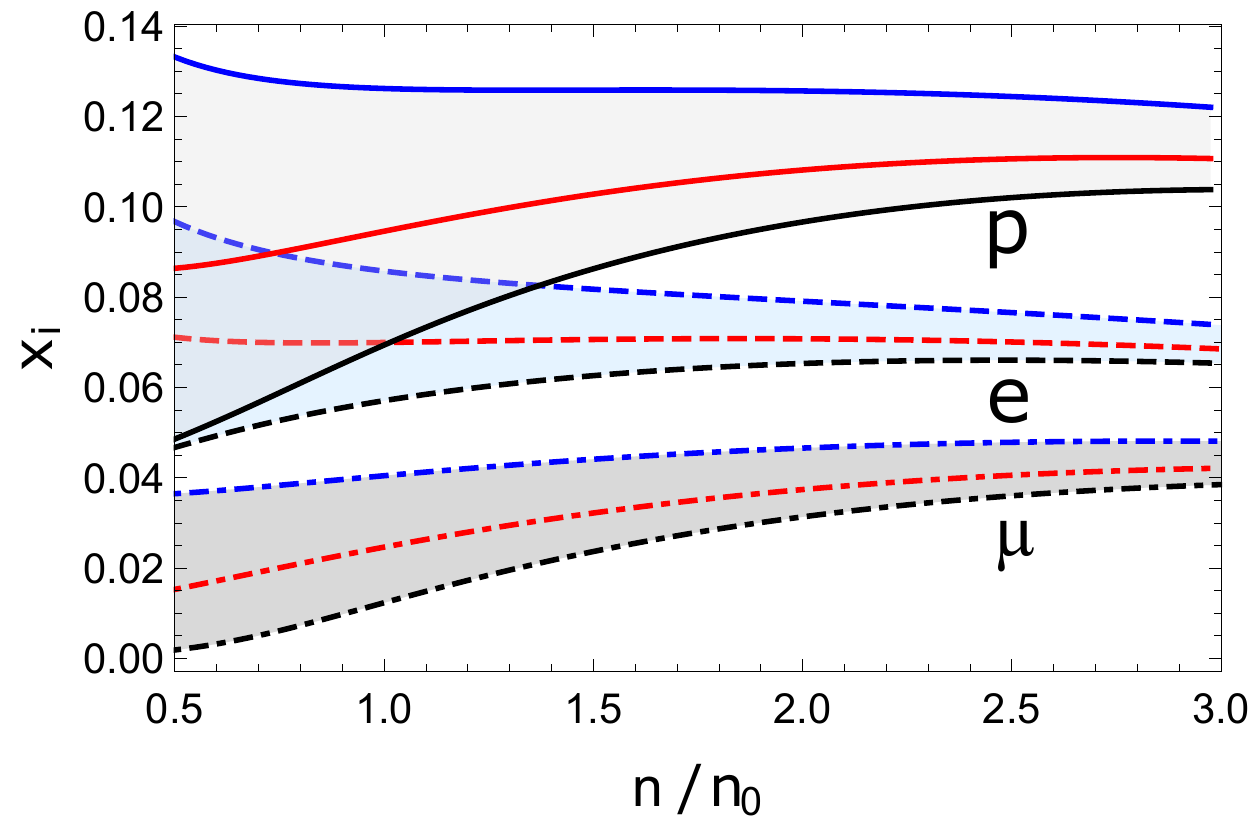}\hspace{0.8cm}\includegraphics[scale=0.6]{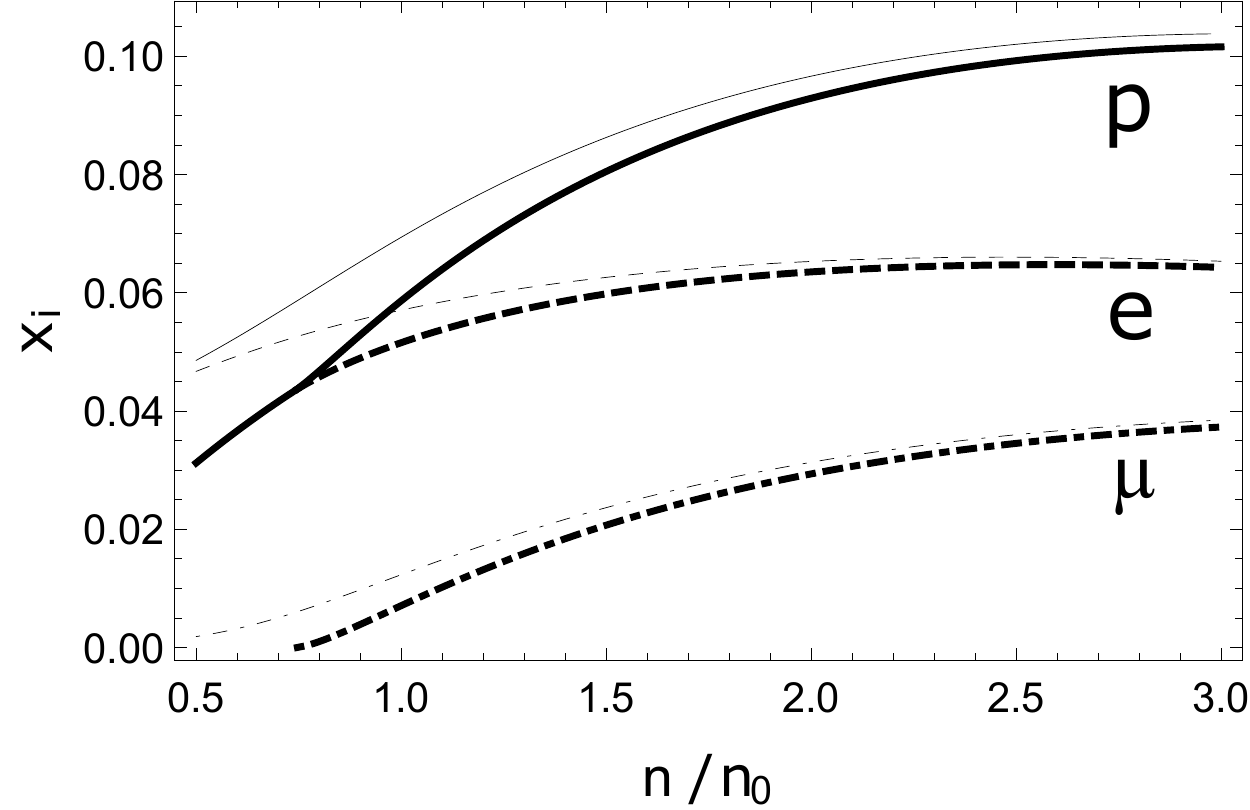}\\[2ex]
\includegraphics[scale=0.84]{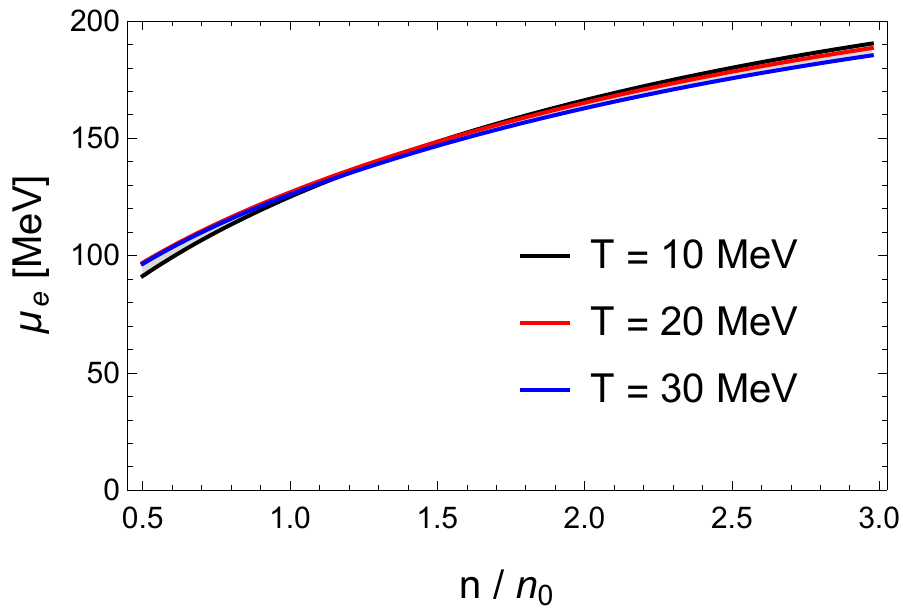}\hspace{0.8cm}\includegraphics[scale=0.6]{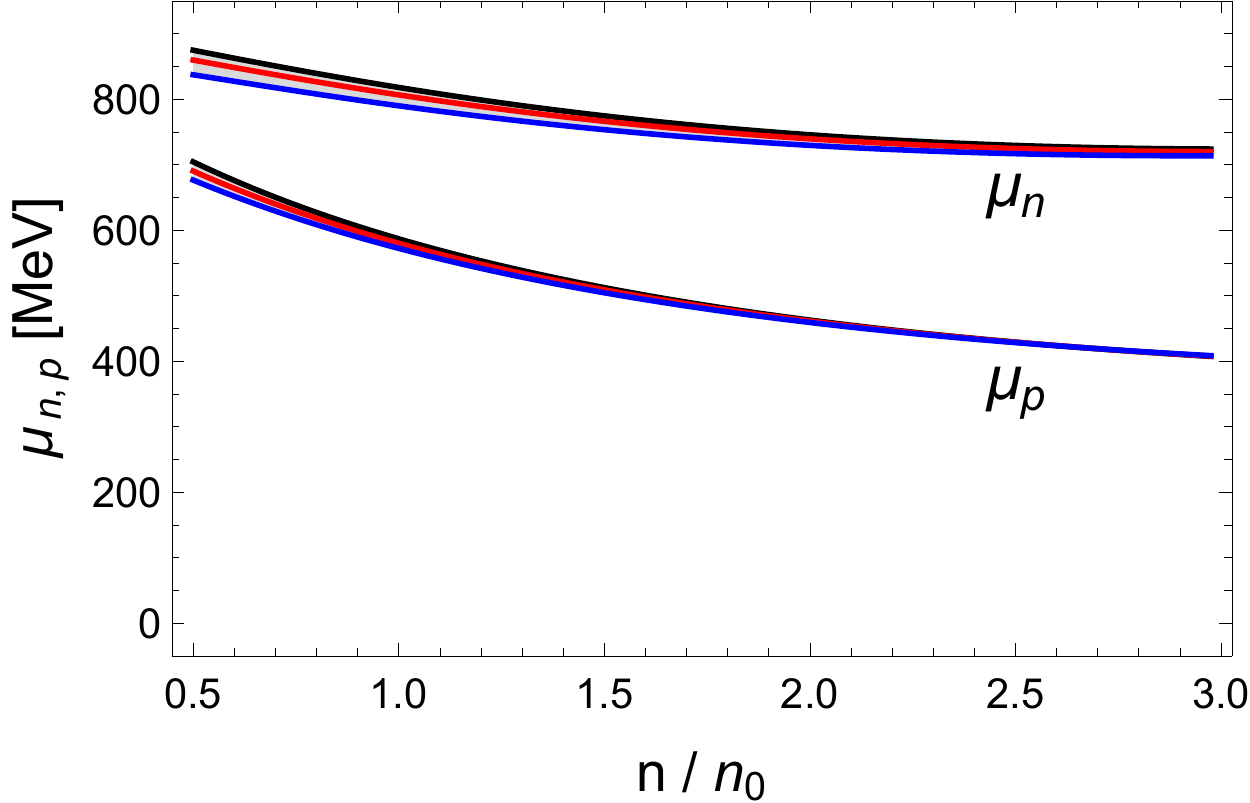}
\caption{\label{fig:FracT} Particle fractions (top) and (effective) chemical potentials (bottom) at three different temperatures $T=10,\,20,\,30$ MeV, calculated using NRAPR (Non-Relativistic Akmal, Pandharipande and Ravenhall \cite{Akmal:1998cf}) Skyrme forces. The upper right panel compares the relative abundances of protons (solid), electrons (dashed), and muons (dot-dashed) at $T=10$ MeV (gray) and $T=0$ (thick black), and shows that the finite temperature results approach the zero temperature limit with increasing density. While the particle fractions vary considerably with temperature, the impact on the chemical potentials is comparatively small. }
\end{figure}
\begin{figure}
\includegraphics[scale=0.6]{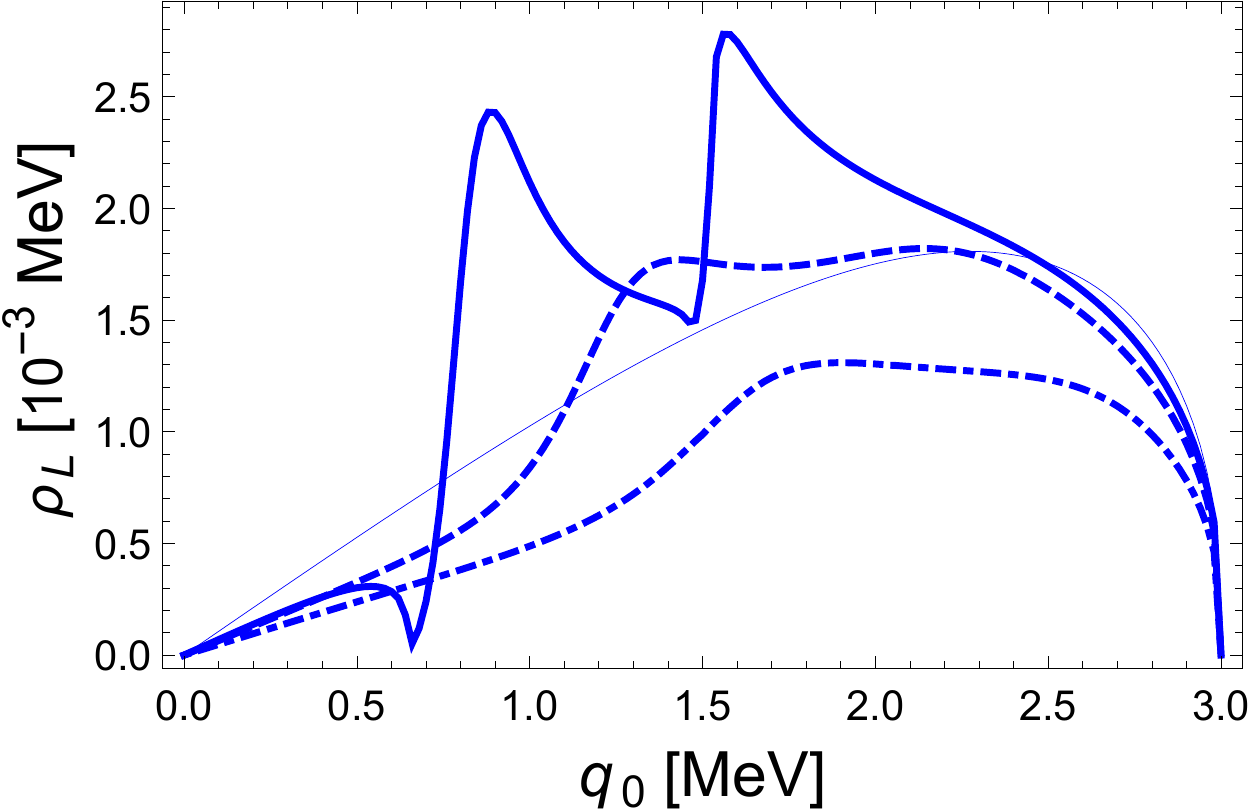}\hspace{0.8cm}\includegraphics[scale=0.82]{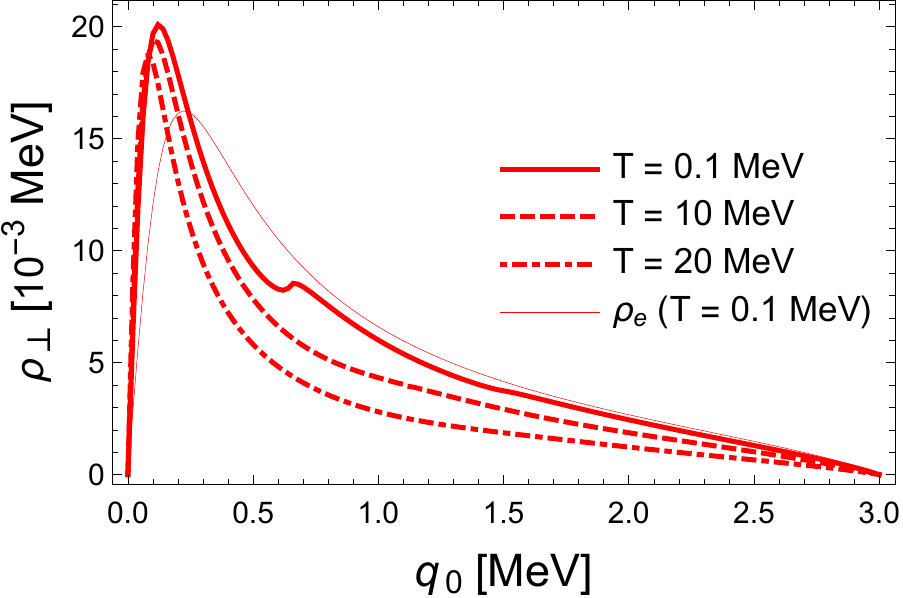}
\caption{\label{fig:RhoMulti} Longitudinal (left) and transverse (right) spectral functions in the EMP plasma at saturation density. Chemical potentials and effective masses are obtained using NRAPR skyrme forces, see Tab. \ref{tab:NRAPR}. The momentum is fixed at $|\boldsymbol{q}|$= 3 MeV, and the continuum contributions due to Landau damping are displayed as a function of the photon energy $q_0$.  The temperatures are set to $T=0.1$ MeV (solid), $T=10$ MeV (dashed), and $T=20$ MeV (dot-dashed). In addition, the thin blue and red lines display the respective spectral functions in the absence of muons and protons. Under degenerate conditions, the longitudinal spectral function displays three characteristic peaks, located roughly at the plasma frequencies $\omega_{0,\,i}$ of each particle species. The transverse spectral function is less sensitive to temperature variations.}
\end{figure}

\noindent Within the RPA it is particularly simple to extend the calculation of the scattering rate to a QED plasma composed of electrons, protons, and muons (EMP plasma): The total photon polarization tensor is simply given by the sum of the individual polarizations, i.e., $\Pi\rightarrow\Pi_e+\Pi_\mu+\Pi_p$, such that the dressed photon propagator reads
\be \label{eq:PropEMP}
D^{\mu\nu}(q)=\frac{1}{\Pi_{00,\,e}(q)+\Pi_{00,\,\mu}(q)+{\Pi}_{00,\,p}(q)-\boldsymbol{q}^{2}}\,g^{\mu0}g^{\nu0}+\frac{1}{\Pi_{\perp,\,e}(q)+\Pi_{\perp,\,\mu}(q)+\Pi_{\perp,\,p}(q)-q^{2}}\,P_{\perp}^{\mu\nu}\,.
\ee
It is easy to check that an expansion in $\alpha_f$ produces all possible combinations of one-loop insertions. The spectral functions $\rho_L$ and $\rho_\perp$ are accordingly obtained from the imaginary part of Eq. \ref{eq:PropEMP}, and can as well be expressed as a sum, $\rho_{j}=-(1/\pi)\,|D_{j}|^2\,\textrm{Im}\,(\Pi_{e,\,j}+\Pi_{\mu,\,j}+\Pi_{p,\,j})$. The results are inserted into the same expressions for $\Gamma_L$ and $\Gamma_\perp$, Eqs. \ref{eq:GammaL} and \ref{eq:GammaP}. If $\Gamma$ is computed from the self energy $\Sigma$ of, say, an electron, it may be interpreted as the sum of individual scattering rates of electrons with all fermion species present in the plasma:  $\Gamma_e=\Gamma_{e,\,e}+\Gamma_{e,\mu}+\Gamma_{e,\,p}$. In each of these channels, the screening receives contributions from all constituents, see Fig. \ref{fig:ScatterMulti}. Naturally, the same applies to the self energy of all other fermions.\newline
The evaluation of electron, muon and proton loops requires for the determination of their respective chemical potentials, and in case of the protons in addition for the determination of the effective mass $m^*_p$. Under degenerate conditions these quantities can be extracted from an energy density functional, as outlined in detail in Ref. \cite{Stetina:2017ozh}, see in particular section IIIA and Appendix B2 therein. In the following the essential steps are summarized. The properties of nuclear matter at a given density are extracted from an energy functional based on Skyrme type interactions \cite{Chamel:2006rc}, 
\begin{equation}
\mathcal{E}[n]=\sum_{T=0,1}\left[\delta_{T,0}\frac{\hbar^{2}}{2m}\tau_{T}+C_{T}^{n}[n]\,n_{T}^{2}+C_{T}^{\tau}n_{T}\,\tau_{T}+C_{T}^{\boldsymbol{j}}\,\boldsymbol{j}_{T}^{2}\right]\,,\label{eq:EpsChamel}
\end{equation}
\noindent where the coefficients $C_{T}^{ {n,\tau,\boldsymbol{j} } }$
are related to standard Skyrme parameters \cite{Chamel:2006rc}. The functional Eq. \ref{eq:EpsChamel} depends on densities $n_a$, kinetic energies $\tau_q$, and currents $\boldsymbol{j}_a$, which in turn are related the quasiparticle occupations via
\begin{equation}\label{eq:functionals}
n_{a}[n_{\boldsymbol{k,}a}]=\int_{k}n_{\boldsymbol{k},a}\,,\hspace{1cm}\tau_{a}[n_{\boldsymbol{k,}a}]=\int_{k}\boldsymbol{k}^{2}n_{\boldsymbol{k},a}\,,\hspace{1cm}\boldsymbol{j}_{a}[n_{\boldsymbol{k,}a}]=\int_{k}\boldsymbol{k}\,n_{\boldsymbol{k},a}\,,
\end{equation}
\noindent with the flavor index $a = n, p$. Isoscalar (T=0) and isovector (T=1) densities  are given by $n_{0}=n=n_{n}+n_{p}$, $n_{1}=n_{n}-n_{p}$ (and similarly for $\boldsymbol{j}$ an $\tau$). Quasiparticle dispersions, effective masses, and residual (density and current) interactions are obtained by taking the derivatives 
\be \label{eq:derivatives}
e_{\boldsymbol{k},a}  =  \frac{\delta\mathcal{E}}{\delta n_{a,\boldsymbol{k}}} = \frac{\hbar^{2}\boldsymbol{k}^{2}}{2m_{a}^{*}}+U_{a}\,,\hspace{1cm}\frac{\hbar^{2}}{2m_{a}^{*}}  :=  \frac{\delta\mathcal{E}}{\delta\tau_{a}}\,,\hspace{1cm}f_{ab}=\frac{\delta^{2}\mathcal{E}}{\delta n_{a,\boldsymbol{k}}\,\delta n_{b,\boldsymbol{k}}}\,,\hspace{1cm}\bar{f}_{ab} \,\delta_{ij} = \frac{\delta^{2}\mathcal{E}}{\delta j_{a}^{i}\,\delta j_{b}^{j}\,.}
\ee
Relations \ref{eq:derivatives} are obviously non-relativistic, and need to be matched with their relativistic counterparts before they can be incorporated into the RPA resummation. In particular the kinetic contributions to single-particle energies and chemical potentials (at zero temperature) are related via
\begin{equation}
e_{\boldsymbol{k},\,(kin,\,rel)}=\sqrt{\boldsymbol{k}^{2}+m^{*}}+m-m^{*}\,,\hspace{1cm}\mu_{(kin,\,rel)}=\sqrt{k_{f}^{2}+m^{*2}}+m-m^{*}\,,\label{eq:EpsMuRPA}
\end{equation}
where adding $\delta m=m-m^*$ ensures that the leading term in a large $m^*$ expansion consists of the \textit{bare} mass plus the non-relativistic expression. The residual quasiparticle interactions will be required for the introduction of induced interactions in the next section. In a partial wave expansion, single particle energies and density-density potentials are related to $l=0$ Landau parameters, while effective masses and current-current potentials correspond to $l=1$ Landau parameters. The latter exhibit a stronger model dependence. At zero temperature, the derivatives Eq. \ref{eq:derivatives} are evaluated at the Fermi surface., i.e. by setting $n_{a,\,\boldsymbol{k}}\rightarrow n_{a,\,0}=\Theta(k_f-|\boldsymbol{k}|)$. Strictly speaking the residual quasiparticle interactions are only valid in the static limit $f_{ab}(\boldsymbol{q}=\boldsymbol{0})$ and $\bar{f}_{ab}(\boldsymbol{q}=\boldsymbol{0})$, where $\boldsymbol{q} = \boldsymbol{k}-\boldsymbol{k^\prime}$ is the momentum transfer in the scattering of two quasi-particles. While it would certainly be desirable to study the momentum dependence of nuclear interactions, the current approach is a reasonable approximation, in particular with regard to the calculation of scattering rates which are dominated by soft momentum exchange.
To reduce the dependency on a single parameter set several modern Skyrme forces recommended in Ref. \cite{Dutra:2012mb}, including KDEv01 \cite{Agrawal:2005ix}, SKRA \cite{SKRA}, SQMC700 \cite{Goriely:2001rbd}, LNS \cite{Cao:2005ac} and NRAPR \cite{Steiner:2004fi} are employed. A comparison of proton fractions and effective masses is shown in Fig. \ref{fig:Skyrmes}.
\begin{table}[t]
\setlength{\belowcaptionskip}{-0.5cm}
\begin{tabular}{|c|c|c|c|c|c|}
\hline 
\textbf{parameter set} & NRAPR & SKRA & SQMC700 & LNS & KDE0v1\tabularnewline
\hline 
\hline 
$n_{c}\,\,[n_{0}]$ & 0.539 & 0.543 & 0.539 & 0.590 & 0.594\tabularnewline
\hline 
$n_{\mu}\,[n_{0}]$ & 0.747 & 0.772 & 0.773 & 0.801 & 0.698\tabularnewline
\hline 
\end{tabular}
\caption{\label{tab:CritDens}Critical densities for the stability of homogeneous nuclear matter and for the onset of muons at zero temperature. The results for $n_c$ are obtained from evaluating condition Eq. \ref{eq:Stability} in $\beta$ equilibrium, employing the Skyrme parameters recommended in Ref. \cite{Dutra:2012mb}. The results for $n_\mu$ are obtained from the conditions $n_\mu$=0 and $m_\mu=\mu_e$.}
\end{table}
Stable homogeneous nuclear matter requires a positive curvature of the groundstate energy density $\mathcal{E}_0$ in the space spanned by $n_{n}$ and $n_{p}$,
\begin{equation}
\frac{\partial^{2}\mathcal{E}_{0}}{\partial n_{n}^{2}}\cdot\frac{\partial^{2}\mathcal{E}_{0}}{\partial n_{p}^{2}}-\frac{\partial^{2}\mathcal{E}_{0}}{\partial n_{n}\partial n_{p}}>0\,.\label{eq:Stability}
\end{equation}
The above condition defines a critical density $n_c$, below which nuclear matter strives to be in a clustered state \cite{Baym:1971ax} \cite{Muller:1995ji} \cite{Li:1997ra}, and which has to be determined for each given model. Note that Eq. \ref{eq:Stability} does not account for electromagnetism. Doing so turns the second-order phase transition at the critical density into a first order one \cite{Lamb:1983djd}. While the nature of the phase transition is not relevant in the present context it should be stressed that the screening mass, being a second order derivative of the energy density, is in any case sensible to the instability and diverges upon approaching $n=n_c$ from above. This behavior will be essential for the discussion in the next section.\newline
At finite temperature, a rigorous evaluation of the derivatives Eq. \ref{eq:derivatives} requires for a self-consistent approach, see e.g. Ref. \cite{Rrapaj:2014yba} and \cite{Tan:2016ypx}. In a first approximation the particle fractions at a given temperature can be obtained from the usual conditions imposed by $\beta$ equilibrium and charge neutrality,
\bea \label{eq:BetaT}
\mu_n(n,x_p,T)-\mu_p(n,x_p,T)=\mu_e(n,x_e,T)=\mu_\mu(n,x_\mu,T)\,,\hspace{1cm}x_p=x_e+x_\mu\,,
\eea
where in each case $\mu$ is obtained by inverting the expressions for the currents ($a=p,\,n$, $l=e,\,\mu$),
\be \label{eq:currents}
x_a\,n=\frac{1}{\pi^2}\int d\boldsymbol{p}\,\boldsymbol{p}^2\,n_f\left[(e_{\boldsymbol{k},\,a}+U_a-\mu_a)/T\right]\,,\hspace{1cm}x_l\,n=\frac{1}{\pi^2}\int d\boldsymbol{p}\,\boldsymbol{p}^2\,n_f\left[(\sqrt{\boldsymbol{p}^2+m_l^2}-\mu_l)/T\right]\,. 
\ee 
This procedure determines one specific $x_p(T)$ for each chosen temperature. Note, that $U_a=U_a(n,\,x_p)$ and $m^*_a=m^*_a(n,\,x_p)$ vary implicitly with temperature, because they depend on $x_p$. The mean-field shift $U_a$ in the single particle energies only matters for the determination of the particle fractions $x_j$. Once they are known, the actual (effective) chemical potentials to be used in a loop calculation are obtained from the same relation Eq. \ref{eq:currents} neglecting $U_a$: as far as the RPA is concerned, nucleons are regarded as free fermions (with effective masses). Particle fractions and chemical potentials at $T=10$ MeV, $T=20$ MeV, and $T=30$ MeV are displayed in Fig. \ref{fig:FracT}. Higher temperatures tend to reduce the difference of $x_p$ and $x_n$ and of $x_e$ and $x_\mu$, in agreement with the findings of Ref. \cite{Tan:2016ypx}. With increasing density or decreasing temperature the particle fractions smoothly approach those obtained at $T=0$. The variation of the chemical potentials is fairly small for all temperatures considered. For future reference chemical potentials and effective masses for three different densities are listed in Tab. \ref{tab:NRAPR}. \newline
\noindent To conclude this subsection the space-like region of $\rho_L$ and $\rho_\perp$ are displayed in Fig. \ref{fig:RhoMulti}. In the degenerate limit $\rho_L$ displays characteristic peaks, located in the vicinity of $v_{f,\,i}\,|\boldsymbol{q}|$, where $v_{f,\,i}$ is the Fermi velocity of a given species. With increasing temperature these structures are "washed out", in agreement with the findings of Ref. \cite{Heiselberg:1993cr}, which attests particular importance to dynamical screening in degenerate matter. The three peaks in the spectrum originate from poles of the (real part of the) longitudinal photon propagator, indicating that there is not one but three plasmon modes, corresponding to the collective response of electrons, muons, and protons \cite{Stetina:2017ozh} \cite{Baldo:2008pb}. As expected, the transverse spectrum is much less structured due to its lack of static screening. For the same reason it is much larger than its longitudinal counterpart at soft energies. The characteristic landscape displayed in Fig. \ref{fig:RhoMulti} defines dynamical screening in the multi-component plasma. Note that we have ignored neutrons entirely; the propagator Eq. \ref{eq:PropEMP} includes only those fermion loops which couple directly to the photon. 
\begin{table}[]
\setlength{\belowcaptionskip}{0.5cm}
\begin{tabular}{|c|c|c|c|c|c|}
\hline 
 & $\mu_{e} = \mu_\mu$ ~{[}MeV{]}~~~  & $\mu_{p}$~ {[}MeV{]}~~~ & $\mu_{n}$~ {[}MeV{]}~~~ & $m_{p}^{*}$~ {[}MeV{]}~~~ & $m_{n}^{*}$~ {[}MeV{]}~~~\tabularnewline
\hline 
\hline 
$n=0.55\,n_{0}$ & 88 & 699 & 872 & 693 & 830\tabularnewline
\hline 
$n=0.65\,n_{0}$ & 97 & 670 & 860 & 663 & 812\tabularnewline
\hline 
$n=n_{0}$ & 122 & 589 & 819 & 575 & 752\tabularnewline
\hline 
$n=2\,n_{0}$ & 162 & 457 & 739 & 419 & 618\tabularnewline
\hline 
\end{tabular}
\caption{Chemical potentials and effective masses calculated in $\beta$ equilibrium at fixed density and zero temperature. At the lower two densities muons are absent. The parameters listed above are extracted using NRAPR Skyrme forces, and matched to their relativistic counterparts. At first glance it may seem surprising that the chemical potentials of the nucleons decrease with increasing density. The difference $\mu-m^*$, however, increases as expected.    }
    \label{tab:NRAPR}
\end{table}
\subsection{Induced interactions} \label{subsec:induced}
\begin{figure}[t]
\includegraphics[scale=0.55]{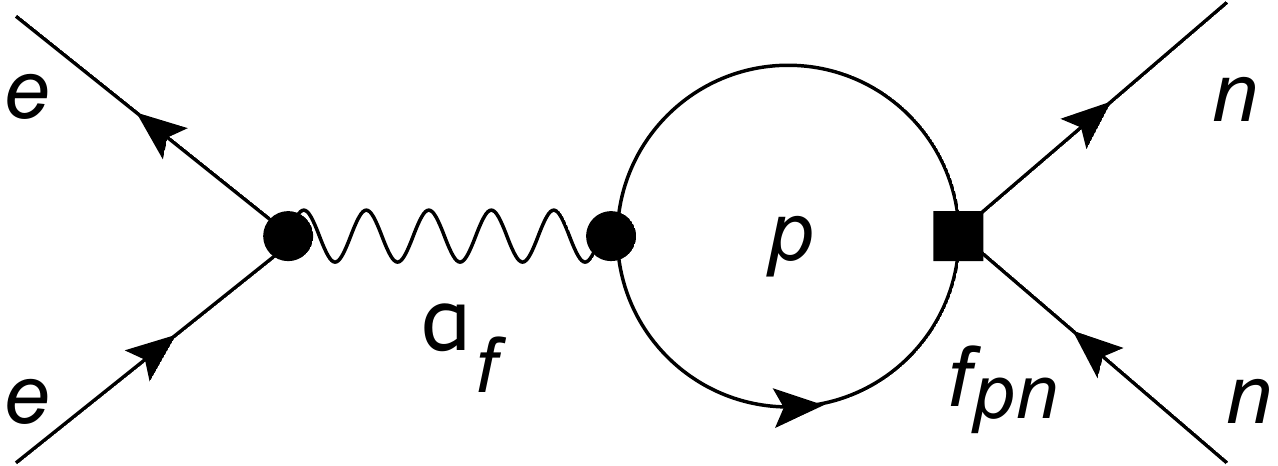}
\caption{\label{fig:inducedEN}Electron-neutron scattering induced by the polarizability of strongly and electromagnetically charged protons. Squared vertices depict strong interaction potentials. These contributions are resummed to obtain the dressed polarization tensor $\tilde{\Pi}_p$, which in turn enters the photon propagator. To connect again with a photon propagator, another proton loop has to be attached on the right hand side. The leading contribution to electromagnetic interactions is thus of order $\alpha_f^2$\,$f_{pn}^2$. }
\end{figure}

\begin{figure}
\includegraphics[scale=0.8]{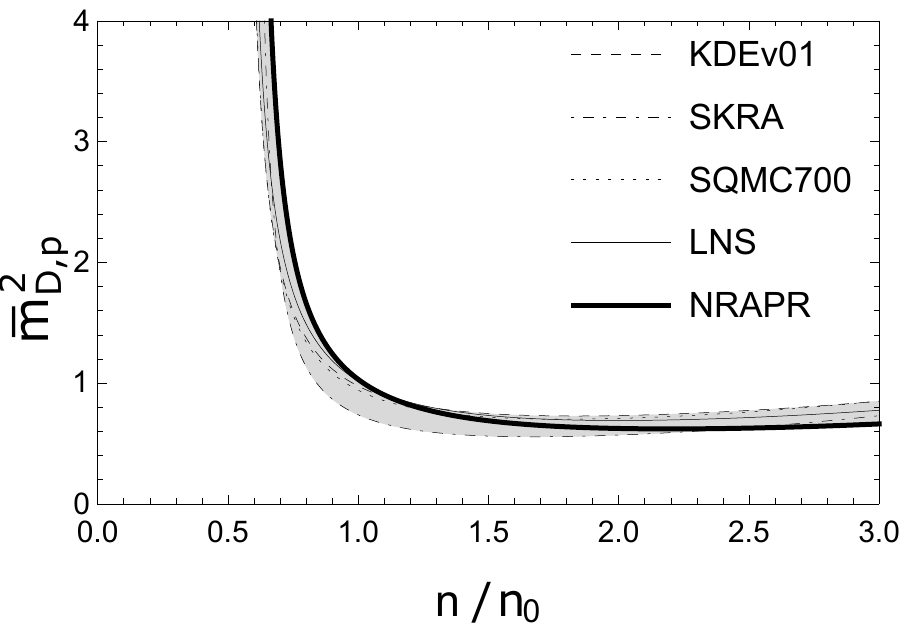}\hspace{0.8cm}\includegraphics[scale=0.6]{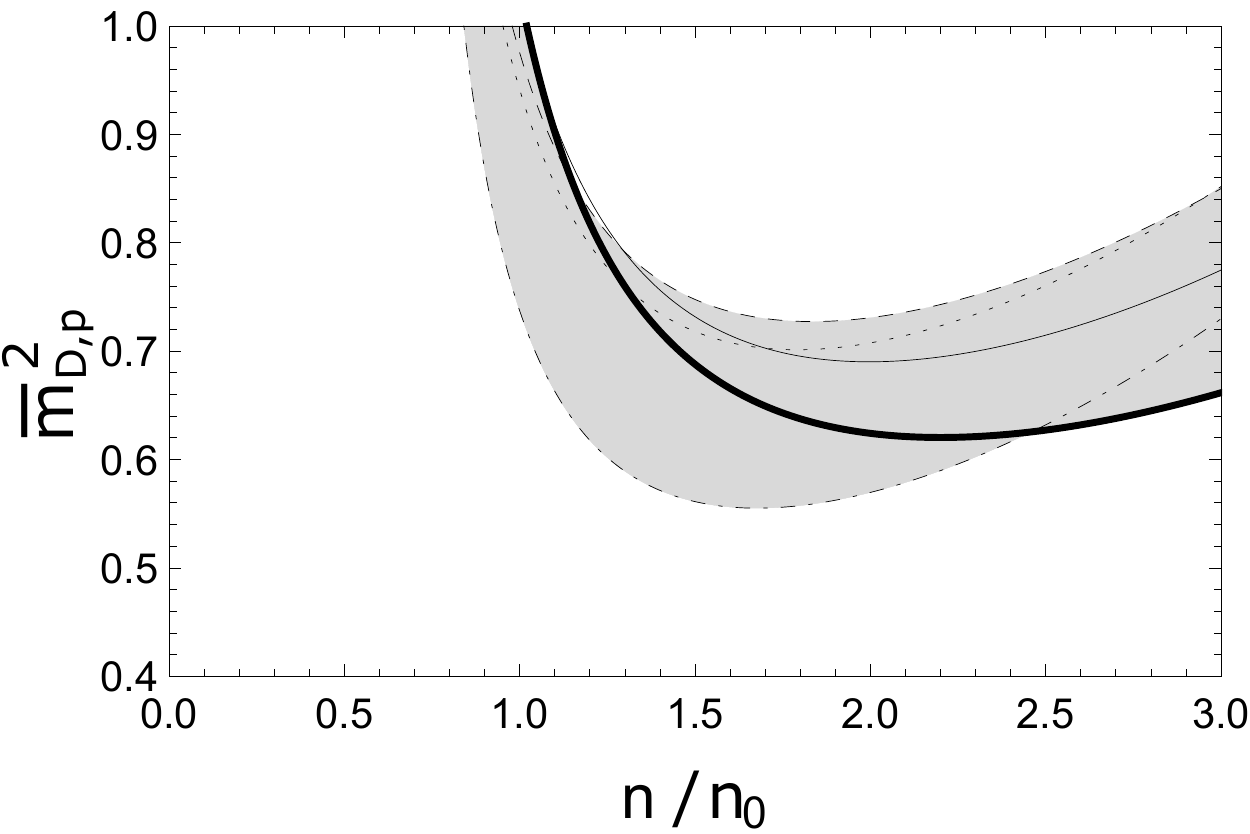}
\setlength{\belowcaptionskip}{-0.5cm}
\caption{\label{fig:Screen} Comparison of the resummed screening mass Eq. \ref{eq:DebyePN} over the one-loop expression $\bar{m}_{D,\,p}=\tilde{m}_{D,\,p}/m^\prime_{D,\,p}$ with $m^{\prime\,2}_{D,p}=\mu\,k_f/\pi^2$, using the Skyrme parameter sets recommended in Ref. \cite{Dutra:2012mb}. To better resolve the various model that constitute the gray band the right hand side displays an enlargement of the left hand side. The screening mass diverges upon approaching the critical density of homogeneous nuclear matter for above. Around $n\sim(1.5-2)\,n_0$ the ratio assumes a minimum, at $n=n_0$ it is close to $1$.}
\end{figure}
\begin{figure}
\includegraphics[scale=0.5]{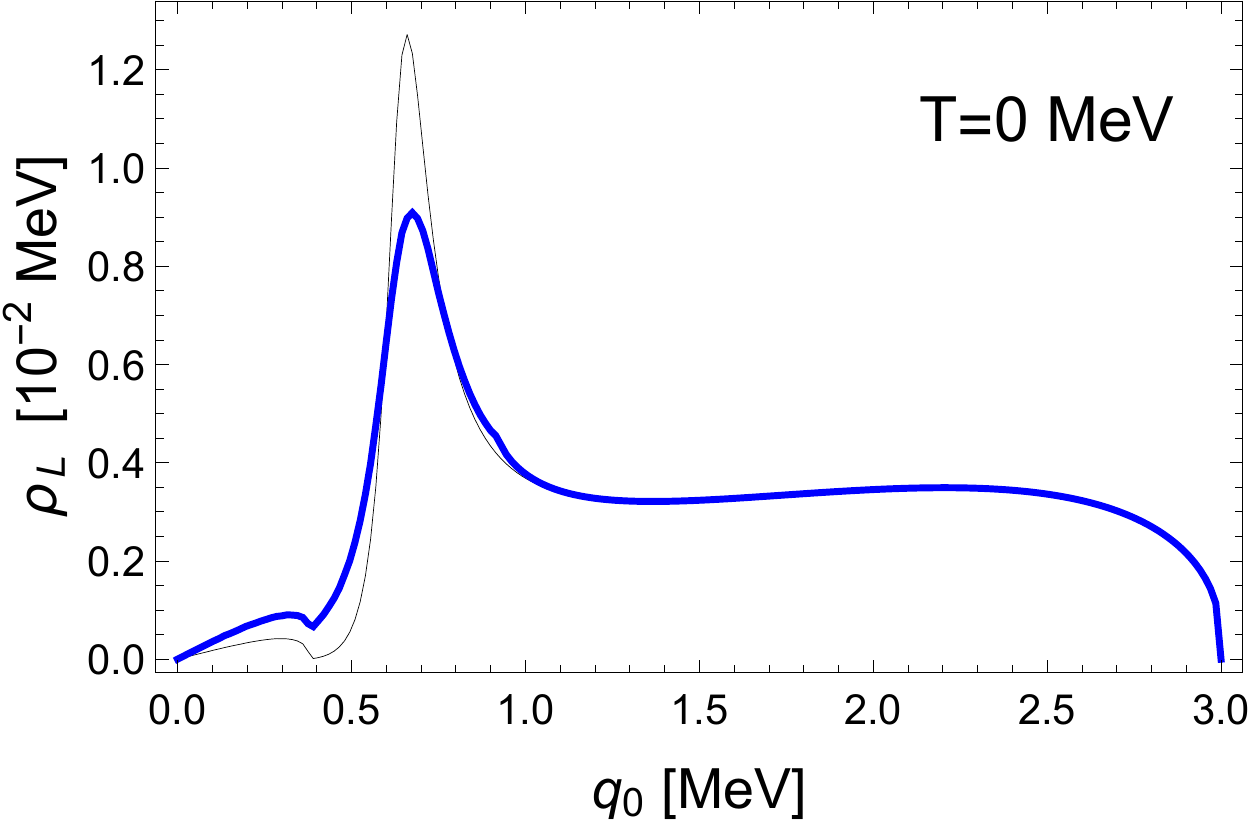}\includegraphics[scale=0.5]{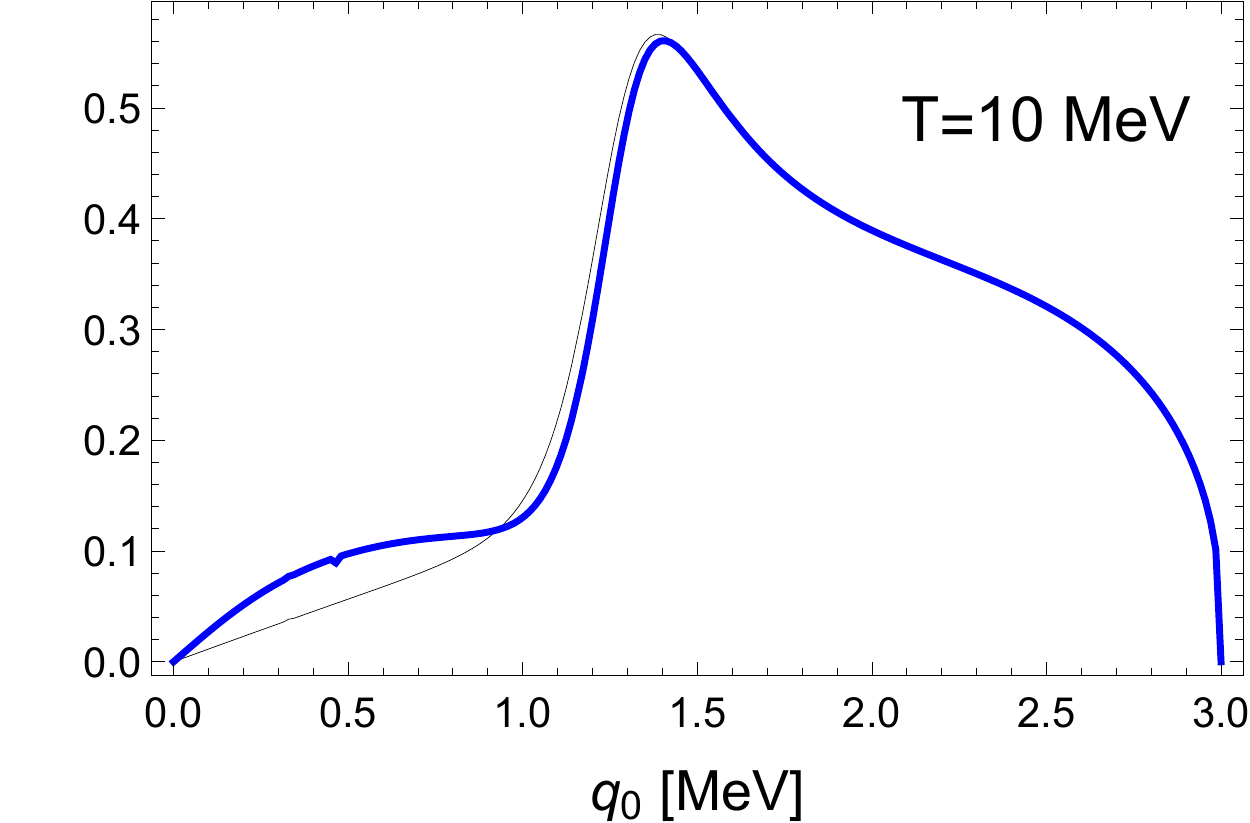}\includegraphics[scale=0.5]{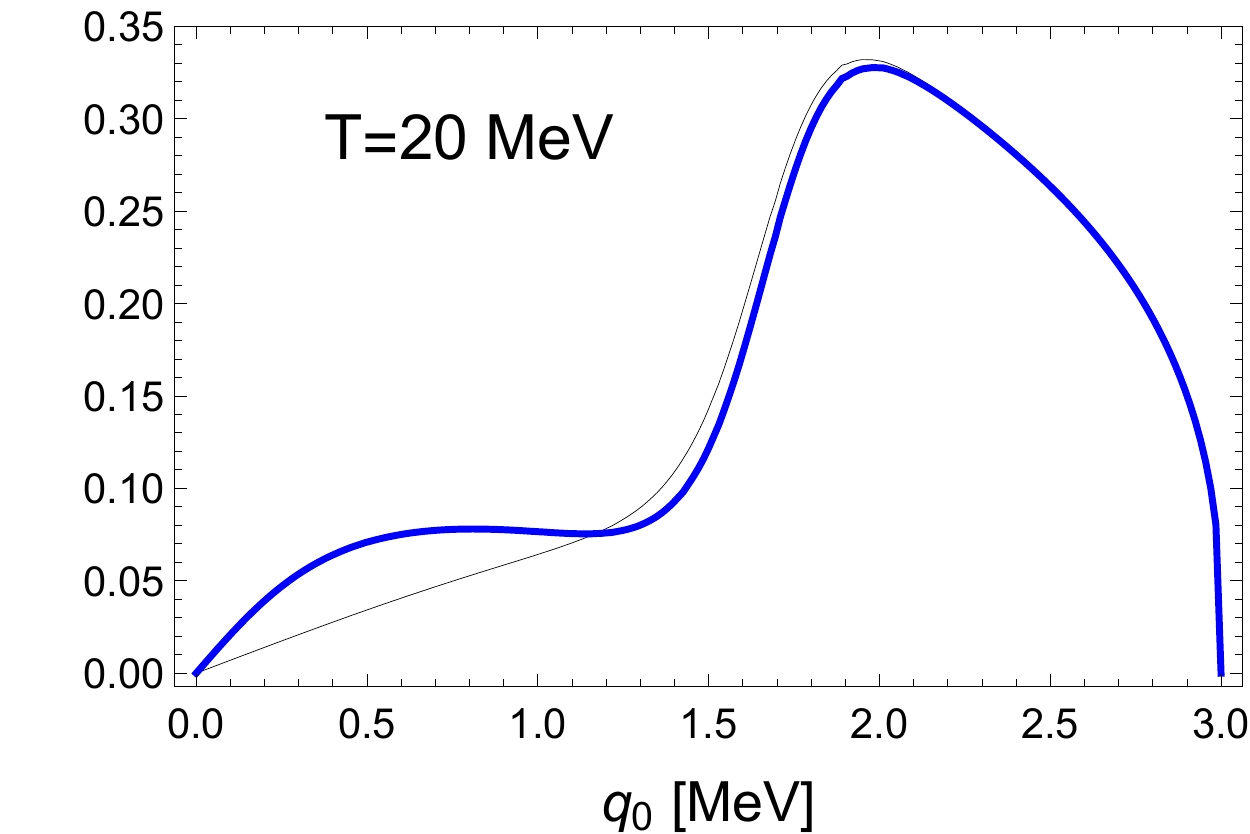}
\caption{\label{fig:RhoLongT} Temperature dependence of the longitudinal photon spectrum including induced interactions (thick blue line) at $n=0.55\,n_0$, calculated using NRAPR Skyrme forces. The momentum is fixed at $|\boldsymbol{q}|=3$ MeV, and the spectra are plotted for space-like energies. The zero temperature case is most relevant for the outer core of neutron stars. The plasma is composed of electrons, protons and neutrons, muons are absent. Thin gray lines show the corresponding spectra ignoring induced interactions. Significant modifications occur at very low energies $q_0$, and, at low temperatures in proximity to the proton peak. the former are particularly relevant for scattering, the latter have repercussions on the spectrum of collective excitations.  }
\end{figure}

\noindent In a final step neutrons are included into the RPA resummation. Neglecting their small magnetic moment, neutrons modify electromagnetic scattering via an interaction induced by the polarizability of strongly and electromagnetically charged protons. The resulting channel for lepton-neutron scattering is depicted in Fig.  \ref{fig:inducedEN}. By definition, induced interactions do not alter the appearance of the photon propagator Eq. \ref{eq:PropEMP}. Using the residual density-density ($f_{ab}$) and current-current ($\bar{f}_{ab}$) potentials from Eq. \ref{eq:derivatives}, polarization effects due to strong interactions are resummed to obtain the polarization function $\tilde{\Pi}_p$, 
\bea  \label{eq:ReProtonFull}
\tilde{\Pi}_{00,\,p} & = & e^2 \frac{\Pi_{00,\,p}^{\prime}\,(1+f_{nn}\,\Pi_{00,\,n}^{\prime})}{1+f_{nn}\,\Pi_{00,\,n}^{\prime}+f_{pp}\,\Pi_{00,\,p}^{\prime}+\Pi_{00,\,p}^{\prime}\,\,\Pi_{00,\,n}^{\prime}(f_{pp}f_{nn}-f_{np}^{2})}\,,\\[3ex]
\tilde{\Pi}_{\perp,\,p} & = & e^2 \frac{\Pi_{\perp,\,p}^{\prime}\,(1+\bar{f}_{nn}\,\Pi_{\perp,\,n}^{\prime})}{1+\bar{f}_{nn}\,\Pi_{\perp,\,n}^{\prime}+\bar{f}_{pp}\,\Pi_{\perp,\,p}^{\prime}+\Pi_{\perp,\,p}^{\prime}\,\,\Pi_{\perp,\,n}^{\prime}(\bar{f}_{pp}\bar{f}_{nn}-\bar{f}_{np}^{2})}\,, \label{eq:ReProtonFullPerp}
\eea 
which replaces $\Pi_p$ in the dressed photon propagator Eq. \ref{eq:PropEMP}. The resummed quantity is consequently of order $e^2$, while the polarization functions $\Pi^\prime$ are independent on $e$. In order to obtain Eqs. \ref{eq:ReProtonFull} and \ref{eq:ReProtonFullPerp} we have assumed that the Lorentz structure of the nuclear potentials $f$ and $\bar{f}$ is identical to that of the photon Eq. \ref{eq:PhotonGauge}, that is, we have projected nuclear interactions on the vector channel, see Ref. \cite{Stetina:2017ozh} for details. This renders the RPA resummation particularly simple, the calculation of axial and mixed correlation functions is not required. In analogy to a relativistic mean field model, one may think of the potentials $f$ and $\bar{f}$ as the static limit of interactions mediated by a massive vector meson, $g_V^2/m_{meson}^2$. The (resummed) Debye mass can be obtained from the static limit of Eq. \ref{eq:ReProtonFull}, and reads
\newline
\be \label{eq:DebyePN}
\tilde{m}_{D,\,p}^{2} = -\tilde{\Pi}_{00,\,p}(q_0,\,\boldsymbol{q}\rightarrow0)=e^2 \frac{m_{D,p}^{\prime\,2}\left(1+m_{D,n}^{\prime 2}f_{nn}\right)}{1+m_{D,n}^{\prime 2}f_{nn}+m_{D,p}^{\prime 2}f_{pp}+m_{D,p}^{\prime 2}m_{D,n}^{\prime 2}(f_{pp}f_{nn}-f_{np}^{2})}\,,
\ee
\newline
with the relativistic definition of the (one-loop) Debye mass $m^{\prime 2}_{D}=\mu\,k_{f}/\pi^2$ (note the missing factor of $e^2$). This expression can alternatively be obtained from the thermodynamic relation $m_D^2= \partial \mu_p / \partial n_p$, assuming that the proton and neutron chemical potential are related (in the present case by $\beta$ equilibrium), $\mu_p=\mu_p(\mu_n)$. The denominator of Eq. \ref{eq:DebyePN} is precisely stability condition Eq. \ref{eq:Stability}, and the static Debye screening consequently diverges upon approaching the critical density $n_c$ from above. This behaviour is illustrated in Fig. \ref{fig:Screen} for several Skyrme parameter sets. The rapid increase of Debye screening due to protons at lower densities will prove to be of great importance for electromagnetic scattering.\newline 
Due to the large mass of protons, induced interactions predominantly modify the longitudinal spectrum, while changes to the transverse spectrum are negligible \cite{Stetina:2017ozh}. The temperature dependence of the longitudinal photon spectrum including induced interactions is depicted in Fig. \ref{fig:RhoLongT}. Since neutrons are incorporated into the polarization function of protons, no additional peak in the vicinity of $v_{f,\,n}\,|\boldsymbol{q}|$ appears. Their presence is manifest in the part of the photon spectrum that is predominantly shaped by protons, at very space-like energies $q_0\leq v_{f,\,p}\,|\boldsymbol{q}|$. At lower temperatures one additionally finds a significant reduction of the height of the proton peak, which is relevant for the spectrum of collective excitations \cite{Stetina:2017ozh}.    


\section{Scattering in the fully degenerate plasma} \label{sec:FullDegen}
\noindent With the preparations of Sec. \ref{sec:OT} in hand the scattering rates can be computed. This section elaborates on the calculation of scattering at strictly zero temperature. Subsection \ref{subsec:approx} discusses approximations to the full one-loop result, subsection \ref{subsec:EMP} considers the multi-component plasma (EMP plasma), and subsection \ref{subsec:ResultsInduced} takes induced interactions into account (EMPN plasma). The generic expressions for the longitudinal and transverse rates Eqs. \ref{eq:GammaL} and \ref{eq:GammaP} can be simplified in fully degenerate matter: The angular dependence of the thermal distribution functions can be eliminated by a shift of the integration variable $\boldsymbol{q}\rightarrow\boldsymbol{p}-\boldsymbol{q}$. Trading momentum and azimuthal angle integration for integrations over the energies $\epsilon_{\boldsymbol{q}}=\sqrt{\boldsymbol{q}^2+m^2}$ and $\epsilon_{\boldsymbol{p}^\prime}$, and taking the degenerate limit of the distribution function $1+n_b(x)\rightarrow\theta(x)$, $n_f(x)\rightarrow\theta(\mu-x)$ one arrives arrives at
\begin{eqnarray}\label{eq:GammaL0}
\Gamma_{L}(\epsilon_{\boldsymbol{p}}) & = & \frac{e^{2}}{4\pi\left|\boldsymbol{p}\right|}\,\Theta\left[\pm(\epsilon_{\boldsymbol{p}}-\mu)\right]\int_{\mu}^{\epsilon_{\boldsymbol{p}}}\,d\epsilon_{\boldsymbol{q}}\,\epsilon_{\boldsymbol{q}}\,\int_{\epsilon^{-}}^{\epsilon+}d\epsilon_{\boldsymbol{p}^{\prime}}\,\epsilon_{\boldsymbol{p}^{\prime}}\,\rho_{L}(\epsilon_{\boldsymbol{p}}-\epsilon_{\boldsymbol{q}},\,\sqrt{\epsilon_{\boldsymbol{p}^{\prime}}^{2}-m^{2}})\,K_{L}\,,\\[3ex]
\Gamma_{\perp}(\epsilon_{\boldsymbol{p}}) & = & \frac{e^{2}}{2\pi\left|\boldsymbol{p}\right|}\,\Theta\left[\pm(\epsilon_{\boldsymbol{p}}-\mu)\right]\int_{\mu}^{\epsilon_{\boldsymbol{p}}}\,d\epsilon_{\boldsymbol{q}}\,\epsilon_{\boldsymbol{q}}\,\int_{\epsilon^{-}}^{\epsilon+}d\epsilon_{\boldsymbol{p}^{\prime}}\,\epsilon_{\boldsymbol{p}^{\prime}}\,\rho_{\perp}(\epsilon_{\boldsymbol{p}}-\epsilon_{\boldsymbol{q}},\,\sqrt{\epsilon_{\boldsymbol{p}^{\prime}}^{2}-m^{2}}\,)\,K_{\perp}\,,\label{eq:GammaP0}
\end{eqnarray}
\newline
where the positive and negative signs correspond to particles and holes respectively. We shall ignore this prefactor for the remainder of this section, keeping in mind that the cases $\epsilon_{\boldsymbol{p}}>\mu$ and $\epsilon_{\boldsymbol{p}}<\mu$ always correspond to the scattering rates of particles and holes. In terms of the new variables, the coefficients in the integrands of longitudinal and transverse rates read
\begin{eqnarray}
K_{L} & = & 1+\frac{1}{2\epsilon_{\boldsymbol{q}}\,\epsilon_{\boldsymbol{p}}}\left(\epsilon_{\boldsymbol{q}}^{2}+\epsilon_{\boldsymbol{p}}^{2}-\epsilon_{\boldsymbol{p}^{\prime}}^{2}+m^{2}\right)\,,\\[2ex]
K_{\perp} & = & 1+\frac{1}{2\epsilon_{\boldsymbol{p}}\,\epsilon_{\boldsymbol{q}}}\left[\epsilon_{\boldsymbol{p}}^{2}+\epsilon_{\boldsymbol{p}^{\prime}}^{2}-\epsilon_{\boldsymbol{q}}^{2}-3m^{2}-\frac{1}{2}\frac{(\epsilon_{\boldsymbol{p}}^{2}+\epsilon_{\boldsymbol{p}^{\prime}}^{2}-\epsilon_{\boldsymbol{q}}^{2}-m^{2})^{2}}{\epsilon_{\boldsymbol{p}^{\prime}}^{2}-m^{2}}\right]\,,
\end{eqnarray}
and the new integration boundaries are
\be 
\epsilon^{\pm}(\epsilon_{\boldsymbol{q}})=\sqrt{(\sqrt{\epsilon_{\boldsymbol{q}}^{2}+m^{2}}\pm\left|\boldsymbol{p}\right|)^{2}+m^{2}}\,. 
\ee
Note that as a result of Pauli blocking all integration boundaries are finite, which greatly simplifies the numerical evaluation. To calculate the energy loss in a multi-component plasma and to take into account induced interactions,the spectral function in Eqs. \ref{eq:GammaL0} and \ref{eq:GammaP0} has to be adapted accordingly.     
\subsection{Approximations to the full one-loop result} \label{subsec:approx}
\noindent  Hard dense loop (HDL) and weak screening approximations are frequently used in the calculation of transport coefficients. In the following these approximations are derived and compared to the full one-loop results Eqs. \ref{eq:GammaL0} and \ref{eq:GammaP0}.   

\subsubsection{Hard Dense Loop approximation}

\noindent The hard dense loop (HDL) approximation takes into account that fermions contributing to scattering in degenerate matter carry hard momenta, and that their interaction rates are dominated by the exchange of soft photons. The definition of ``soft" depends on the Fermi momenta $k_{f,i}$, and therefor on the masses of the fermions thermalized in the plasma. Expanding Eqs. \ref{eq:GammaL} and \ref{eq:GammaP} accordingly, and using $\epsilon_{\boldsymbol{p}^{\prime}}\simeq\epsilon_{\boldsymbol{p}}-\boldsymbol{v}\cdot\boldsymbol{q}$, where $\boldsymbol{v}=\boldsymbol{p}/\epsilon_{p}$ is the velocity of the fermions, one finds to leading order 
\begin{equation}\label{eq:GammaHDL}
\Gamma_{L}(\epsilon_{\boldsymbol{p}})+\Gamma_{\perp}(\epsilon_{\boldsymbol{p}})=e^{2}\int\frac{d^{3}\boldsymbol{q}}{(2\pi)^{2}}\,\left[1+n_{b}(\boldsymbol{v}\cdot\boldsymbol{q})-n_{f}^{-}(\epsilon_{\boldsymbol{p}}-\boldsymbol{v}\cdot\boldsymbol{q})\right]\,\bigg[\rho_{L}(\boldsymbol{v}\cdot\boldsymbol{q}\,,\,\left|\boldsymbol{q}\right|)+\boldsymbol{v}^{2}\left(1-\text{cos}\,\theta^{\,2}\right)\,\rho_{\perp}(\boldsymbol{v}\cdot\boldsymbol{q}\,,\,\left|\boldsymbol{q}\right|)\,\bigg]\,.
\end{equation}
The assembly of the appropriate approximations of the spectral functions Eqs. \ref{eq:SpecL} and \ref{eq:SpecPerp} requires for the HDL expressions of  $\Pi_{00}$ and $\Pi_\perp$, see e.g. Appendix A1 of Ref. \cite{Stetina:2017ozh} for a derivation,
\begin{equation} \label{eq:RePiHDL}
\Pi_{\textrm{HDL}}^{00}(q)=-m_{D}^{2}\left[1-\frac{1}{2}\, x\,\text{log}\, f(x)\right]\,,\,\hspace{1em}\hspace{1em}\Pi_{\perp,\,\textrm{HDL}}(q)=\frac{1}{2}m_{D}^{2}v_{f}^{2}x\left[x+\frac{1}{2}\left(1-x^{2}\right)\text{log}\, f(x)\right]\,.
\end{equation}
From the above expressions one obtains the imaginary parts
\begin{equation} \label{eq:ImPiHDL}
\text{Im}\,\Pi_{\text{HDL}}^{00}(q)=-\frac{\pi}{2}m_{D}^{2}\, x\,\Theta(1-x^{2})\,,\,\hspace{1em}\hspace{1em}\text{Im}\, \Pi_{\perp,\,\text{HDL}}(q)=-\frac{\pi}{4}m_{D}^{2}\, v_{f}^{2}\, x\left(1-x^{2}\right)\Theta(1-x^{2})\,,
\end{equation}
with the dimensionless variable $x=q_{0}/(v_{f}\left|\boldsymbol{q}\right|)$, the standard definition of the relativistic Debye mass $m_{D}^{2}=e^{2}\mu k_{f}/\pi^{2}$, the Fermi velocity $v_{f}=k_{f}/\mu$ and the function $f(x)=\left|(x+1)/(x-1)\right|$. The step function $\Theta(1-x^2)$ restricts the spectral function to the domain in which Landau damping operates, i.e. $q_0< v_f\,|\boldsymbol{q}|$ \footnote{\noindent This is the approximate boundary in the soft region. When the full one-loop polarization tensor is used, Landau damping resides in the region $0\leq q_0\leq -\mu+\sqrt{\mu^2+|\boldsymbol{q}|^2+2k_f\mu}$  for $|\boldsymbol{q}|<2 k_f$.}. Using $u=\boldsymbol{v}\cdot\boldsymbol{q}$ as integration variable Eq. \ref{eq:GammaHDL} simplifies to
\begin{equation} \label{eq:GammaHDL2}
\Gamma_{L}(\epsilon_{\boldsymbol{p}})+\Gamma_{\perp}(\epsilon_{\boldsymbol{p}})=\frac{e^{2}}{(2\pi)}\frac{1}{\left|\boldsymbol{v}\right|}\int_{0}^{B}du\,\int_{0}^{\infty}\left|\boldsymbol{q}\right|d\left|\boldsymbol{q}\right|\left[\rho_{L}(u\,,\,\left|\boldsymbol{q}\right|)+\left(\boldsymbol{v}^{2}-\frac{u^2}{\boldsymbol{q}^2}\right)\,\rho_{\perp}(u\,,\,\left|\boldsymbol{q}\right|)\right]\,.
\end{equation}
where the upper boundary is either determined by kinematics or Pauli blocking,  $B=\textrm{min}\left(\left|\boldsymbol{v}\right|\left|\boldsymbol{q}\right|,\left|\mu-\epsilon_{\boldsymbol{p}}\right|\right)$. The absolute value used in the integration boundary takes care of particles and holes. Examples for calculations based on the HDL approximation of the scattering rate include degenerate quark plasmas \cite{Heiselberg:1993cr} \cite{Baym:1990uj} \cite{Baym:1991qf}  \cite{Heiselberg:1992ha}, degenerate electron systems \cite{Shternin:2006uq}, and warm neutron star crusts \cite{Harutyunyan:2016rxm}.
\subsubsection{Weak screening approximation}
\noindent The HDL result Eq. \ref{eq:GammaHDL2}  still requires a numerical evaluation. Following Refs. \cite{LeBellac:1996kr} \cite{Vanderheyden:1996bw} \cite{Manuel:2000mk} analytic insights can be obtained by evaluating Eq. \ref{eq:GammaHDL2} in close proximity to the Fermi surface. To do so one may approximate $\left|\boldsymbol{v}\right|$ by $v_{f}$, and expand the polarization functions for $x\ll1$. The fundamental difference between longitudinal and transverse scattering now becomes evident: plasmons exchange is screened by the Debye mass, and the longitudinal rate can be evaluated in the \textit{static} limit without subtleties. The leading contribution reads
\be\label{eq:GLongA}
\Gamma_{L}\simeq\frac{e^{2}}{(4\pi)}\frac{m_{D}^{2}}{v_f^2}\int_{0}^{\left|\mu-\epsilon_{\boldsymbol{p}}\right|}u\, du\,\int_{0}^{\infty}d\left|\boldsymbol{q}\right|\,\frac{1}{(m_{D}^{2}+\boldsymbol{q}^{2})^{2}}=\frac{e^2}{32}\frac{1}{m_{D}v_{f}^{2}}\left(\mu-\epsilon_{\boldsymbol{p}}\right)^{2}\,.
\ee
The above result corresponds essentially to an interaction rate calculated using a Thomas-Fermi screened interaction. Transverse photons on the other hand exhibit no static screening, $\Pi_{\perp}(q_{0}\rightarrow0)=0$. The leading contribution to $\Pi_\perp$ is linear in $q_{0}$ and imaginary, see Eq. \ref{eq:ImPiHDL}. As it turns out, implementing this simple form of \textit{dynamical} screening into the transverse spectral function is sufficient to obtain a finite result for the scattering rate at order $u^2$, as long as the temperature is strictly zero. It reads
\be\label{eq:GPerpA}
\Gamma_{\perp}\simeq\frac{e^{2}}{(2\pi)}m_{D}^{2}\,\int_{0}^{\left|\mu-\epsilon_{\boldsymbol{p}}\right|}du\,\int_{0}^{\infty}d\left|\boldsymbol{q}\right|\frac{4\,v_f^2\, u\,}{16\boldsymbol{q}^{4}+\pi^{2}\,m_{D}^{4}\,v_f^2\,u^{2}/\boldsymbol{q}^{2}}=\frac{e^2}{12\pi}v_{f}\left|\epsilon_{\boldsymbol{p}}-\mu\right|\,.
\ee
\begin{figure}[t]
\begin{framed}
\includegraphics[scale=0.5]{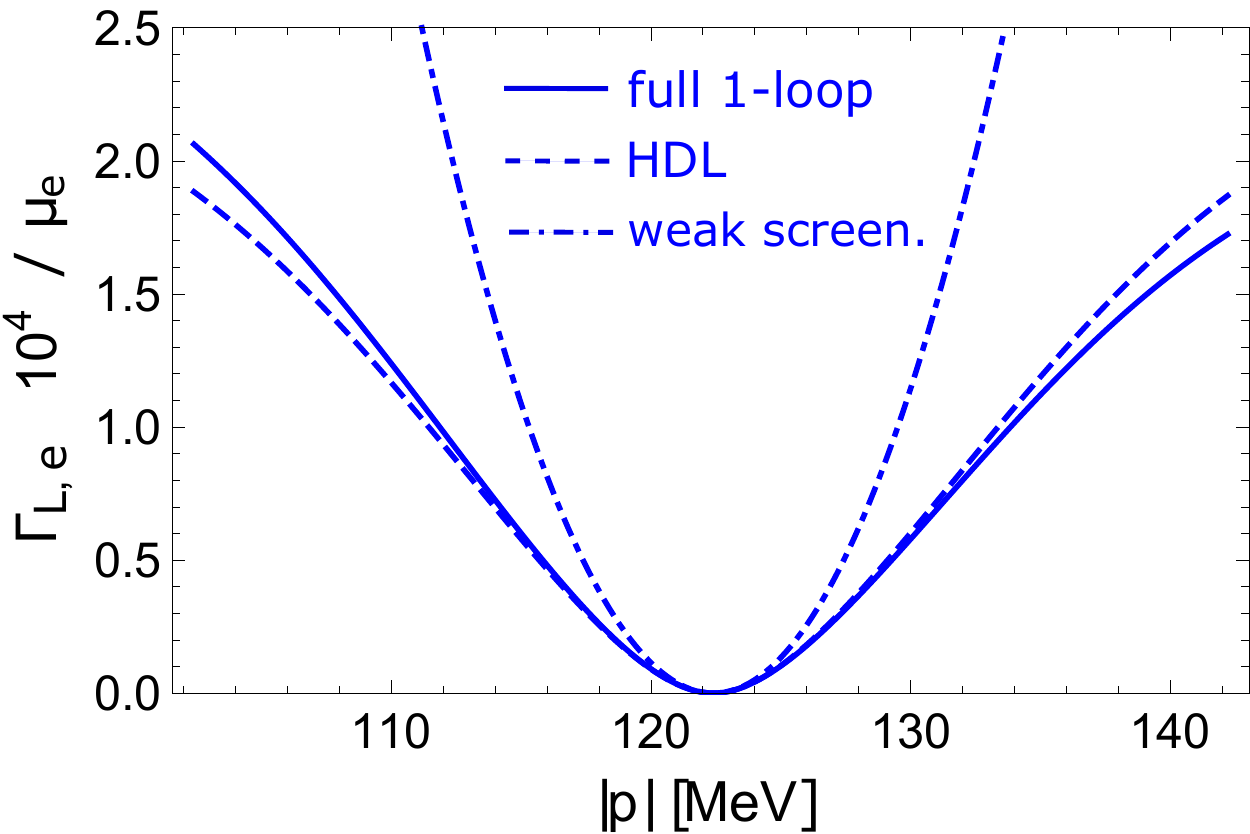}\hspace{1.4cm}\includegraphics[scale=0.5]{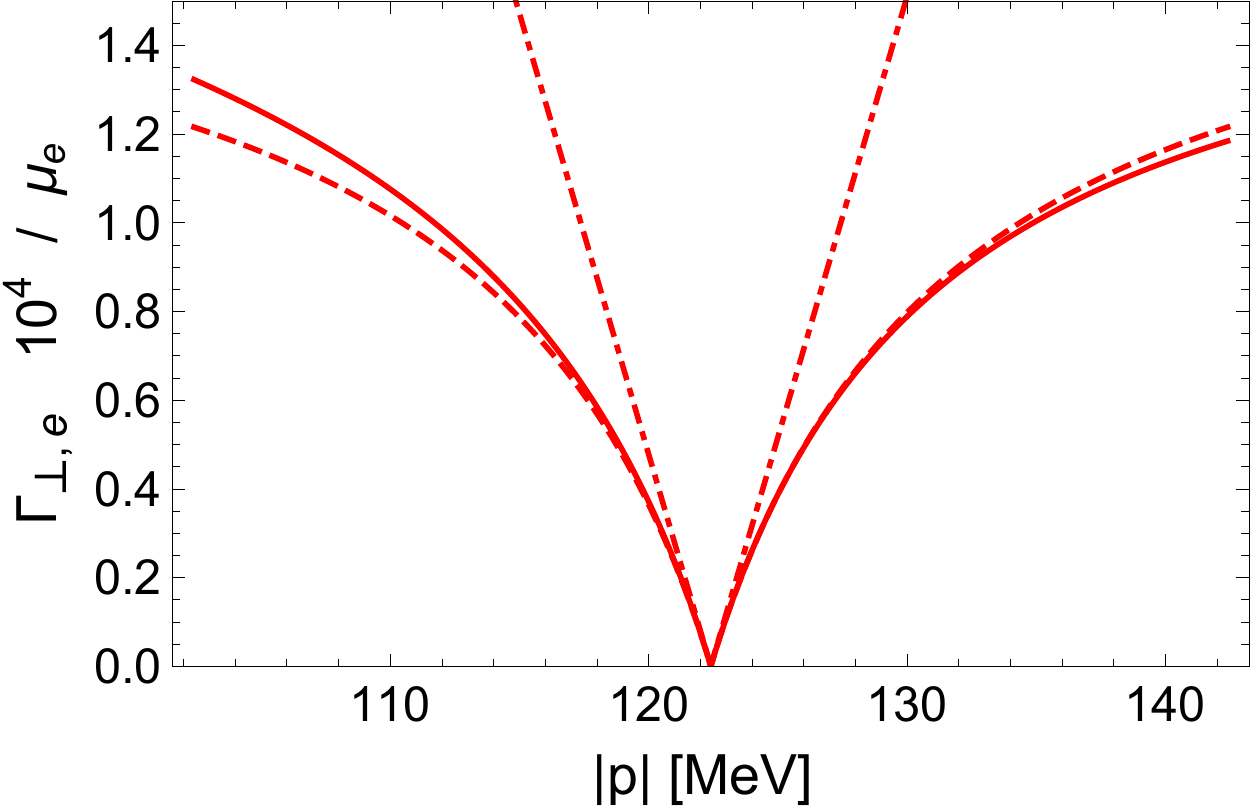}\\[2ex]
\includegraphics[scale=0.5]{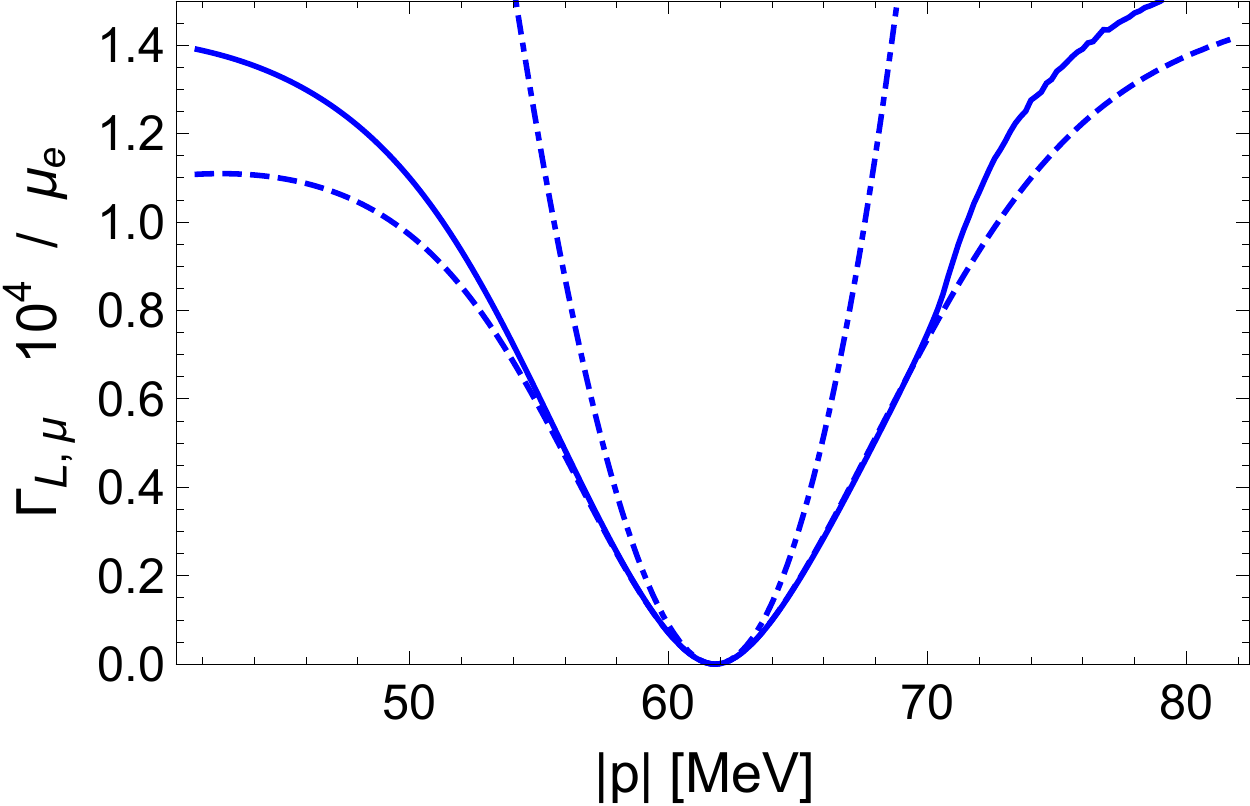}\hspace{1.4cm}\includegraphics[scale=0.5]{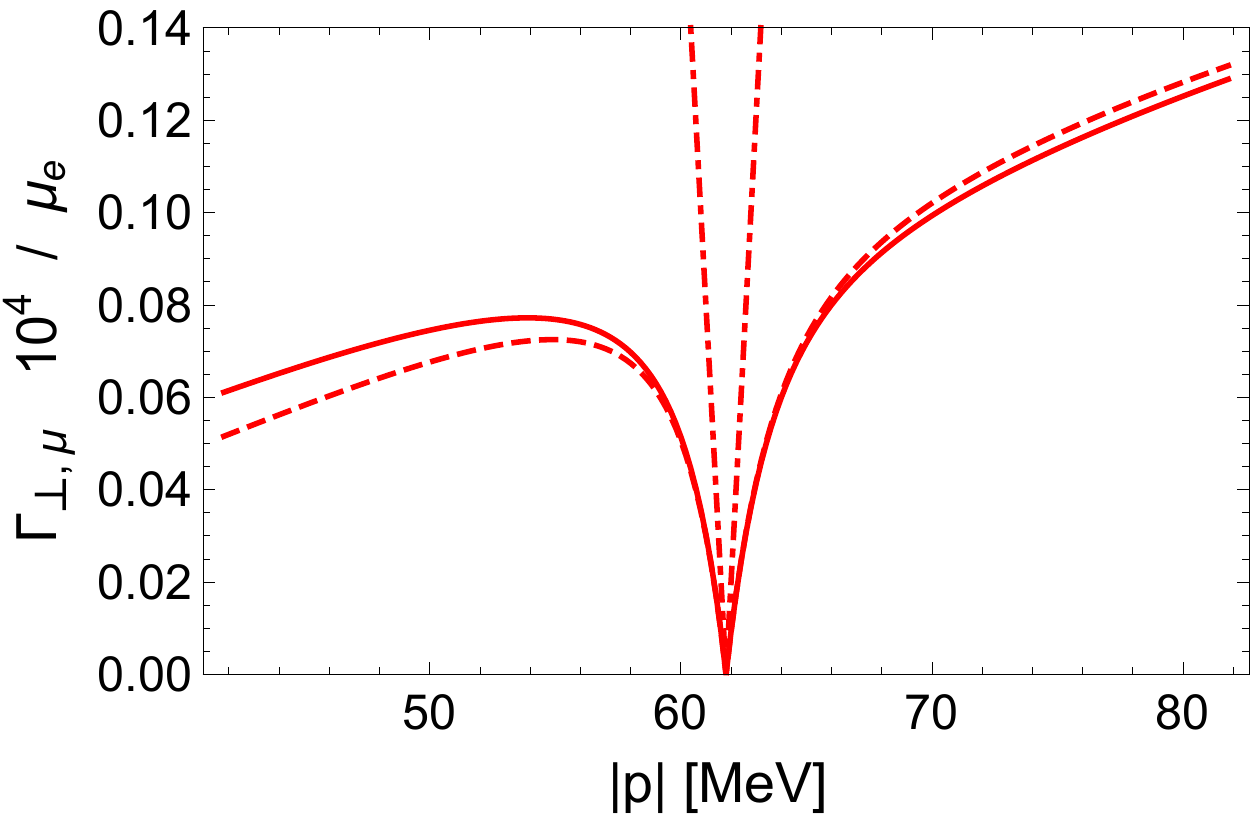}\\[2ex]
\hrule\vspace{0.2cm}
\includegraphics[scale=0.34]{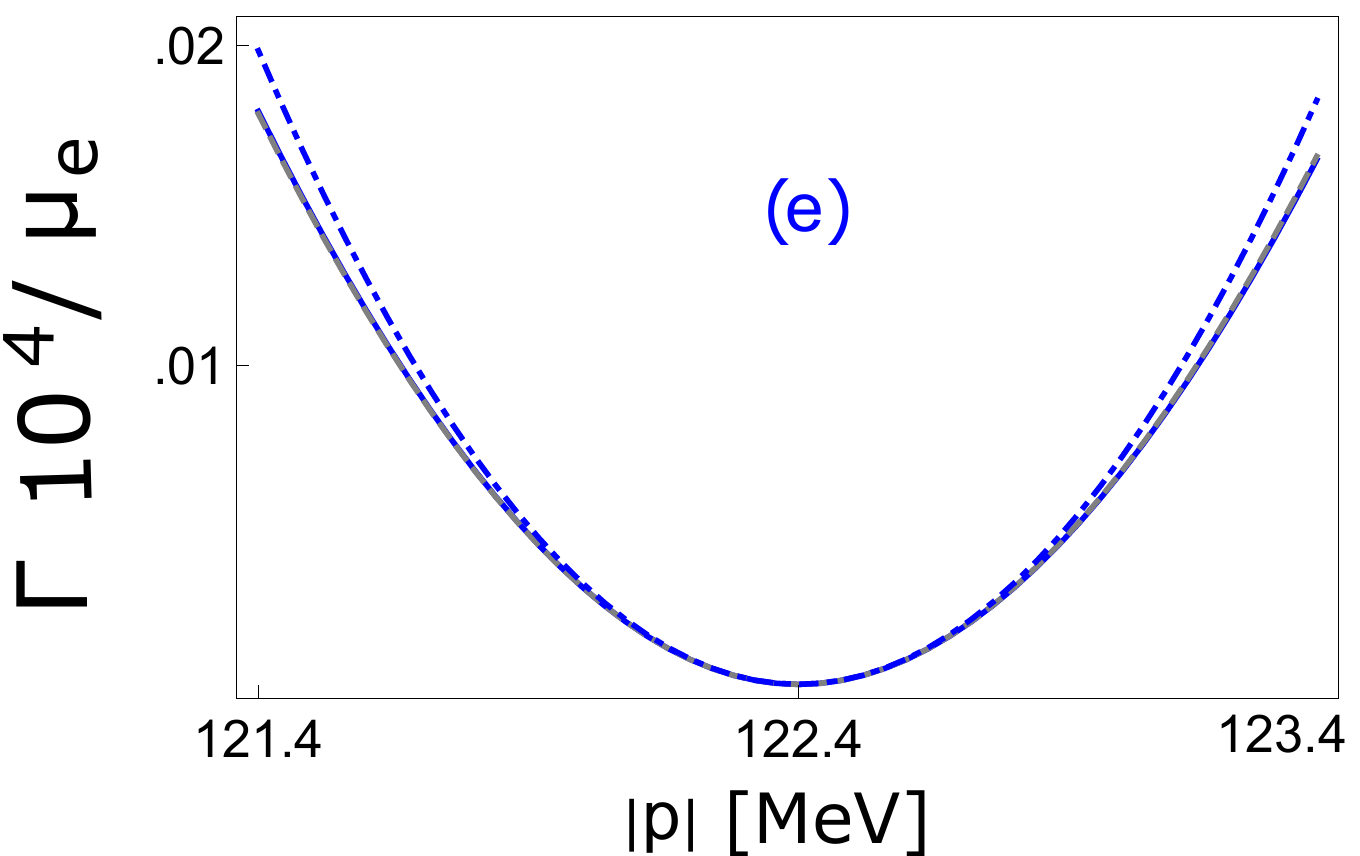}\,\,\,\includegraphics[scale=0.34]{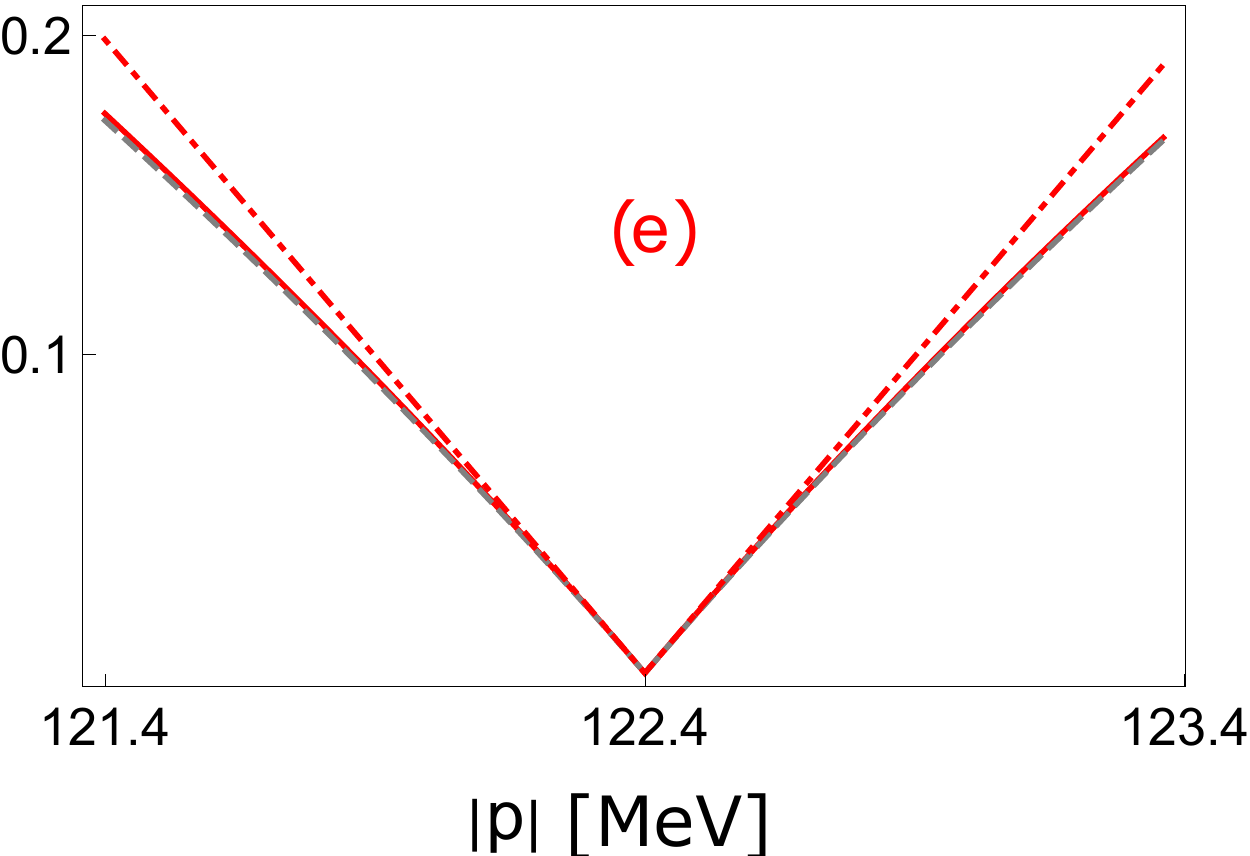}\,\,\,\includegraphics[scale=0.335]{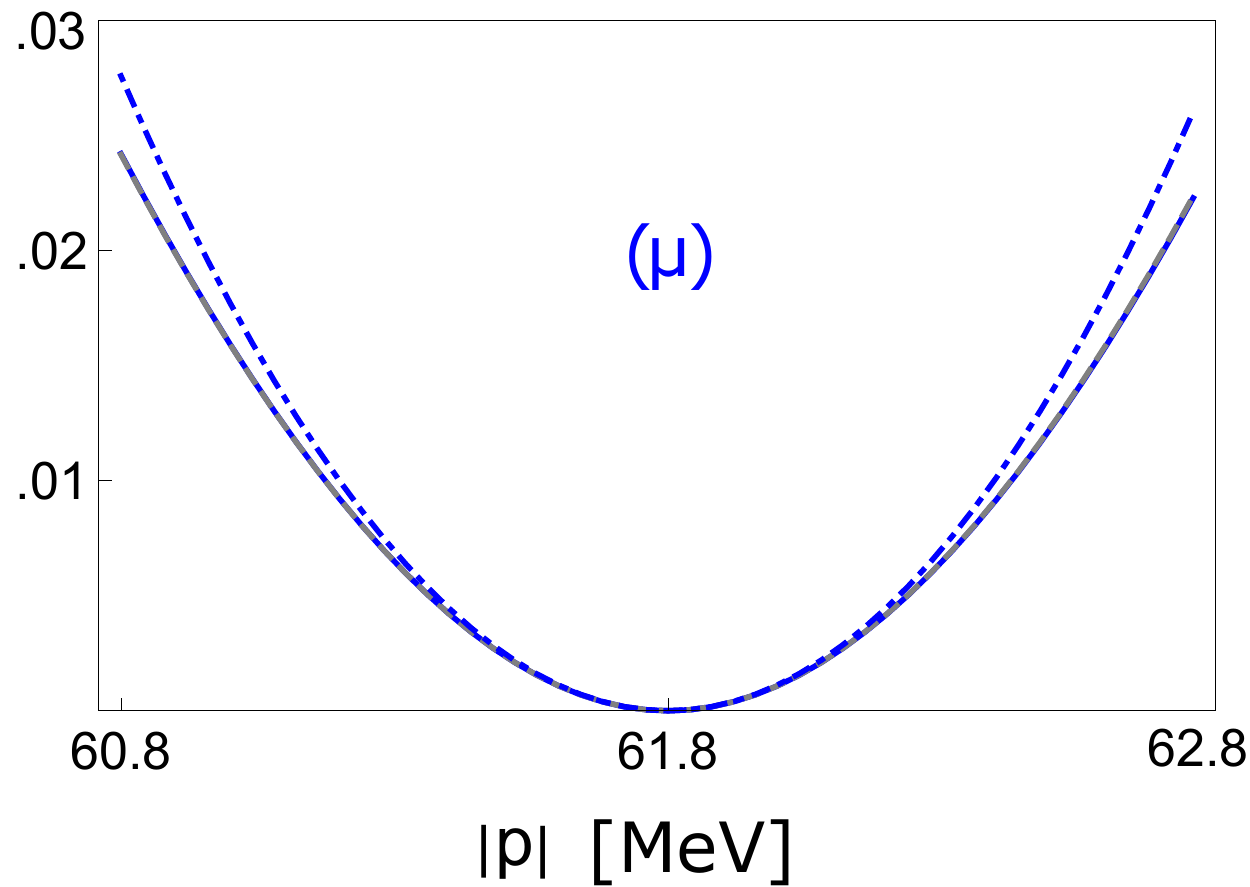}\,\,\,\includegraphics[scale=0.335]{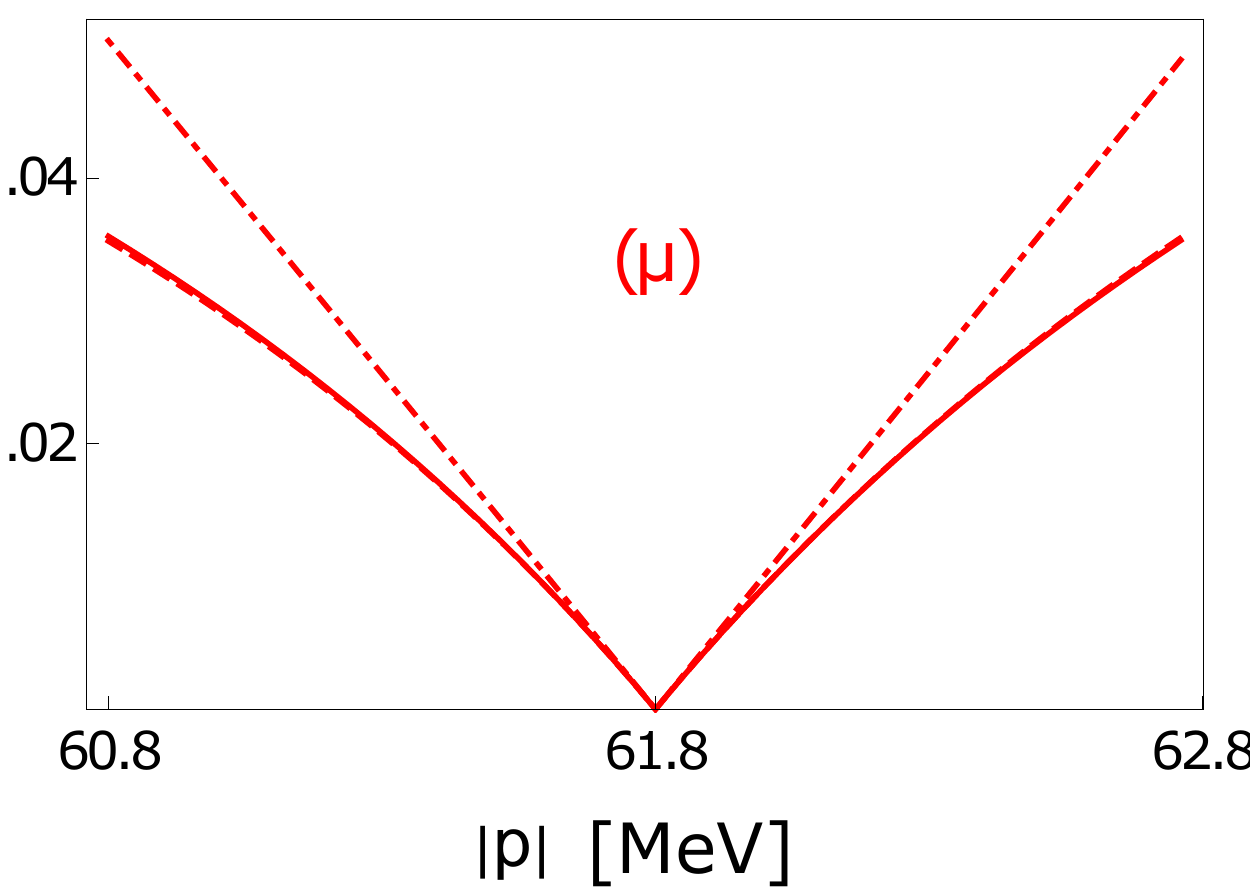}
\end{framed}
\setlength{\belowcaptionskip}{-8pt}
\caption{\label{fig:ScreenComp} Longitudinal (blue) and transverse (red) scattering rates of electrons and muons at saturation density and zero temperature, calculated using NRAPR Skyrme forces, see Tab. \ref{tab:NRAPR} for a list of parameters. Dashed lines display the HDL approximations, dot-dashed lines display the weak screening approximation. $\Gamma$ is divided by the chemical potential to obtain dimensionless results, and multiplied by a factor of $10^4$ ( $\mu_e/10^4\sim 0.012$). In fully degenerate plasmas $\Gamma_{L}$ and $\Gamma_{\perp}$ are exactly zero at the Fermi surface, for $|\boldsymbol{p}|< k_f$ the rates describe to the damping of holes, for  $|\boldsymbol{p}|> k_f$ they describe the damping of particles. The weak screening results Eqs. \ref{eq:GLongA} and  \ref{eq:GPerpA} capture the momentum dependence in close proximity to the Fermi surface, (see bottom four figures). The transverse rate of muon scattering exhibits sizeable deviations from the full result, HDL results represent an excellent approximation.  }
\end{figure}
\noindent Both results, first obtained in Ref. \cite{Manuel:2000mk}\footnote{Mind the difference of a global factor of 2, stemming from the definition of $\Gamma$, Eq. \ref{eq:OTheorem} here and Eq. 19 in Ref. \cite{Manuel:2000mk}.}, highlight the distinct importance of electric and magnetic interactions for heavy and light particles in close proximity to the Fermi surface. In non-relativistic systems $v_f\ll1$, and longitudinal scattering becomes increasingly important. In the ultra-relativistic limit transverse scattering dominates. The weak screening approximation has been applied in the calculation of transport phenomena in white dwarfs and neutron star cores, see e.g., \cite{Shternin:2008es}, \cite{Shternin:2018jop}, \cite{Potekhin:1999yv}. \newline A comparison of weak-screening approximation, hard dense loop approximation, and full one-loop results is shown in Fig. \ref{fig:ScreenComp} for electrons and muons, which resemble ultra-relativistic and mildly-relativistic particles respectively. In each case the scattering rates are plotted in a range of $\pm$ 20 MeV around $|\boldsymbol{p}|=k_f$, where the quasiparticles are stable. It should be noted that the energies of particles and holes available for scattering are typically of the order of the temperature. The fully degenerate case is supposed to closely resembles conditions encountered in the core of neutron stars, where temperatures are well below $1$ MeV, while lepton chemical potentials are of the order of $100$ MeV. 
The calculation of scattering rates with $|\epsilon_{\boldsymbol{p}}-\mu|\gg1$ MeV  is therefore a somewhat academic exercise. Equations \ref{eq:GammaL0} and \ref{eq:GammaP0} determine the rate at which a particle or hole of a given momentum $\boldsymbol{p}$, added to the system ``by hand", scatters with particles of the Fermi sea. The information whether such particles are available in the system is not inherent. To see this, take a look at Eqs. \ref{eq:FermiRateP} and \ref{eq:FermiRateH}, where the thermal distribution functions have been reorganized to reproduce the results of Fermi's golden rule: no Fermi distributions for the initial particles with energies $\epsilon_{\boldsymbol{p}}$ remain. It is nevertheless interesting to study the impact of dynamical screening with increasing distance from the Fermi surface, in particular in light of the subsequent discussion of partially degenerate plasmas, where the momentum exchange can be significantly higher. \newline 
Each rate displayed in Fig. \ref{fig:ScreenComp} is essentially determined by two ingredients: the parameters of the scattering particle, i.e., its mass, momentum, and chemical potential, and the spectral function of the photon. The former determines the phase-space available for scattering (encoded in the integration boundaries of Eqs. \ref{eq:GammaL0} and \ref{eq:GammaP0}), and specific features of the integrand, in particular the $\boldsymbol{v}^2$ dependence of the transverse rate (see Eq. \ref{eq:GammaHDL}). The latter determines the screening in each channel. For ultra-relativistic particles (in the present case electrons) transverse scattering is indeed found to dominate in close proximity to the Fermi surface, as projected by Heiselberg and Pethick \cite{Heiselberg:1993cr}. With increasing $|\epsilon_{\boldsymbol{p}}-\mu|$, however, longitudinal scattering catches up quickly and both rates become equally important. Muons interpolate between the two extremes of ultra-relativistic and non-relativistic particles: very close to the Fermi surface ($|\epsilon_{\boldsymbol{p}}-\mu|<1$ MeV) their transverse rates are larger, at about $1$ MeV distance both rates are essentially equal in magnitude, and even further away from the Fermi surface the longitudinal channel becomes dominant. Naturally the characteristics of the rates are density dependent, and muon scattering becomes similar to electron scattering with increasing density. The HDL approximation works remarkably well in all four cases, up to about $10$ MeV distance from the Fermi surface. The weak-screening approximation can be applied in a range of $|\boldsymbol{p}|=(k_f\pm 1)$ MeV, though sizeable deviations occur for muons in the transverse channel.    
%
\subsection{Scattering in the EMP plasma } \label{subsec:EMP}
\begin{figure}
\includegraphics[scale=0.475]{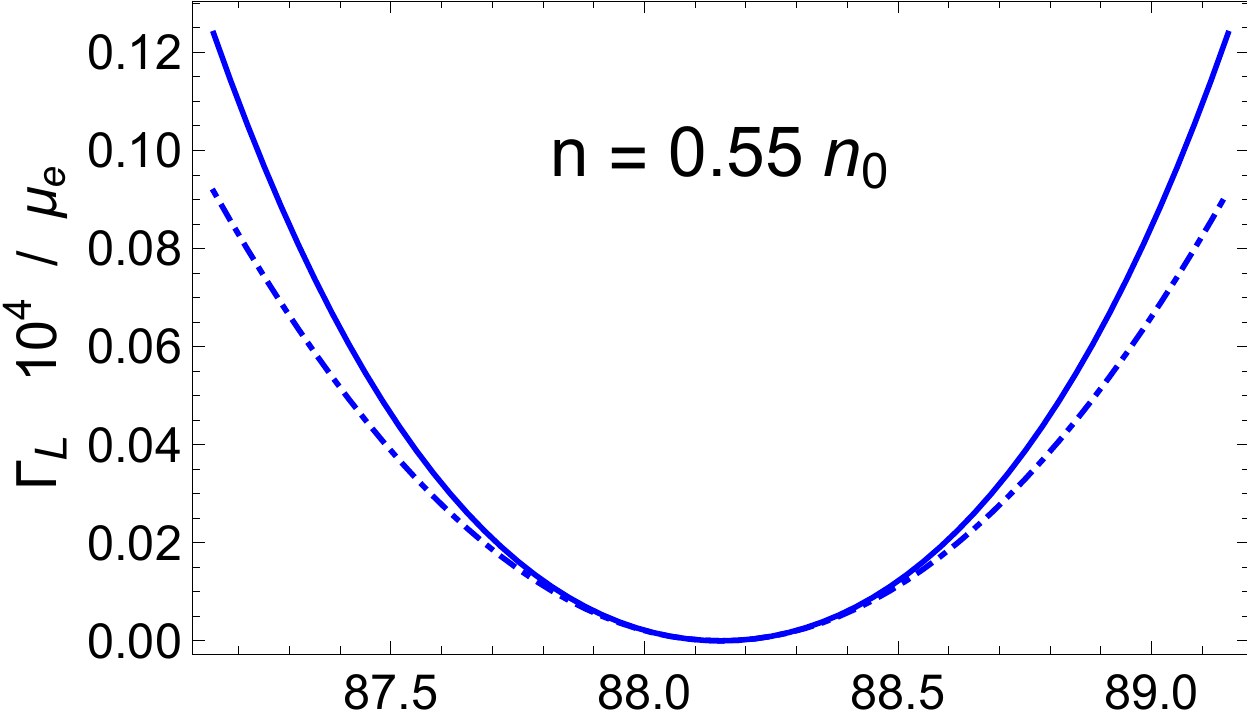}\,\,\,\includegraphics[scale=0.45]{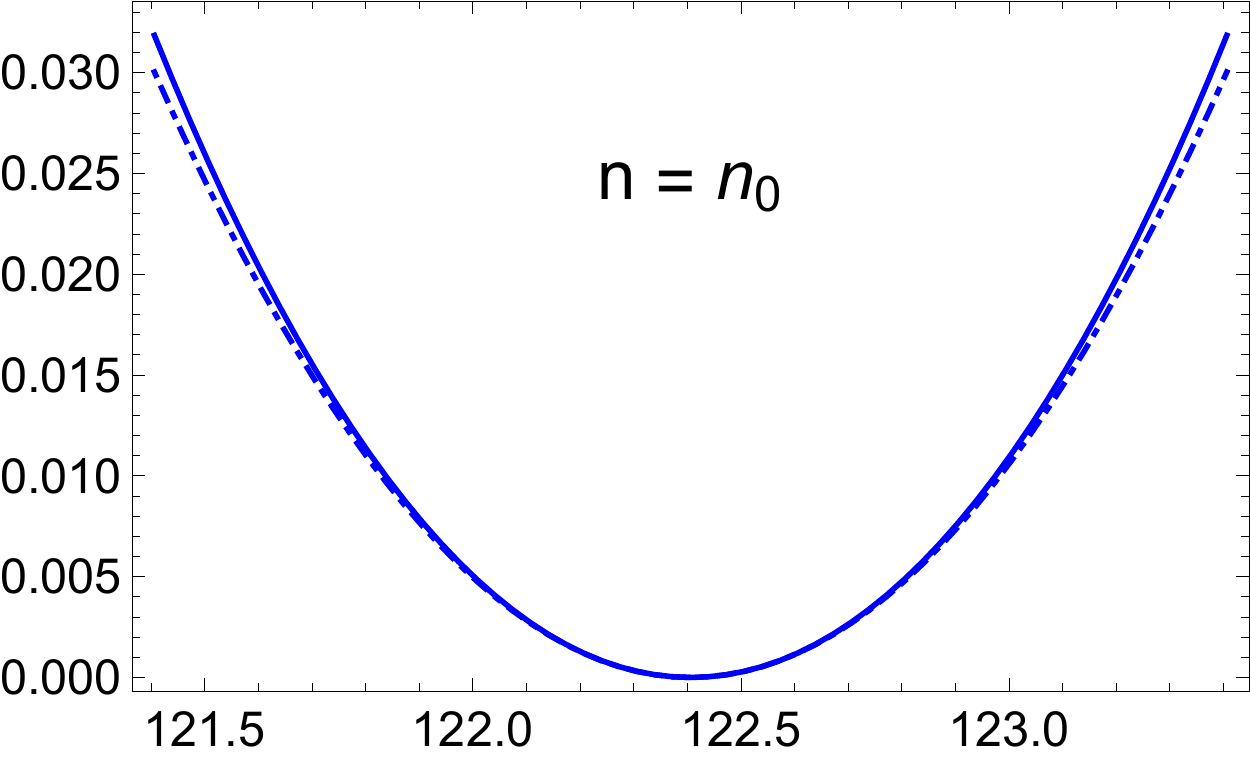}\,\,\,\includegraphics[scale=0.45]{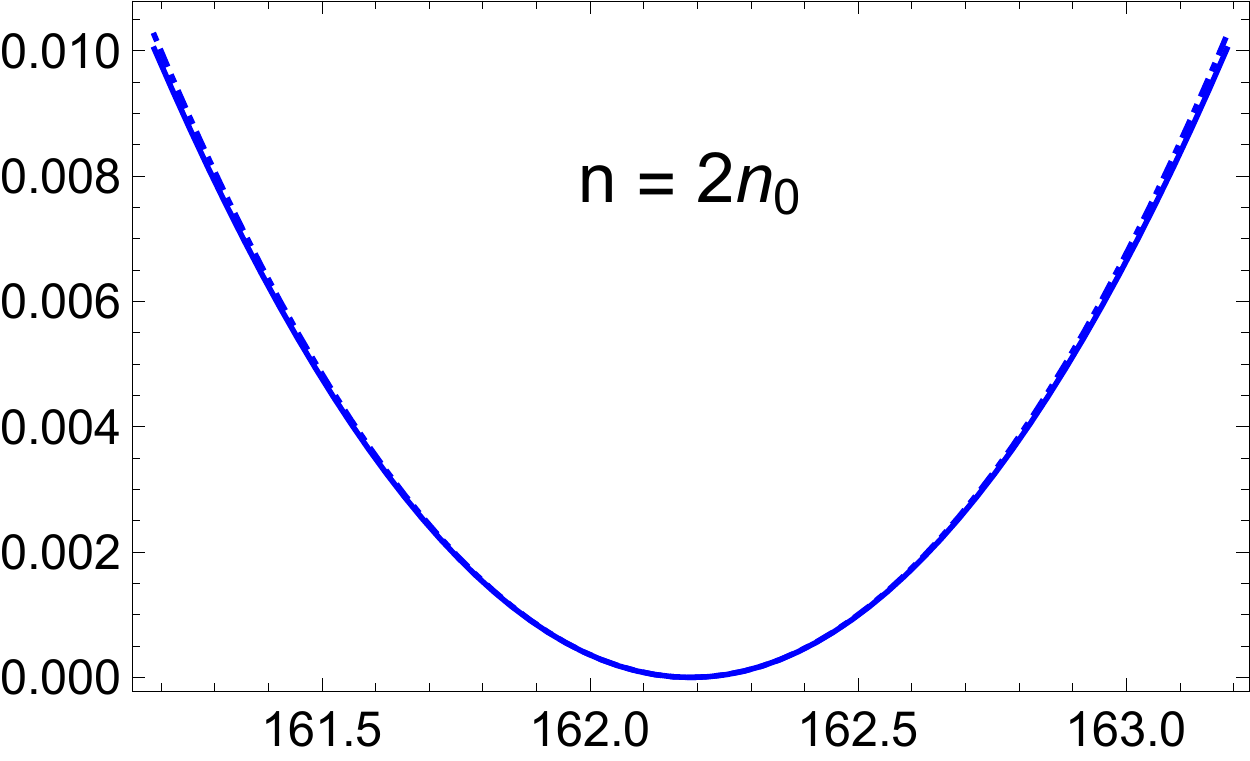}\\[2ex]
\includegraphics[scale=0.45]{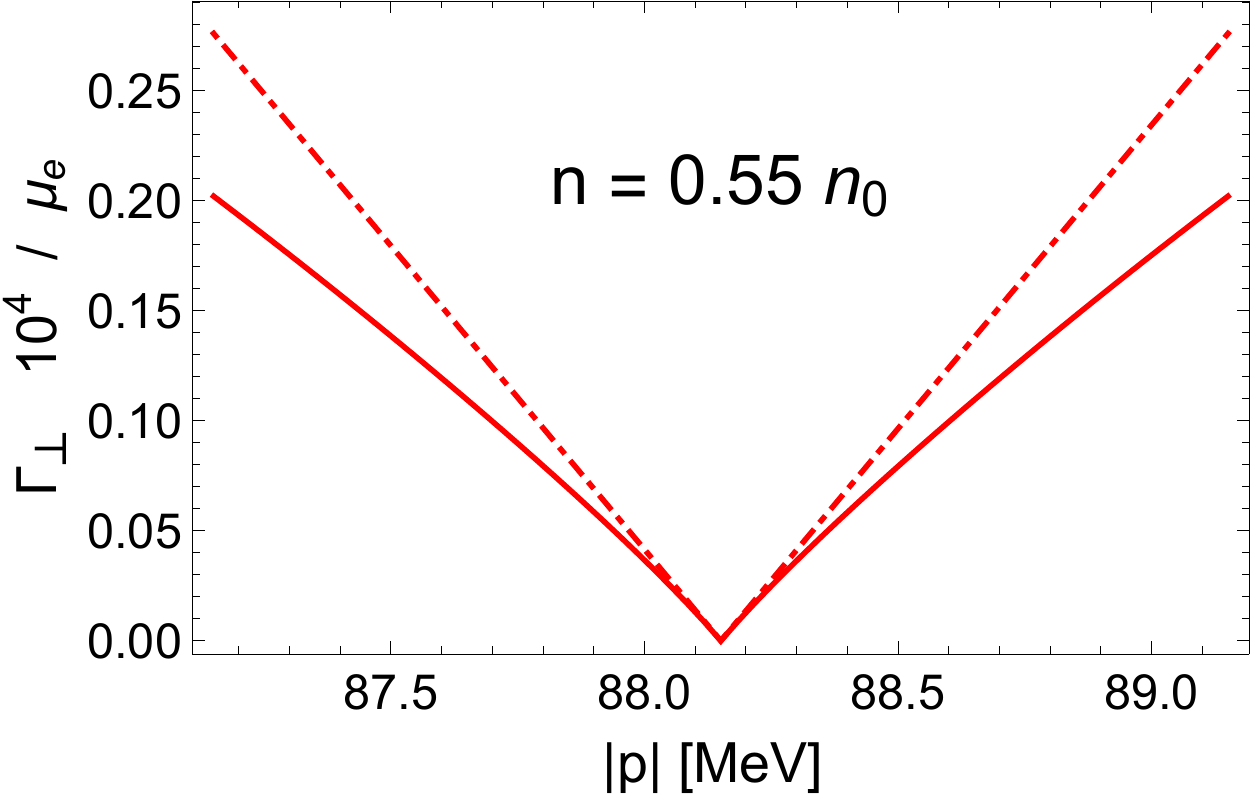}\,\,\,\includegraphics[scale=0.45]{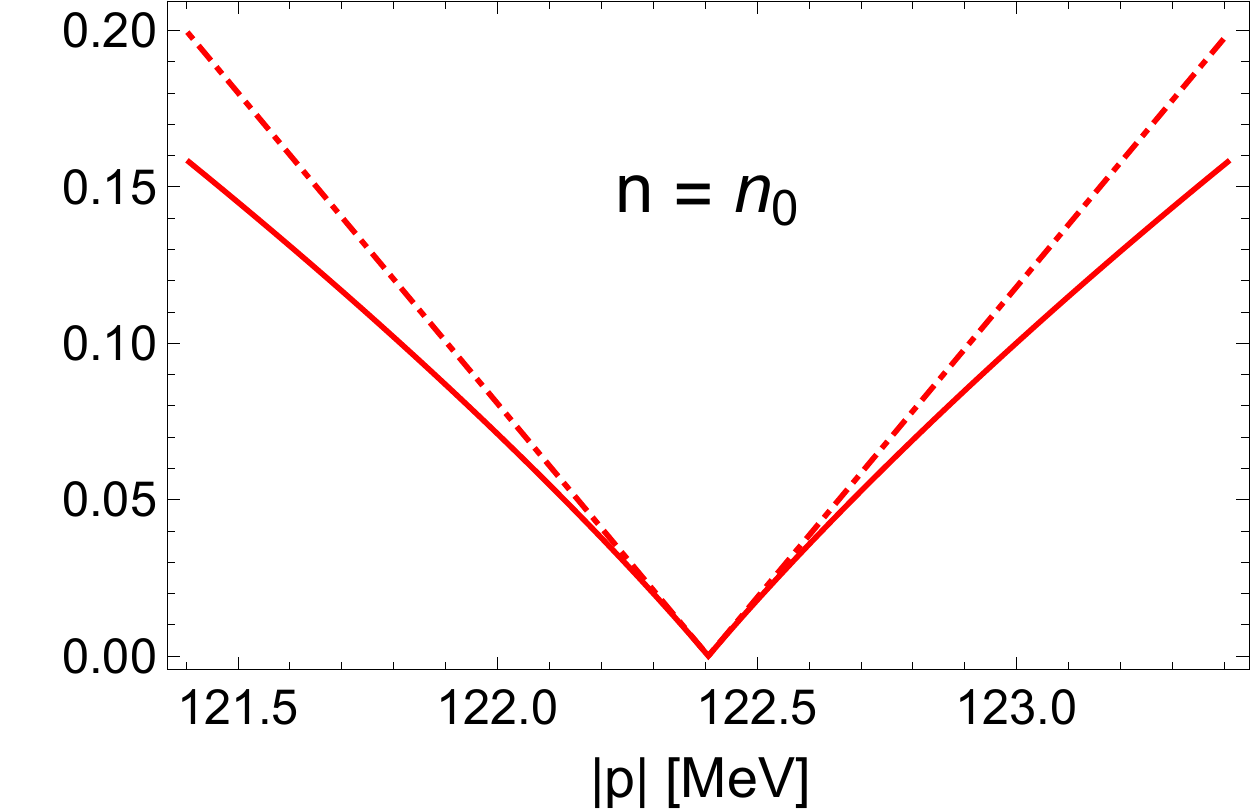}\,\,\,\,\includegraphics[scale=0.45]{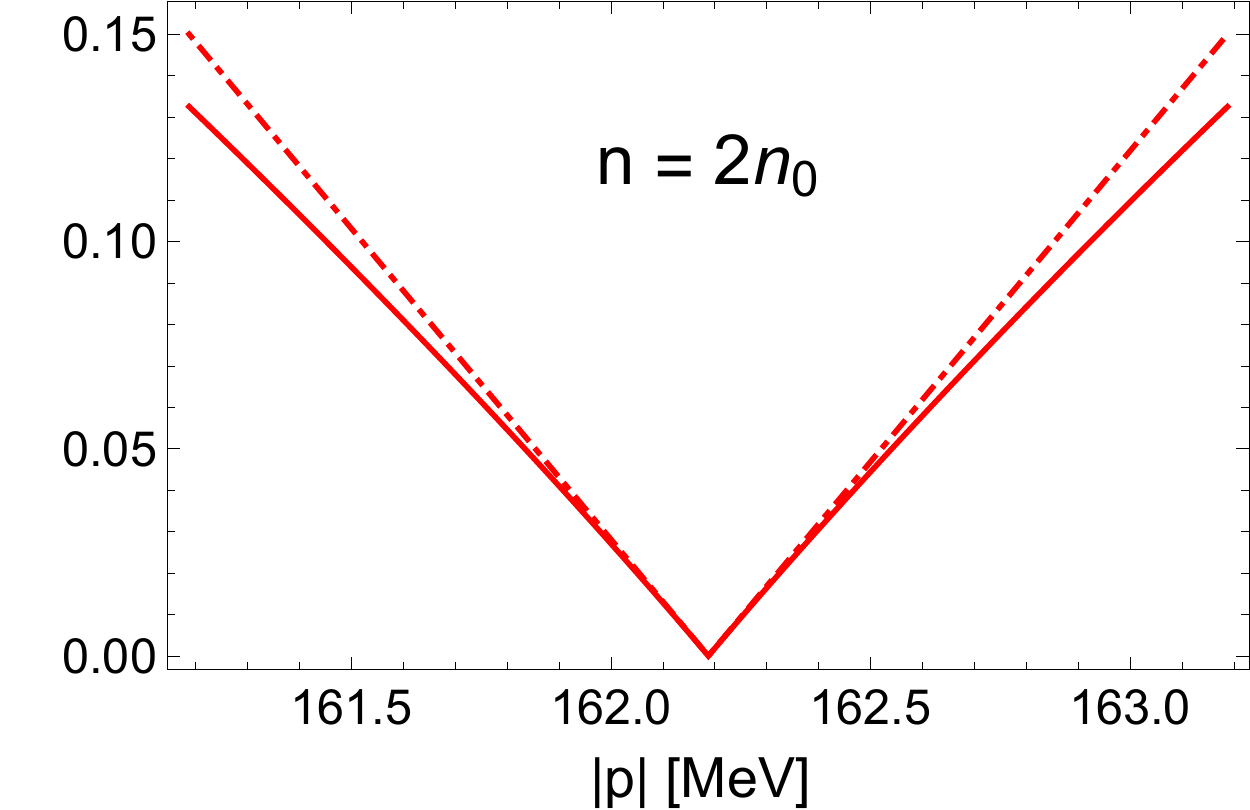}
\caption{\label{fig:WeakFull} Full one-loop results (solid) and weak-screening results (dot-dashed) in a range of ($k_f\pm1$) MeV, for three different densities. The relevant parameters are listed in Tab. \ref{tab:NRAPR}. The lowest density corresponds to a region close to the crust-core boundary of neutron stars, where muons are absent. At lower densities dynamical screening becomes important, even for particles with momenta $|\boldsymbol{p}|\sim k_f$, see also Fig. \ref{fig:RhoLongStat}. The longitudinal rates increase by more than two orders of magnitude upon approaching the crust-core boundary region from deeper inside the core. The transverse rates decrease, although by a much smaller amount.}
\end{figure}
\begin{figure}
\includegraphics[scale=0.55]{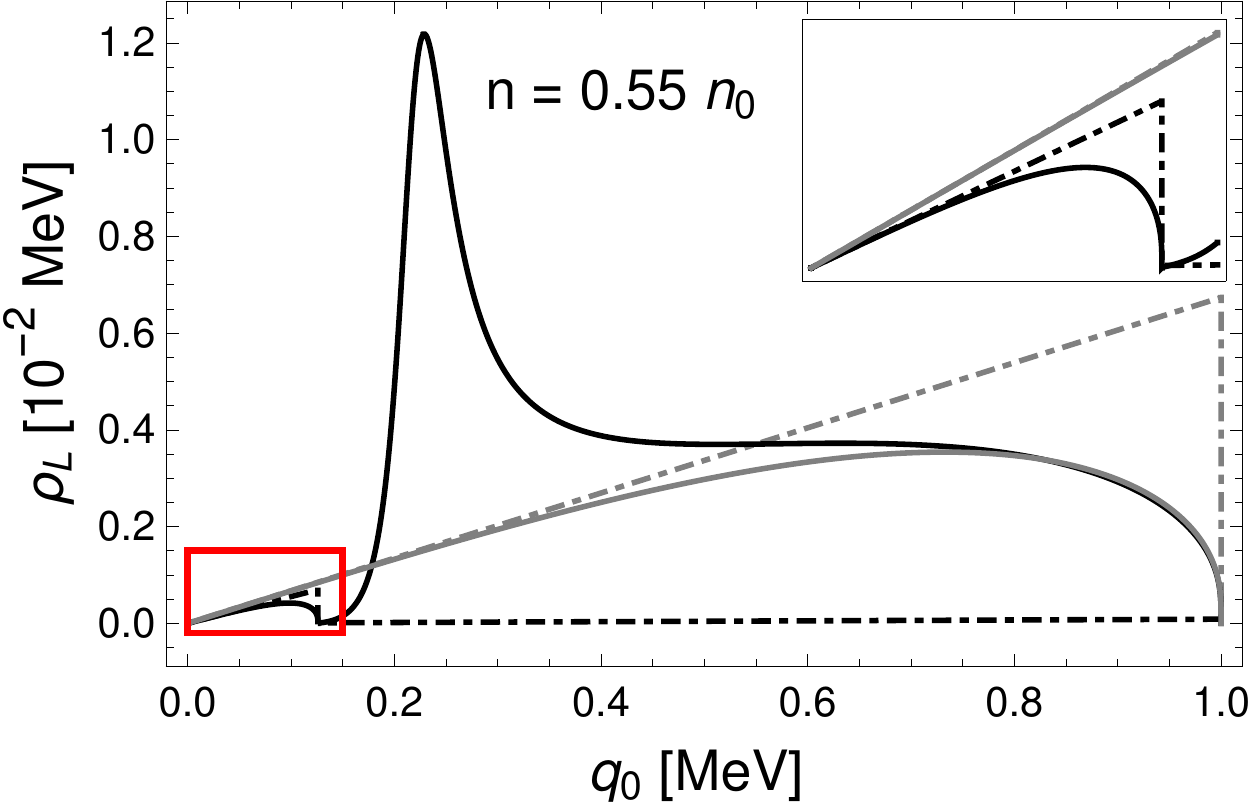}\hspace{1.5cm}\includegraphics[scale=0.55]{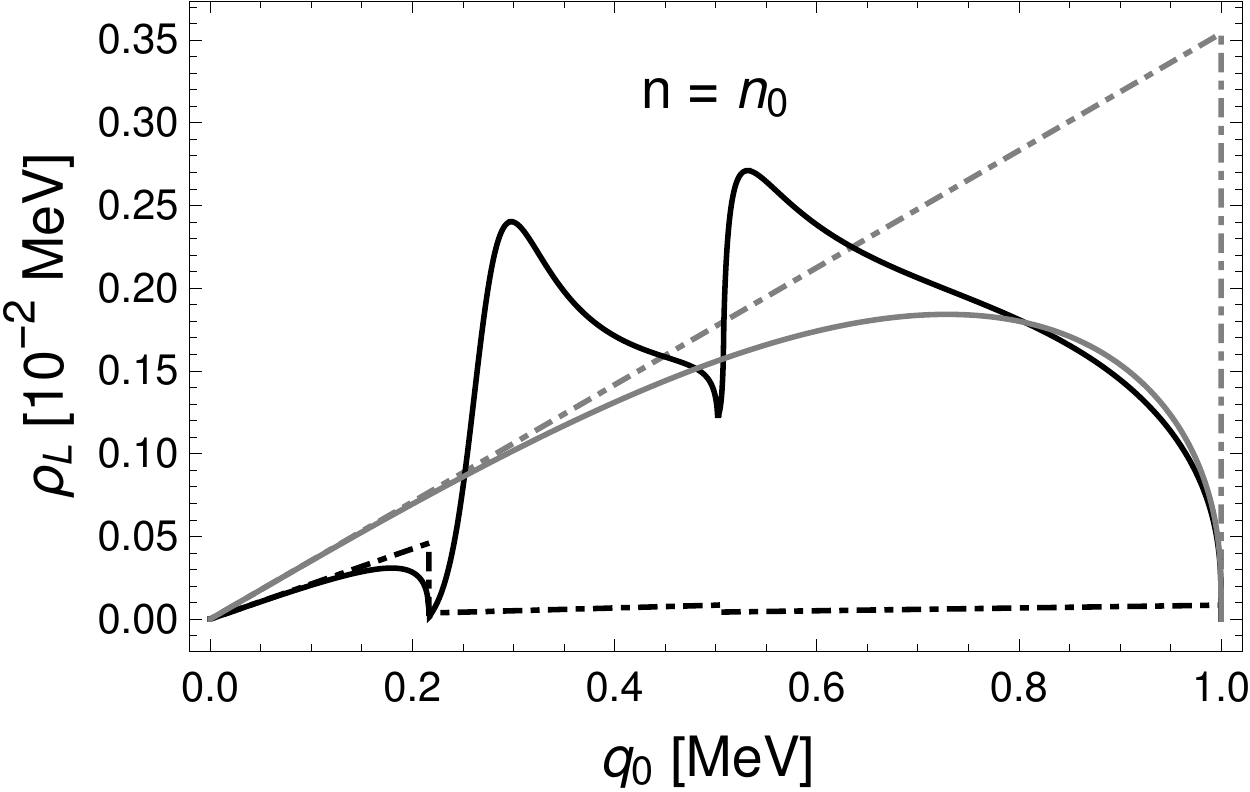}
\setlength{\belowcaptionskip}{-8pt}
\caption{\label{fig:RhoLongStat} Density dependence of the photon spectrum, illustrated by the example of $\rho_L$. Full one-loop results (solid) and  weak-screening approximation (dot-dashed) of $\rho_L$ in the multi-component plasma (black) and a pure electron plasma (gray) are shown. In each case the photon momentum is fixed at $|\boldsymbol{q}|=1$ MeV. At very low energies $q_0<v_{f,\,p}|\boldsymbol{q}|$ ($|\boldsymbol{q}|\ll k_f$) the spectrum can properly be described by $\rho_L$ as given by Eq. \ref{eq:RhoWeak}. Collisions with higher energy transfer $q_0$ probe the spectral function in a region where it becomes highly dynamical, and where the weak screening approximation is practically useless. As the Fermi velocity of the protons increases with density, the accuracy of the weak screening results improves. 
}
\end{figure}
\begin{figure}
\begin{framed}
\,\,\includegraphics[scale=0.5]{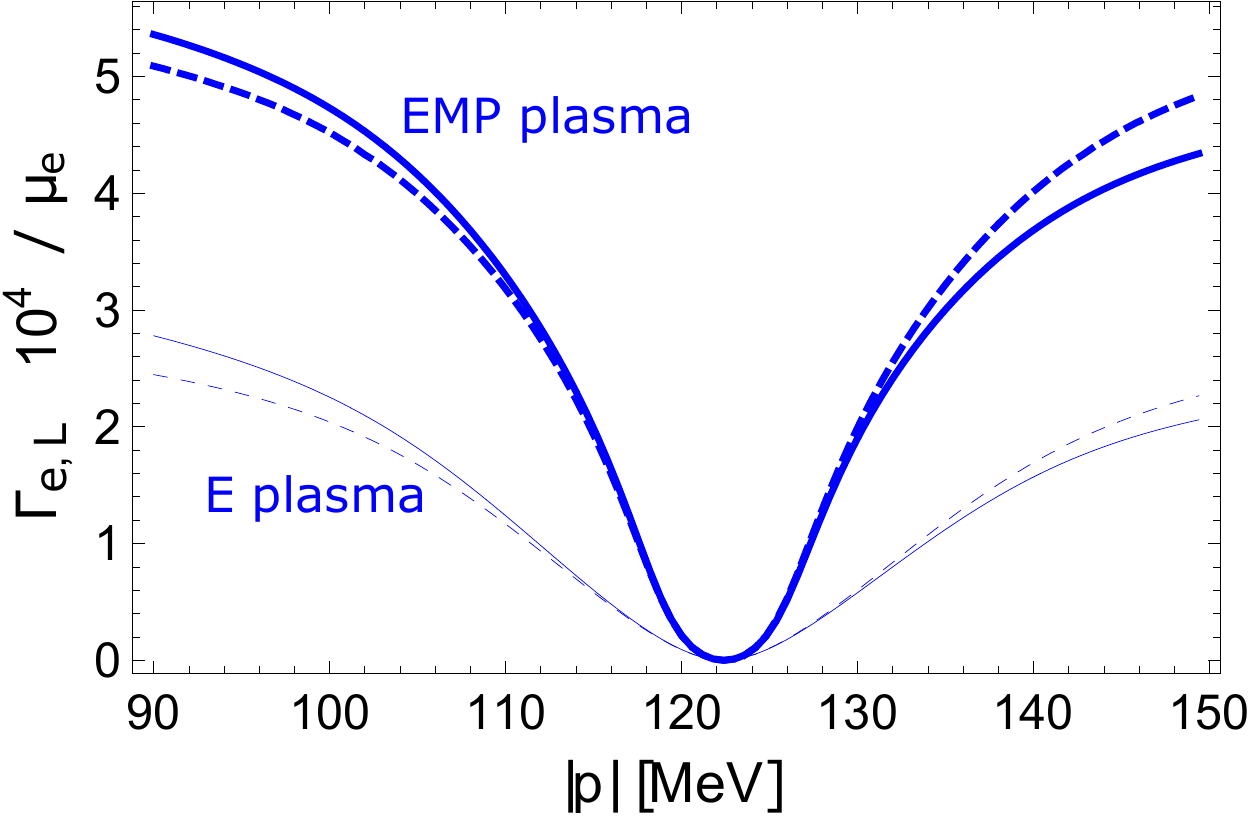}\hspace{1.2cm}\includegraphics[scale=0.51]{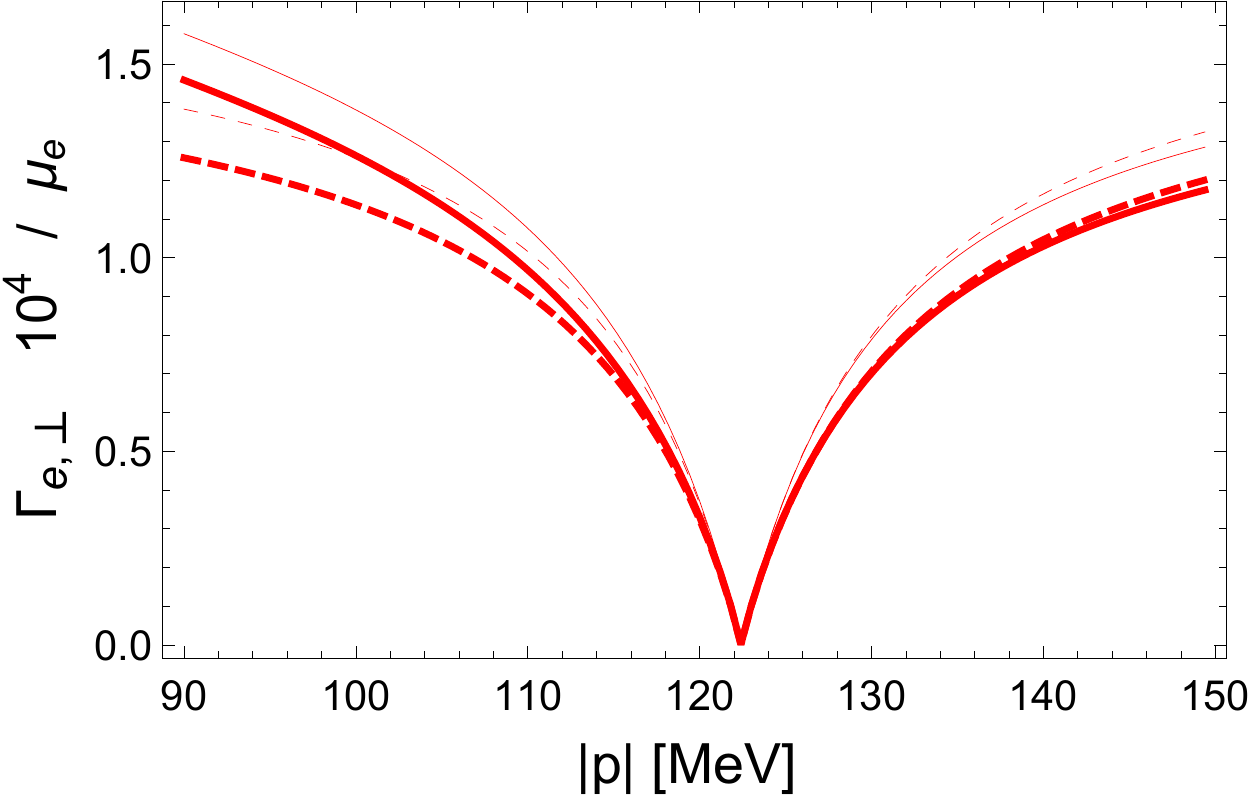}\\[2ex]\includegraphics[scale=0.5]{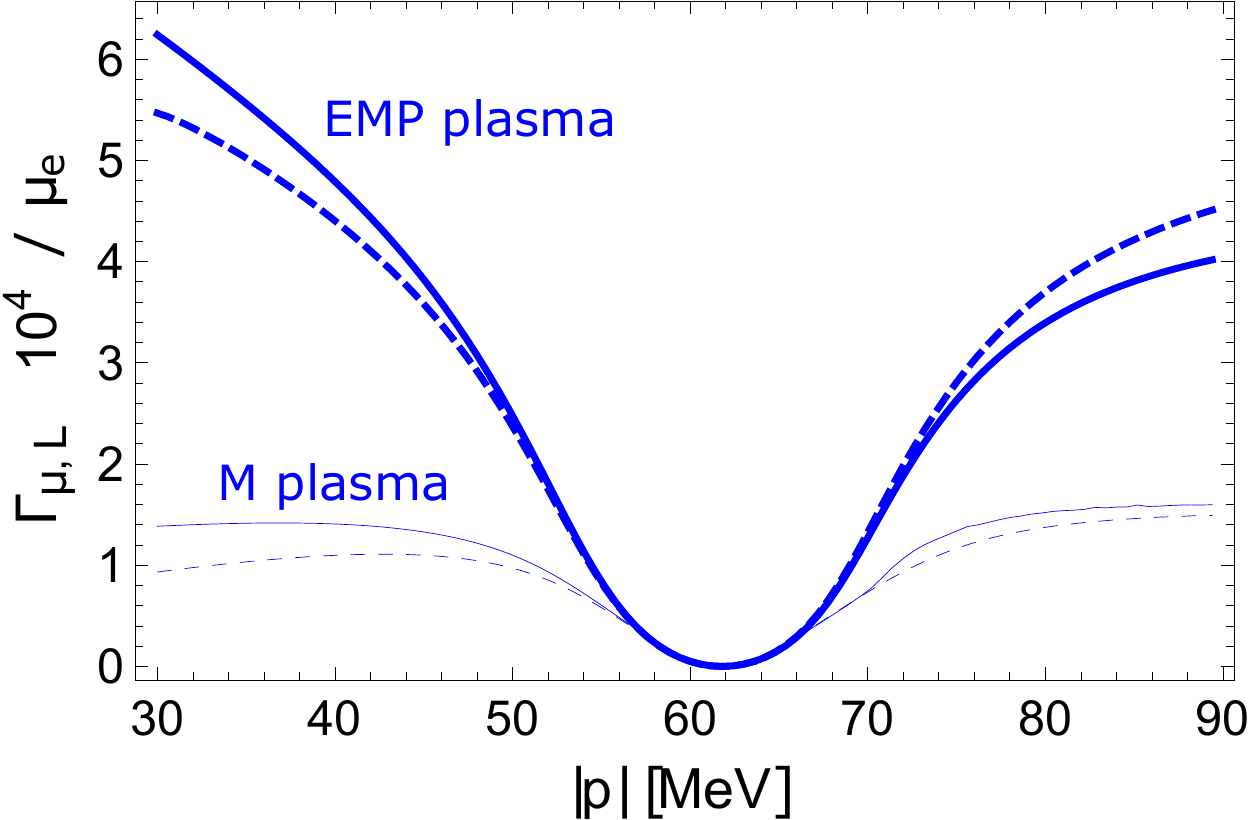}\hspace{1.2cm}\includegraphics[scale=0.51]{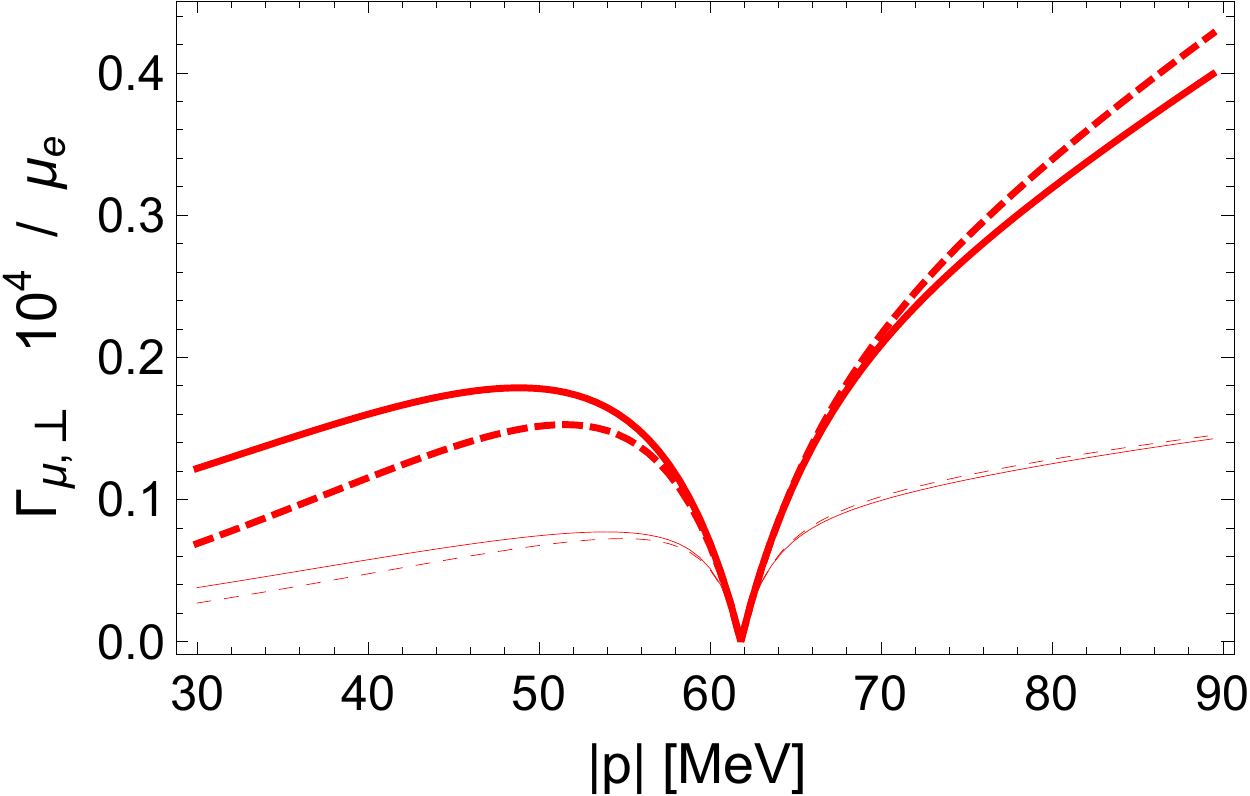}\\[2ex]
\hrule\vspace{0.2cm}
\includegraphics[scale=0.33]{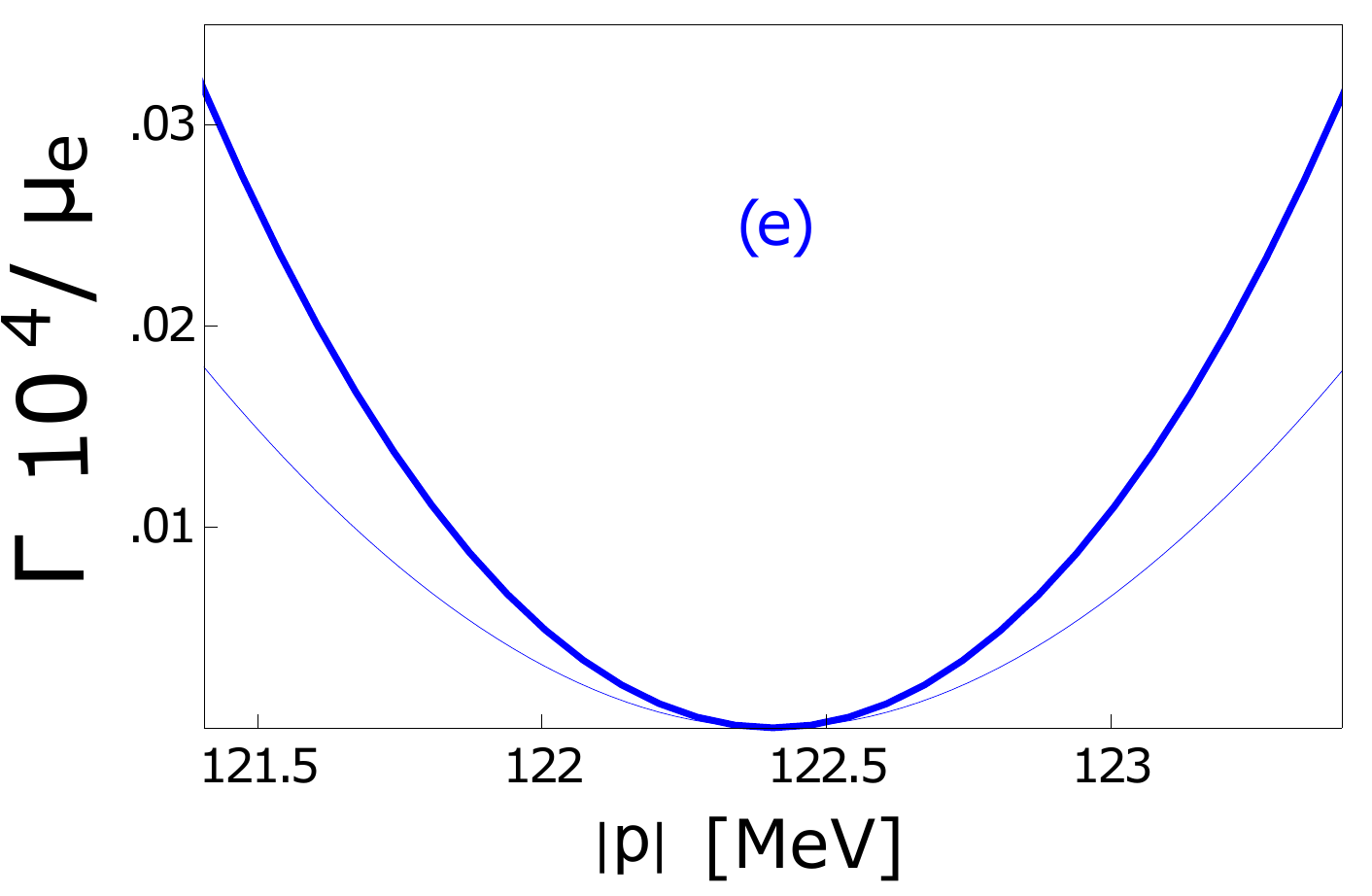}\,\,\,\includegraphics[scale=0.33]{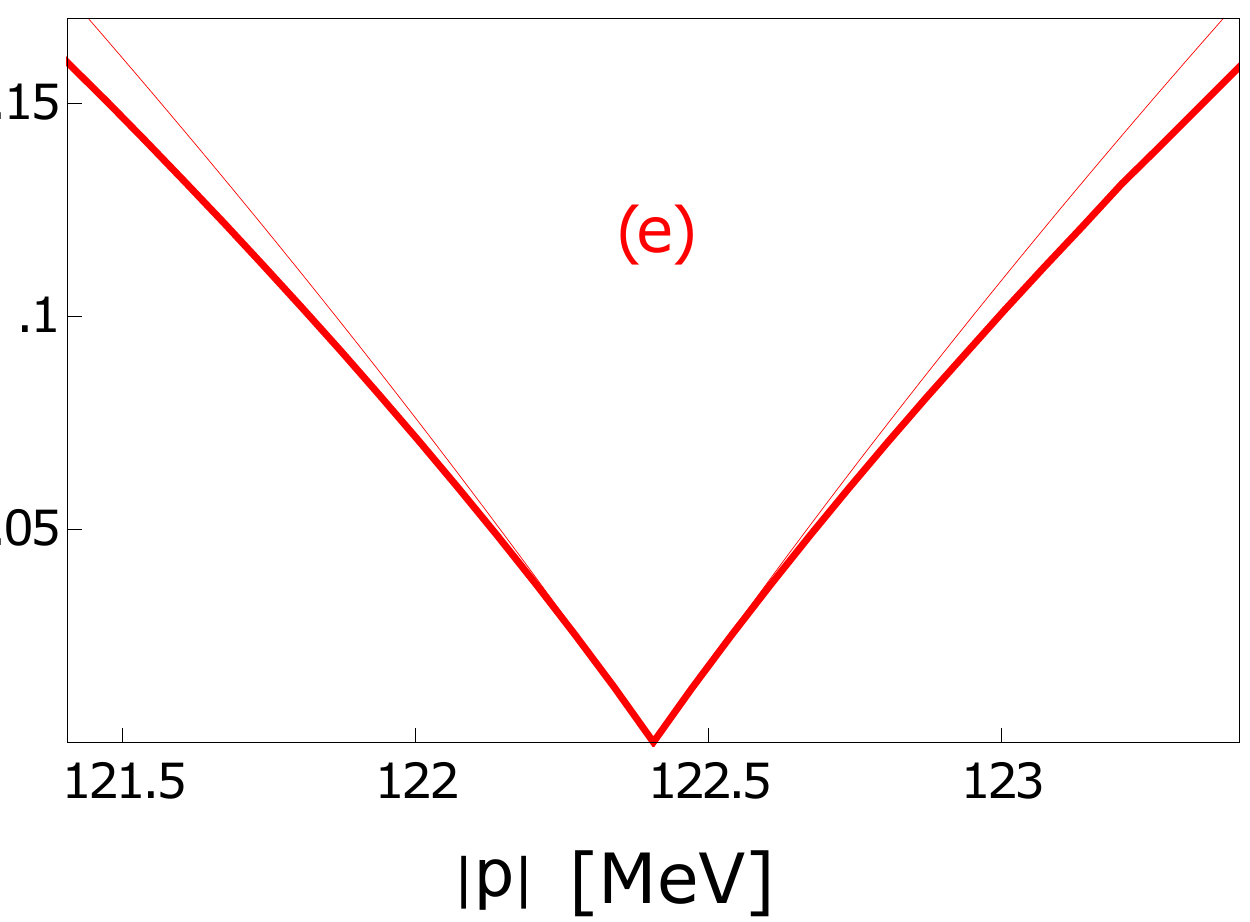}\,\,\,\includegraphics[scale=0.33]{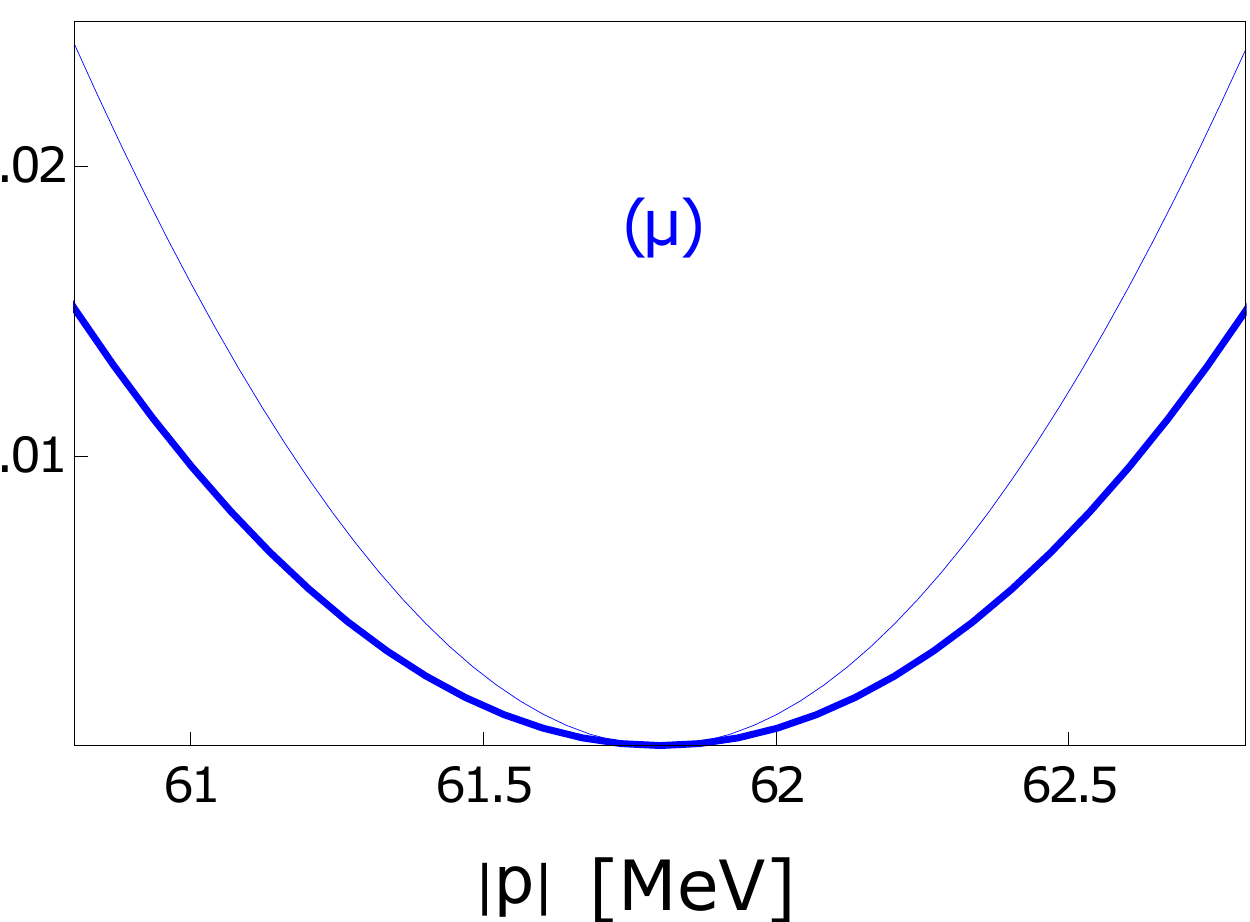}\,\,\,\includegraphics[scale=0.33]{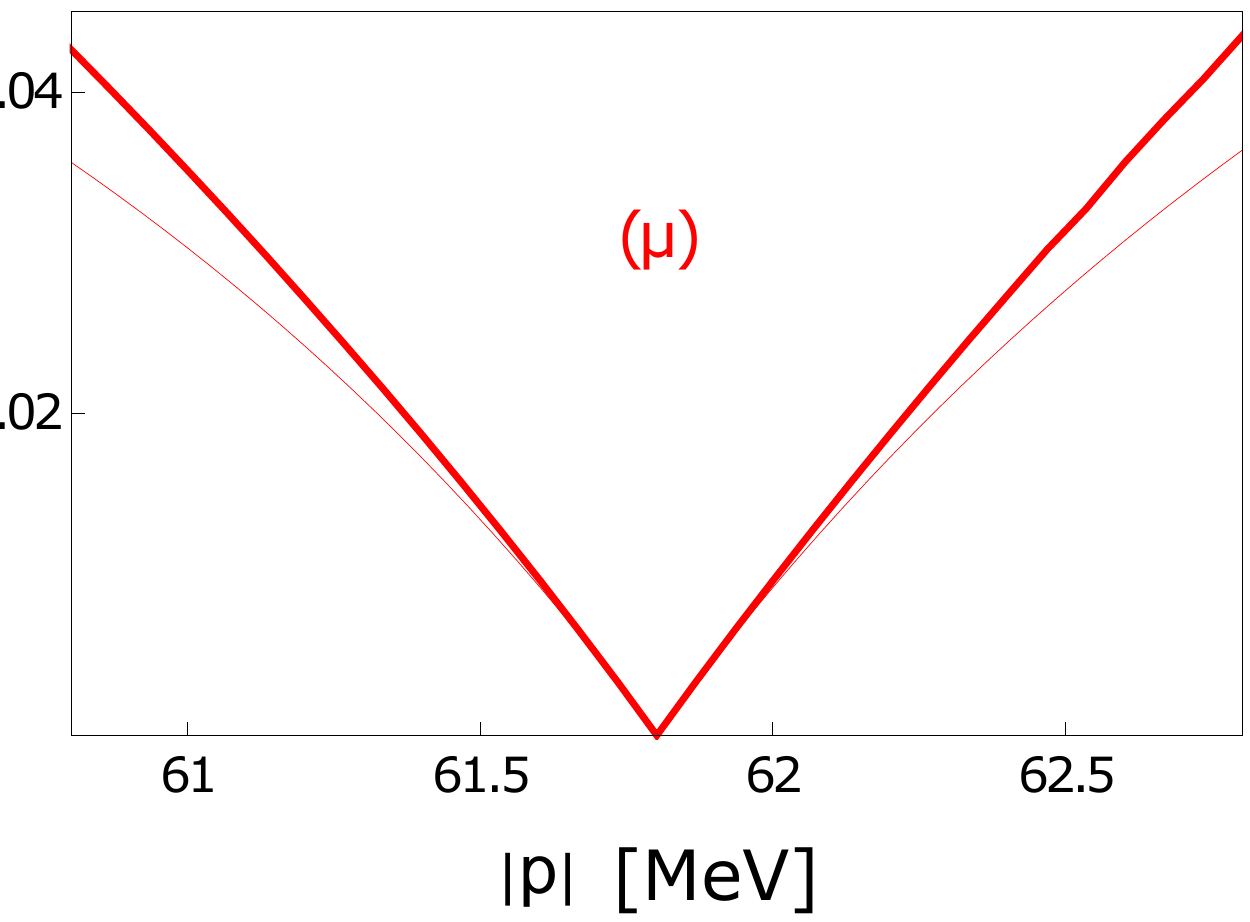}
\end{framed}
\setlength{\belowcaptionskip}{-10pt}
\caption{\label{fig:EMPComp} Comparison of longitudinal and transverse scattering rates in an EMP  plasma (thick) and a single-component plasma (thin) at saturation density and zero temperature. Relevant parameters can be found in Tab. \ref{tab:NRAPR}. Dashed lines indicate the corresponding HDL approximations. Further away from the Fermi surface scattering rates increase significantly in the EMP plasma, with the exception of transverse electron scattering, which decrease due to dynamical screening effects, see also Fig. \ref{fig:EEScat}. Though less pronounced, the impact of additional plasma constituents in close proximity to the Fermi surface is still sizeable (lower four figures). Electrons and muons, however, react in opposite manner: longitudinal (transverse) rates are increased (decreased) for electrons, while they are decreased (increased) for muons. }
\end{figure}
\begin{figure}
\begin{centering}
\includegraphics[scale=0.5]{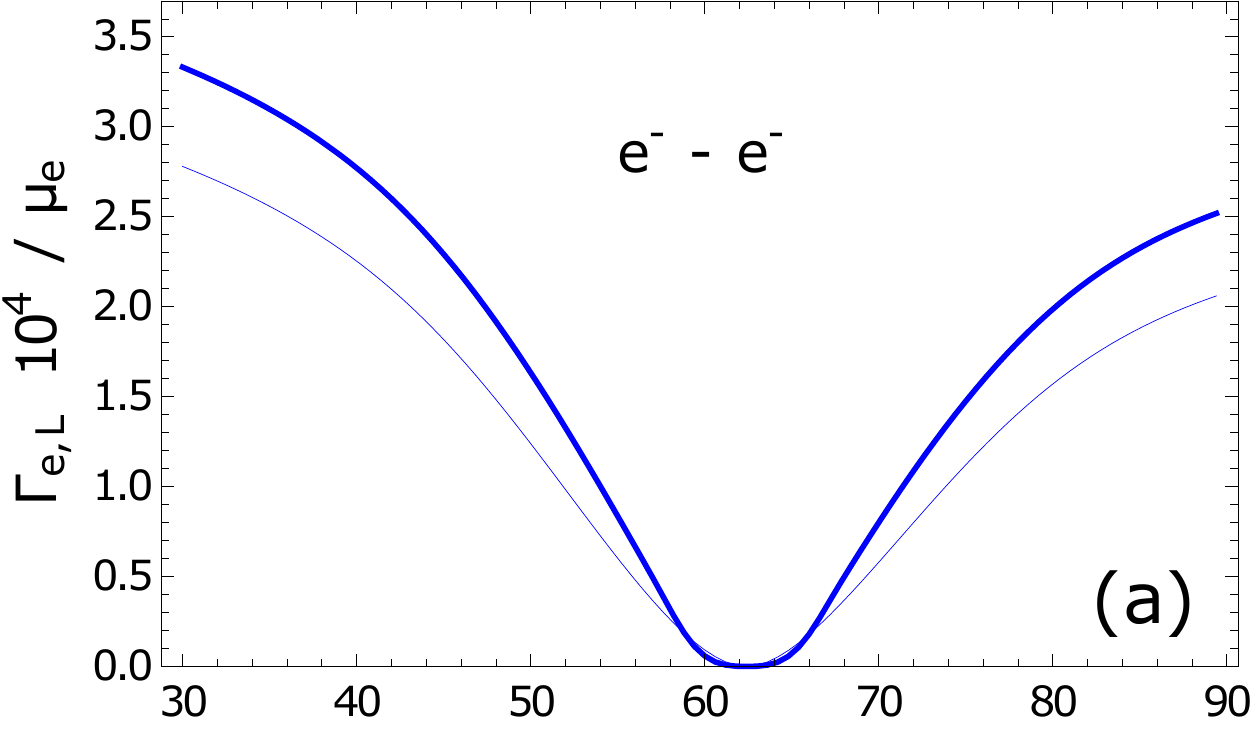}\,\,\includegraphics[scale=0.49]{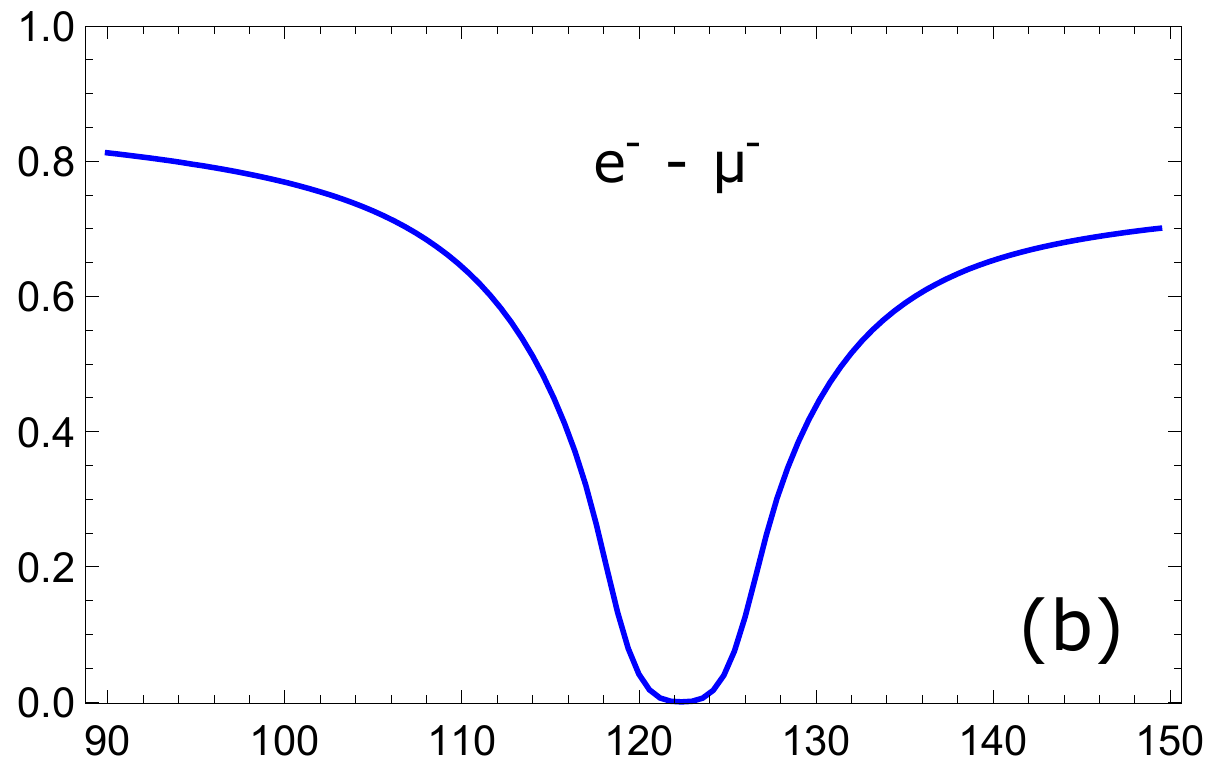}\,\,\includegraphics[scale=0.49]{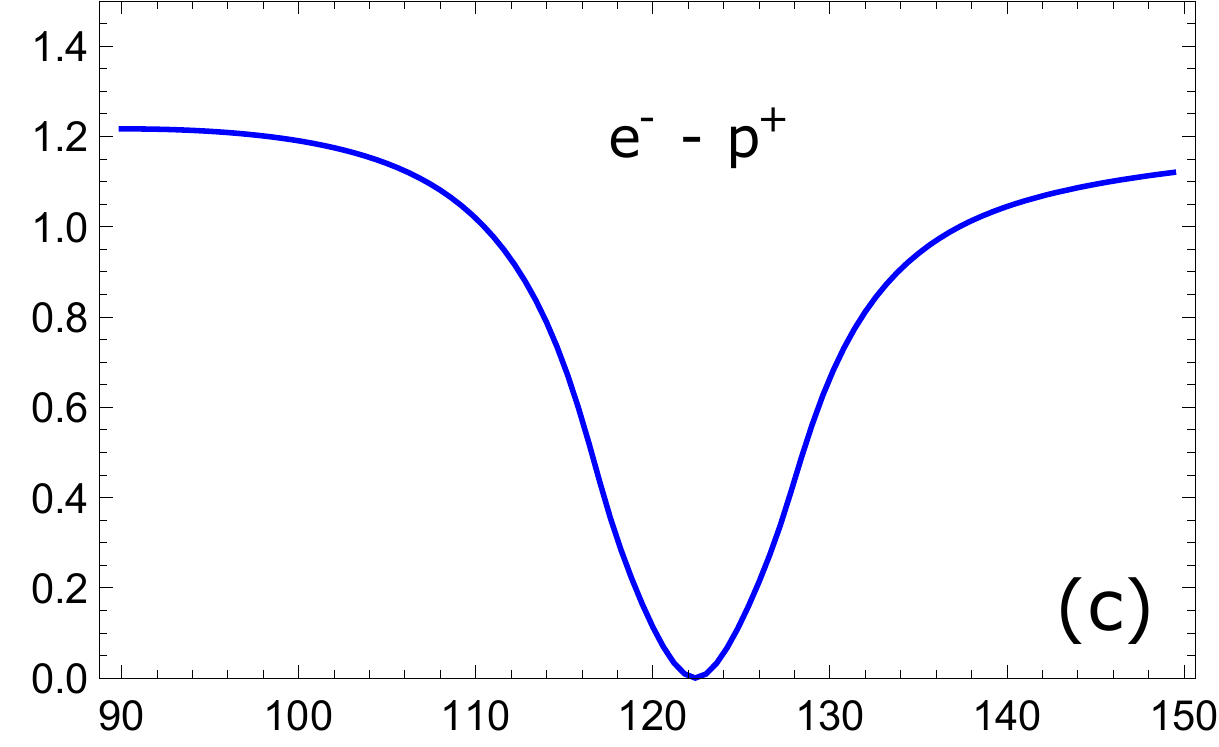}\\[2ex]\includegraphics[scale=0.5]{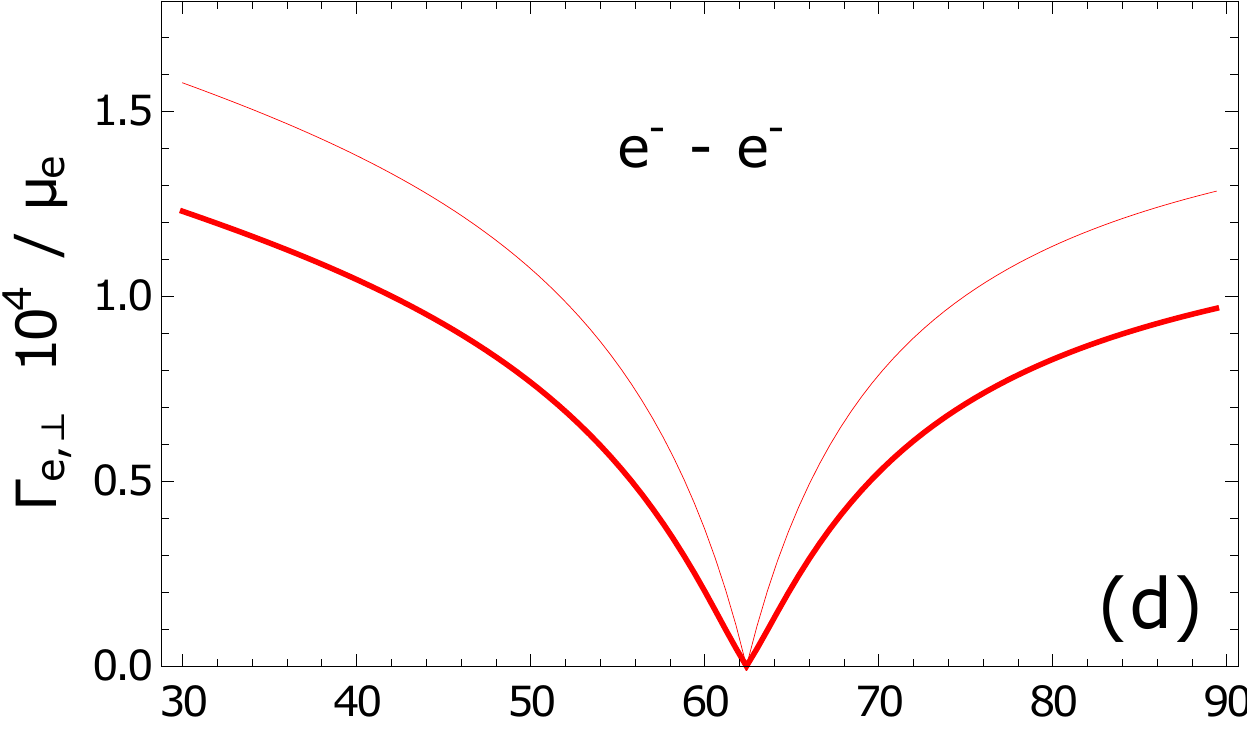}\,\includegraphics[scale=0.5]{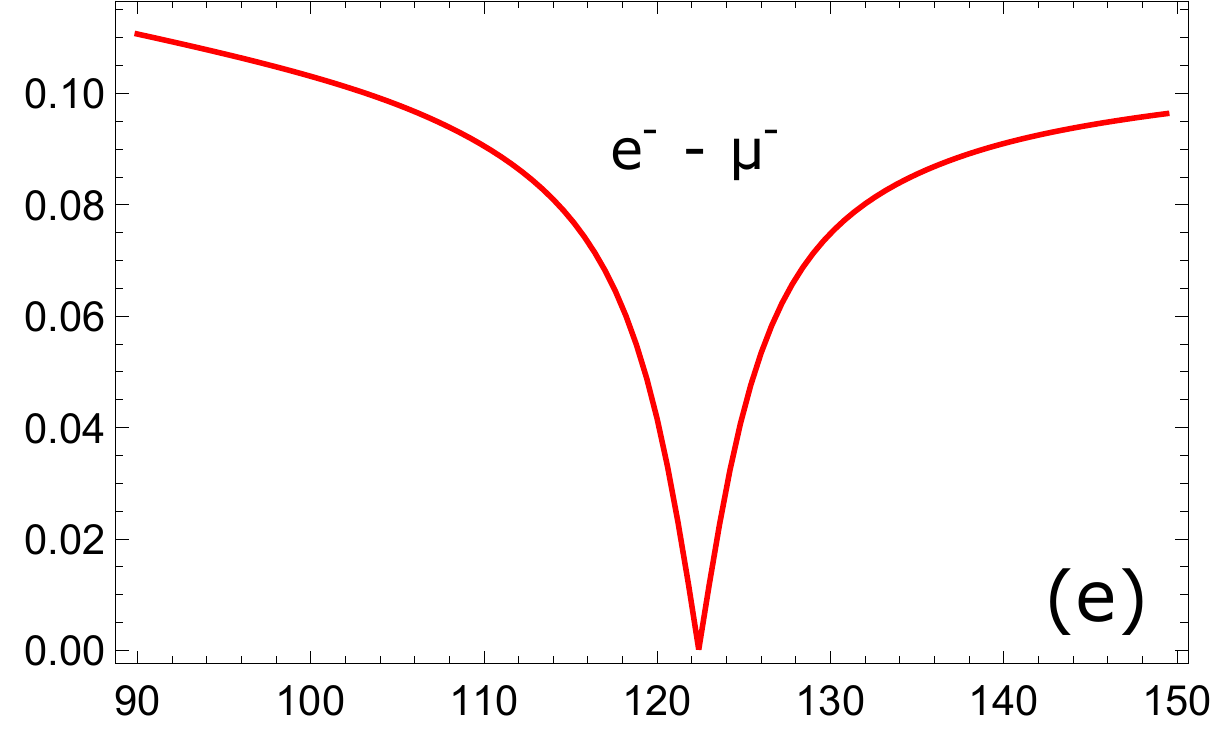}\,\includegraphics[scale=0.5]{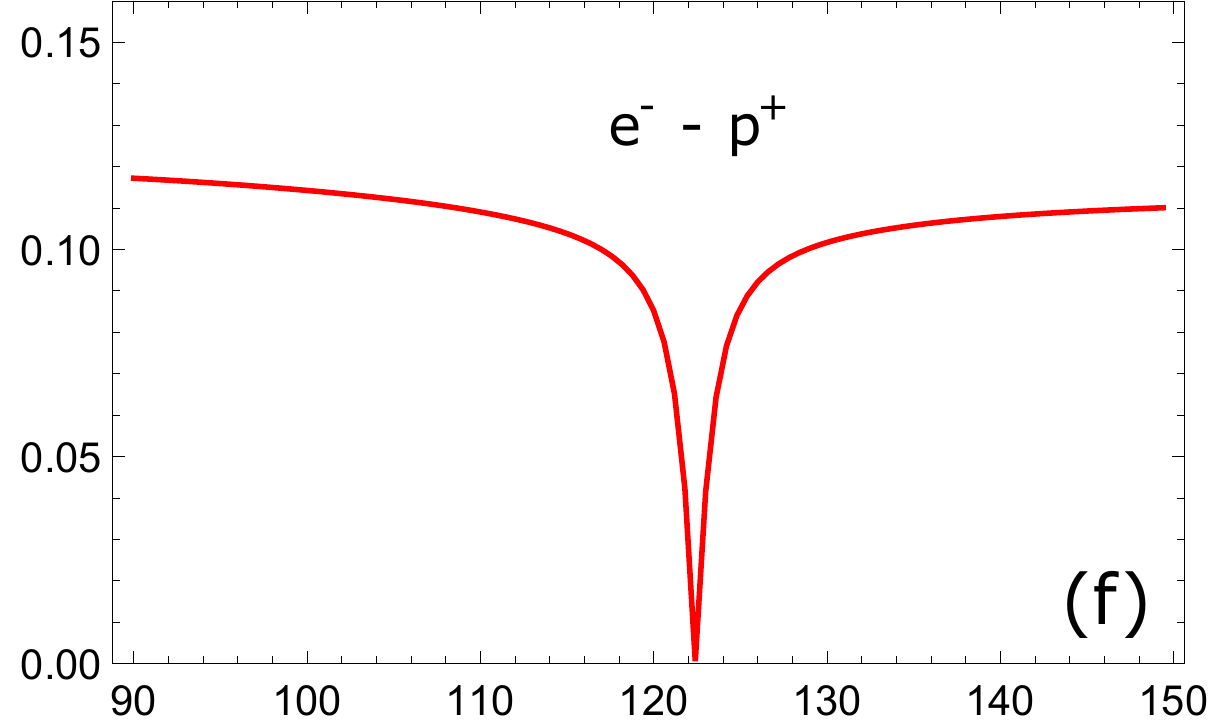}\\[2ex]
\includegraphics[scale=0.515]{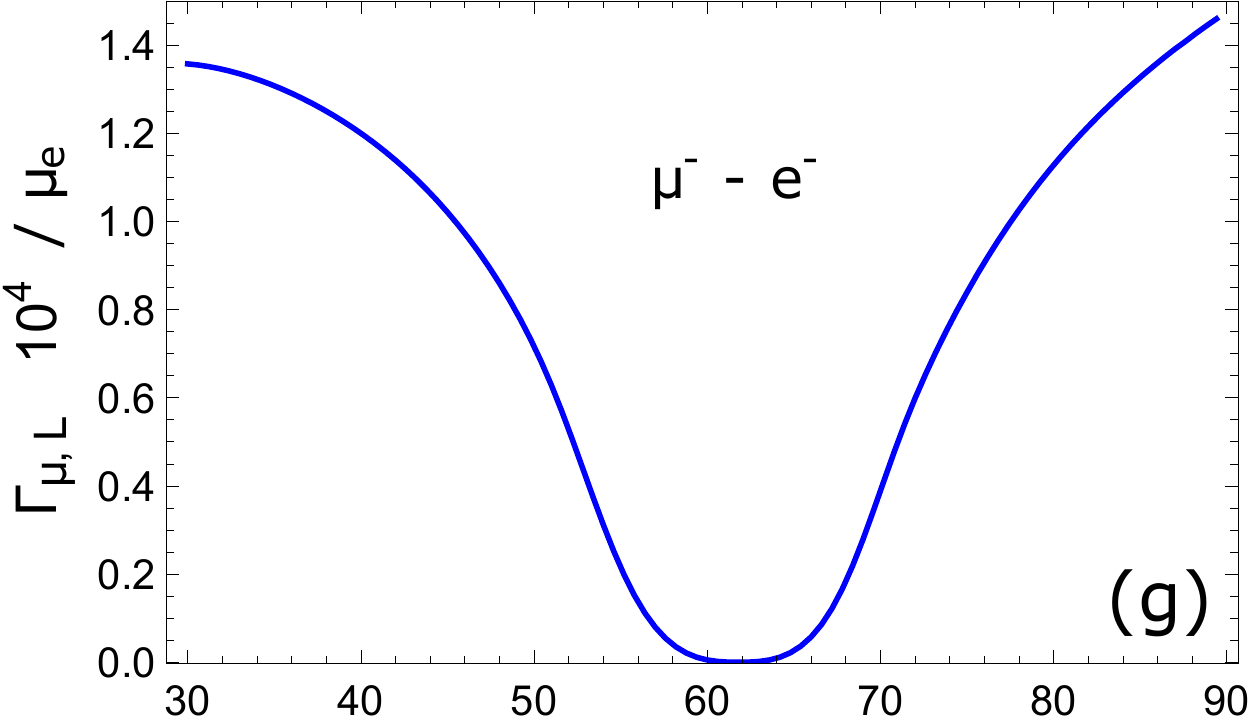}\,\,\includegraphics[scale=0.47]{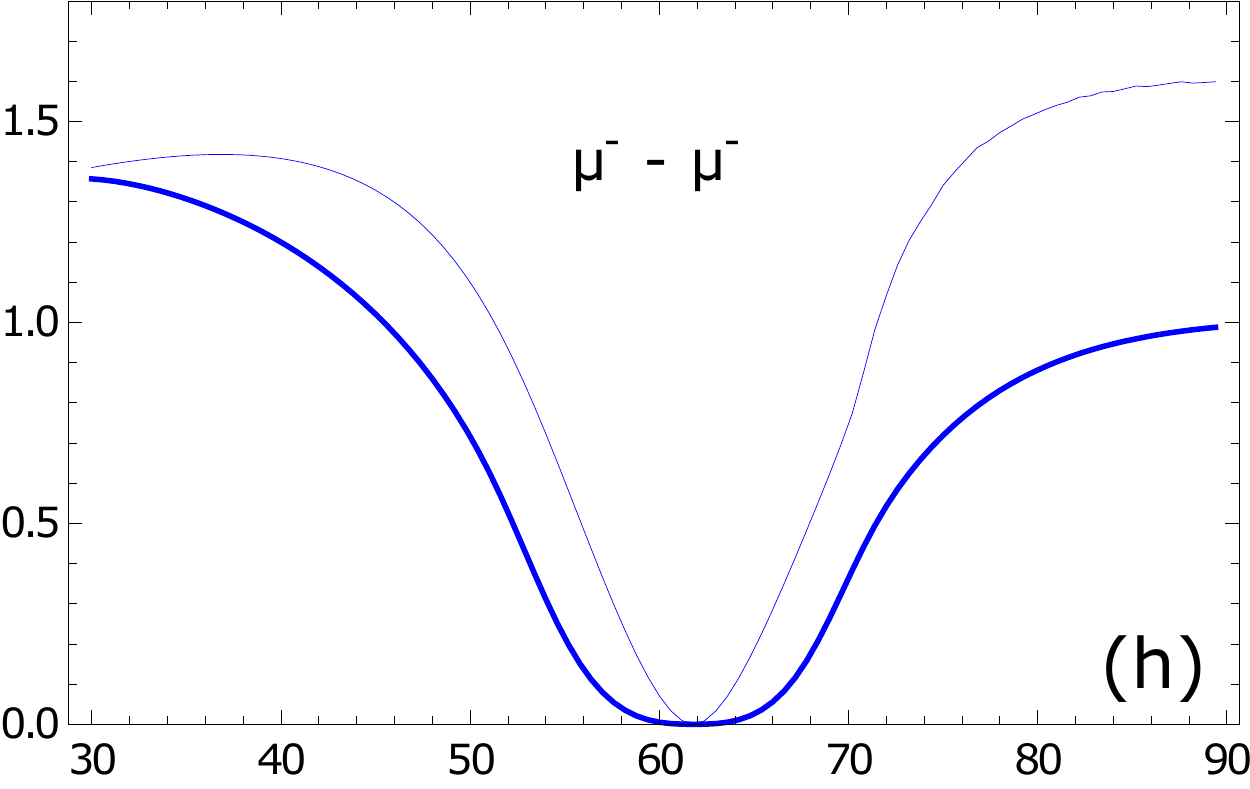}\,\,\includegraphics[scale=0.47]{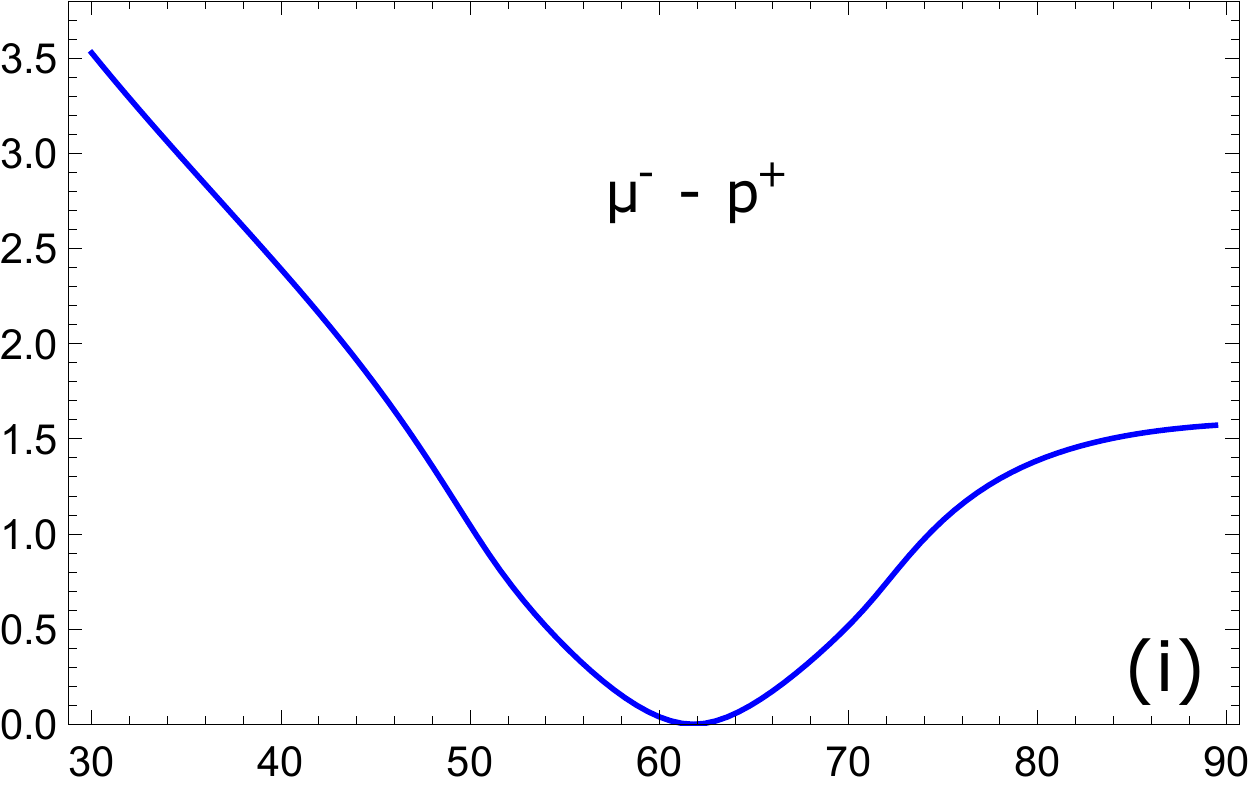}\\[2ex]
\includegraphics[scale=0.515]{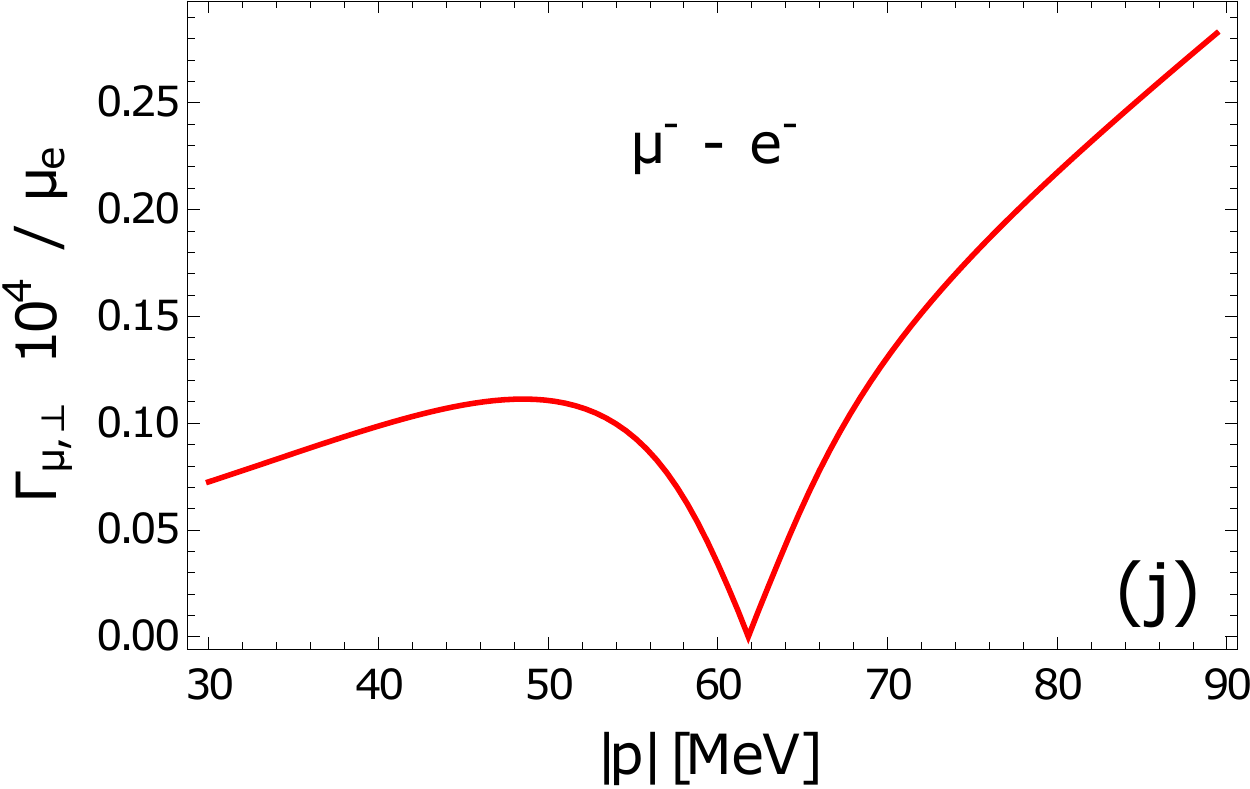}\,\includegraphics[scale=0.5]{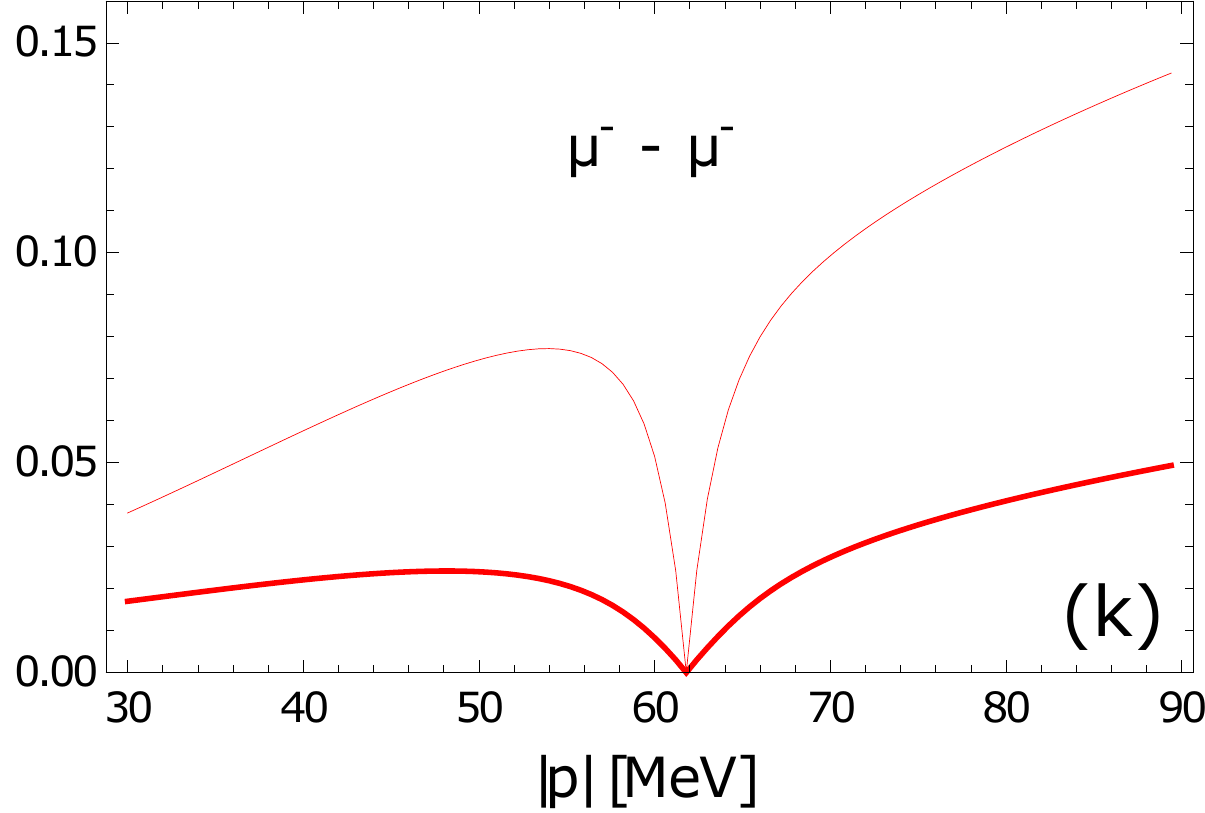}\,\includegraphics[scale=0.5]{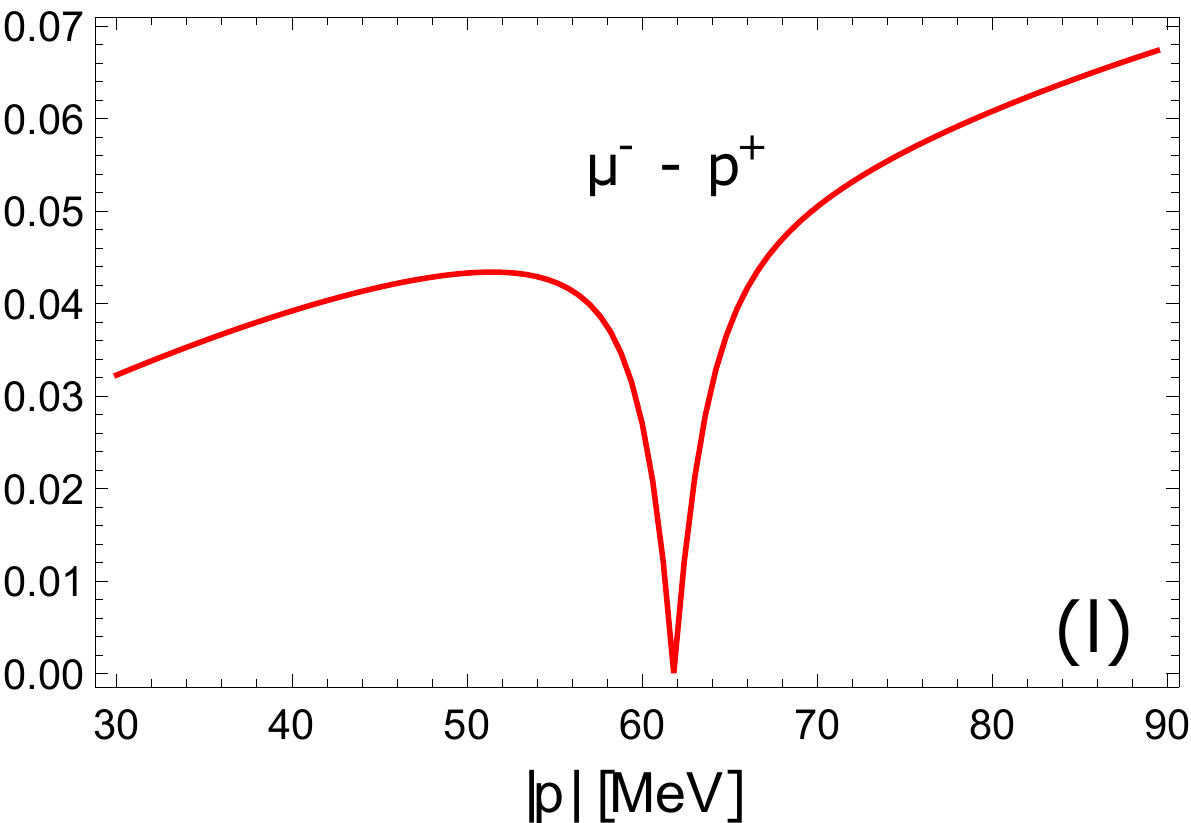}
\end{centering}
\caption{\label{fig:EEScat} Partial rates of longitudinal and transverse scattering of electrons and muons with other electrons, muons, and protons in a fully degenerate plasma at saturation density and zero temperature. Parameters according to Tab. \ref{tab:NRAPR}. The calculation of each contribution takes into account, that interactions are screened by \textit{all} constituents of the plasma. Thin lines display the scattering rates in the fictitious single component plasmas, see also Fig. \ref{fig:EMPComp}. The total interaction rate of electrons is clearly dominated by electron-electron scattering, although electron-muon and electron-proton scattering yield sizable contributions in the longitudinal channel. Screening effects of muons and protons \textit{increase} the rate of electron-electron scattering in the \textit{longitudinal} channel, while they \textit{decrease} it in the \textit{transverse} channel. In the latter case the reduction outweighs the small gains from transverse collisions with muons and protons, such that total transverse rate of electrons is effectively reduced in the EMP plasma.\newline The energy loss of muons results predominantly from longitudinal scattering, except for a small region around $k_f$, where the curvature of the longitudinal rate remains flat, and the linear increase of $\Gamma_\perp$ becomes the dominant feature. With increasing energy transfer, the longitudinal rates of muons become comparable to those of electrons. Compared to scattering in a ``muon plasma", screening reduces the rates in both channels. In the transverse channel, however, muon-electron scattering dominates such that the total transverse rate is in fact larger in the EMP plasma, see Fig. \ref{fig:EMPComp}.}
\end{figure}
\noindent  
To calculate the scattering rate of electrons and muons in a degenerate EMP plasma the photon spectrum derived in subsection \ref{subsec:multi} is employed. Correlations with nuclear interactions are ignored for the moment. The main purpose of this subsection is the study of the relative importance of new scattering channels arising from collisions with the respective other plasma constituents. It is instructive to revisit the weak screening approximation, which, as before, can be obtained from the HDL result Eq. \ref{eq:GammaHDL2} upon replacing the longitudinal and transverse photon spectra by 
\be \rho_L\rightarrow\frac{1}{2}\sum_i\frac{\,x_i\,m_{D,\,i}^2\,\Theta_i}{\left(\boldsymbol{q}^2+\sum_i\,m_{D,\,i}^2\right)^2}\,,\hspace{1cm}      \rho_\perp\rightarrow\frac{3}{4}\sum_i\frac{x_i\,\omega_{0,\,i}^2\,\Theta_i}{\boldsymbol{q}^4+(3\,\pi / 4)^2\,\sum_i \left(x_i\,\omega_{0,\,i}^2\,\Theta_i\right)^2}\,, \label{eq:RhoWeak}
\ee
with the plasma frequencies $\omega_{0,\,i}^2=e^2\,k_{f,\,i}^3/(3\pi^2\mu_{i})$, the parameter $x_i=q_0/ (v_{f,\,i}\,|\boldsymbol{q}|)$, and the kinematic restriction for Landau damping $\Theta_i=\Theta(1-x_i)$. The damping rate of quasi-particles with $|\boldsymbol{p}|\sim k_f$ is calculated analogously to Eqs. \ref{eq:GLongA} and \ref{eq:GPerpA}. For a given lepton species $l=\{e,\,\mu\}$ scattering with all other constituents of the plasma (including themselves) one finds
\be\label{eq:LongEMPAn}
\Gamma_{L,\,l}\sim\frac{e^2}{32}\,\frac{1}{v_{f,\,l}}\,\left(\epsilon_{\boldsymbol{p}}-\mu_l\right)^2\,\frac{1}{M^3}\,\left(\sum_i \frac{m_{D,\,i}^2}{v_{f,\,i}}\right)\,,\hspace{1cm}\epsilon_{\boldsymbol{p}}=\sqrt{\boldsymbol{p}^2+m_l^2}\,,\hspace{1cm}M=\sqrt{\sum_i m_{D,\,i}^2}\,,
\ee
where, depending on the density, $i=\{e,\,p\}$ or $i=\{e,\,\mu,\,p\}$. To leading order, the \textit{transverse} rate retains its form Eq. \ref{eq:GPerpA} in the multi-component plasma, i.e. remains ignorant of other plasma constituents, see e.g. Fig. \ref{fig:EMPComp}. A comparison with the full one-loop results for $\Gamma(\boldsymbol{p})$ in a range $|\boldsymbol{p}|=(k_f\pm 1)$ MeV is shown in Fig. \ref{fig:WeakFull}. The accuracy of the weak-screening approximation increases with the density. At lower densities weak-screening results tend to underestimate longitudinal rates, and overestimate transverse rates. The transverse rates are particularly sensitive to higher order screening effects. The discrepancy in the longitudinal channel can be understood by taking a closer look at the photon spectrum, see Fig. \ref{fig:RhoLongStat}. At very low energies $q_0\ll v_{f,\,p}|\boldsymbol{q}|$  the longitudinal spectrum is predominantly shaped by Landau damping of protons, which is captured correctly by the approximate expression \ref{eq:RhoWeak}. As the Fermi velocity $v_{f,\,p}$ decreases with the density, the domain in which Eq. \ref{eq:RhoWeak} describes the photon spectrum correctly becomes smaller, and the range of $|\epsilon_{\boldsymbol{p}}-\mu|$ in which reliable results for the rates can be obtained decreases accordingly.\newline 
Full one-loop and HDL results for electrons and muons with momenta $(k_f\pm 30)$ MeV are displayed in Fig. \ref{fig:EMPComp}, and compared to the corresponding rates in a \textit{single component} plasma. The HDL results remain an excellent approximation for quasiparticles with momenta in a range of roughly $|\boldsymbol{p}|\sim(k_f\pm 10)$ MeV, in the range of $|\boldsymbol{p}|\sim(k_f\pm 1)$ MeV they are virtually indistinguishable from full one-loop results. In addition, Fig. \ref{fig:EEScat} shows a decomposition of $\Gamma_{L,\,\perp}$ into the partial energy losses inflicted by collisions with electrons, muons, and protons. The \textit{longitudinal} rates of electrons increase  strongly in the multi-component plasma, both, far away, and in close proximity to the Fermi surface. This happens to a small extent because of screening effects, see Fig. \ref{fig:EEScat} (a), and predominantly because of contributions from electron-muon (b) and electron-proton (c) scattering. The rates of electrons in the \textit{transverse} channel, in contrast, are clearly dominated by electron-electron scattering (d), contributions from electron-muon (e) and electron-proton (f) scattering are small. Screening effects of heavy fermions \textit{reduce} the rate of transverse electron-electron scattering, and the small gains from collisions with muons and protons cannot compensate this reduction. The net result is, that the transverse damping rate of electrons is reduced in the EMP plasma for any given momentum $|\boldsymbol{p}|$. In total, electron scattering remains dominated by the exchange of transverse photons close to the Fermi surface, but becomes dominated by the exchange of longitudinal plasmons further away, see Fig. \ref{fig:EMPComp}. Muons close to the Fermi surface exhibit the opposite characteristics of electrons: screening in the EMP plasma reduces (increases) the magnitude of longitudinal (transverse) rates, and transverse rates remain the dominant contribution. Further away, the rates in both channels increase strongly. Particularly important in this respect are muon-proton scattering (i) for the longitudinal channel, and muon-electron scattering (j) for the transverse channel. All things considered, scattering rates of fermions in close proximity to their respective Fermi surfaces in an EMP plasma are clearly dominated by transverse electron-electron scattering.

\subsection{Scattering in the EMPN plasma}
\label{subsec:ResultsInduced}

\begin{figure}[t]
\hspace{0.6cm}\includegraphics[scale=0.5]{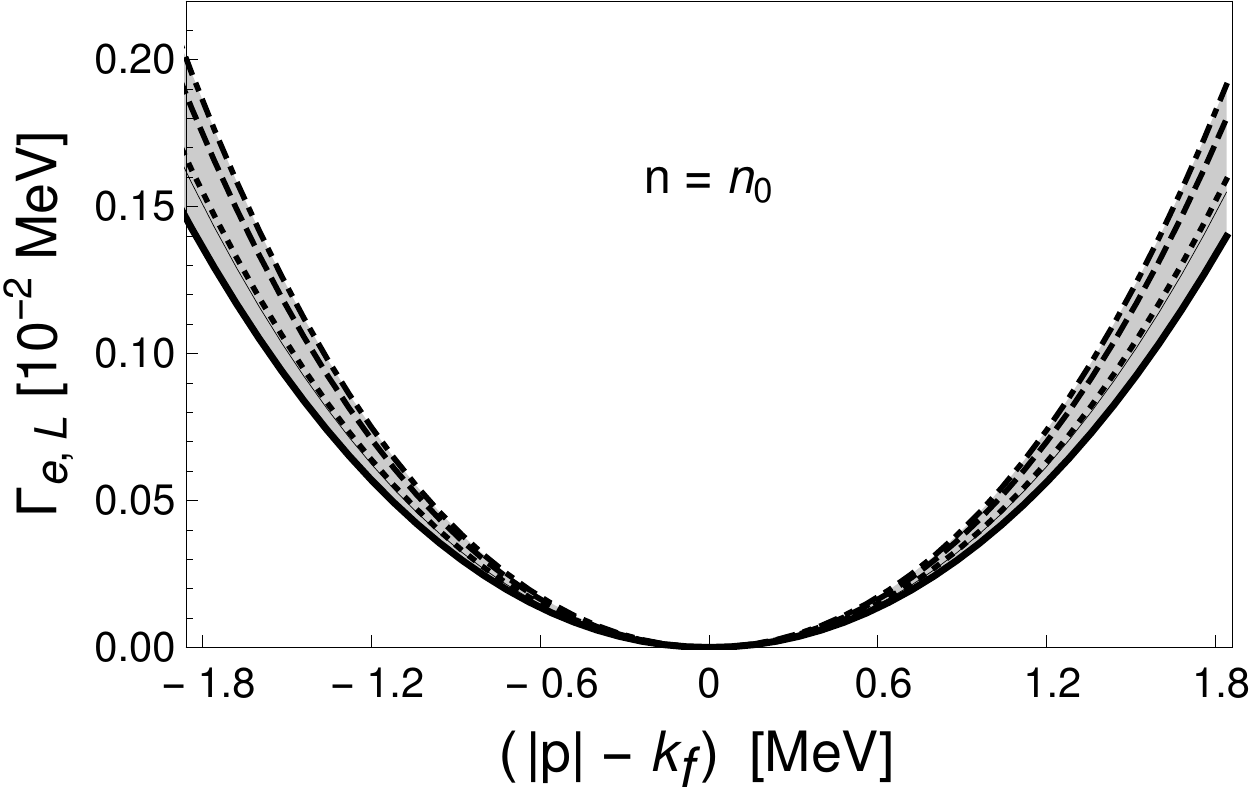}\hspace{1cm}\includegraphics[scale=0.5]{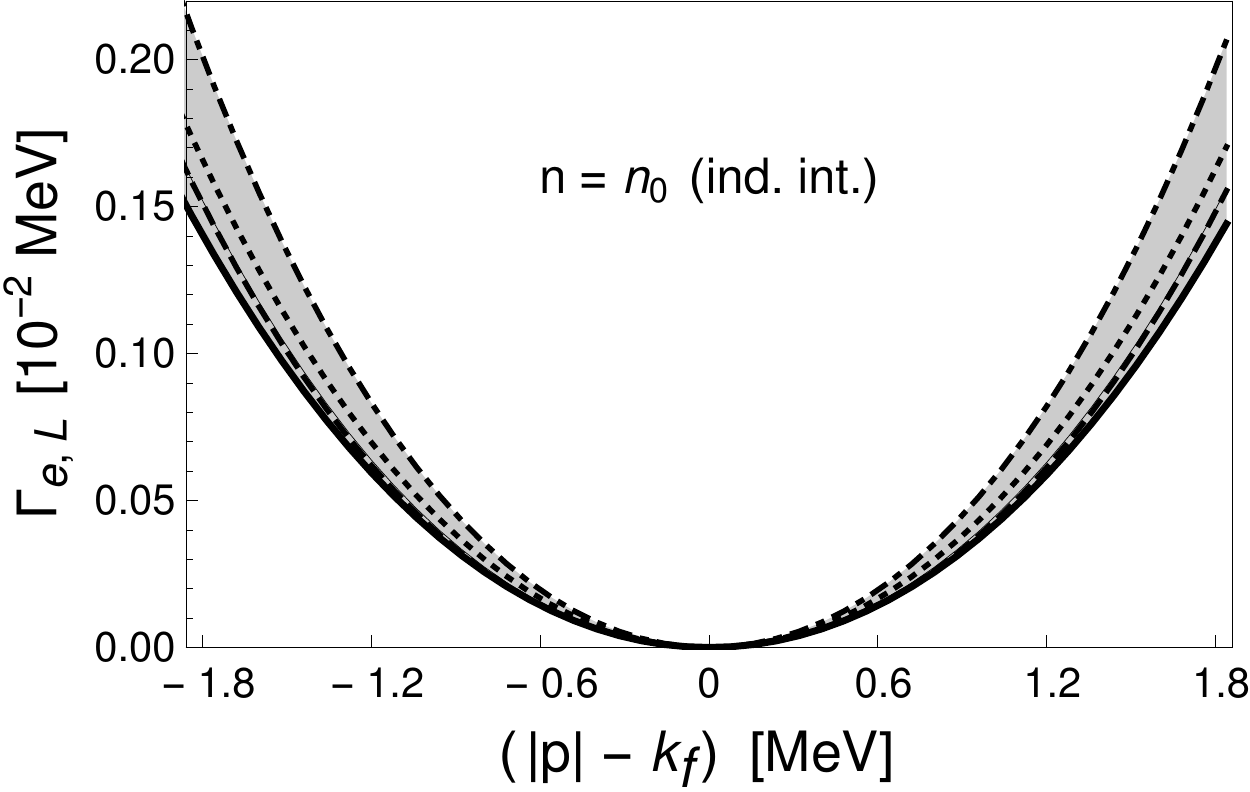}\\[2ex]
\hspace{0.6cm}\includegraphics[scale=0.5]{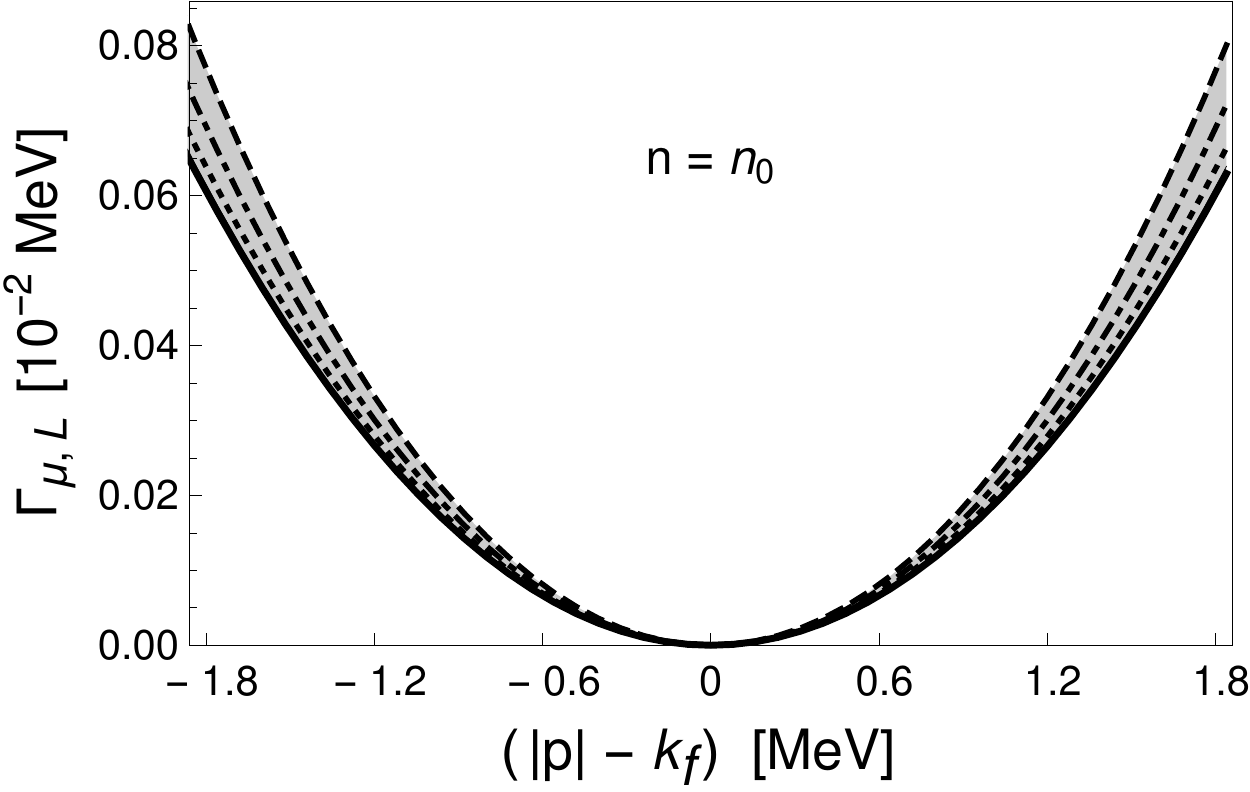}\hspace{1cm}\includegraphics[scale=0.5]{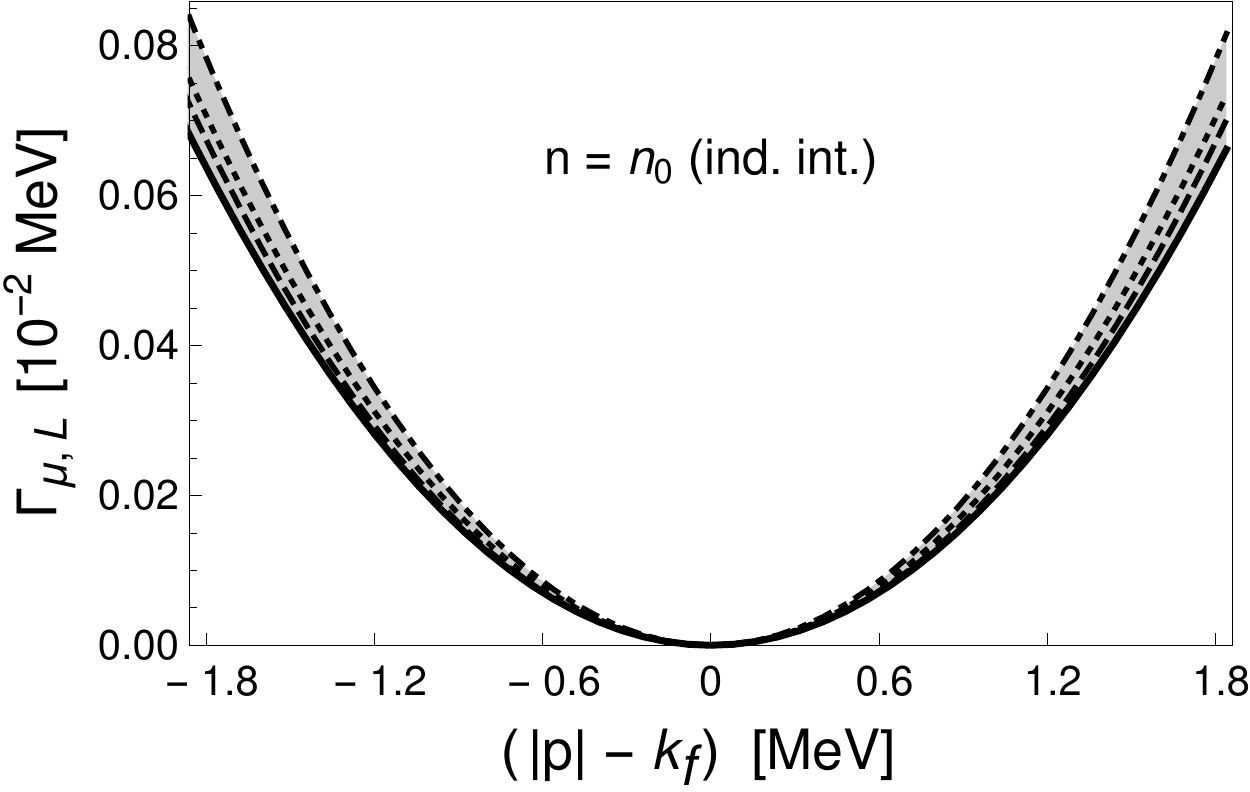}
\caption{\label{fig:ComparisonN0} Comparison of longitudinal interaction rates of electrons and muons, without (left) and with (right) induced interactions. The same Skyrme models as in Fig. \ref{fig:Screen} are used, parameters can be found in Tab. \ref{tab:NRAPR}. At fixed density, the Fermi momenta of electrons calculate to slightly different values in each model, and the rates are plotted in a range of $\pm2$ MeV around these values. To facilitate a comparison the results are not normalized over $\mu_e$ as in the previous figures. At saturation density, the bands on the left and right hand side overlap almost perfectly, indicating that the impact of induced interactions is minimal.}
\end{figure}
\begin{figure}
\begin{centering}
\,\,\includegraphics[scale=0.56]{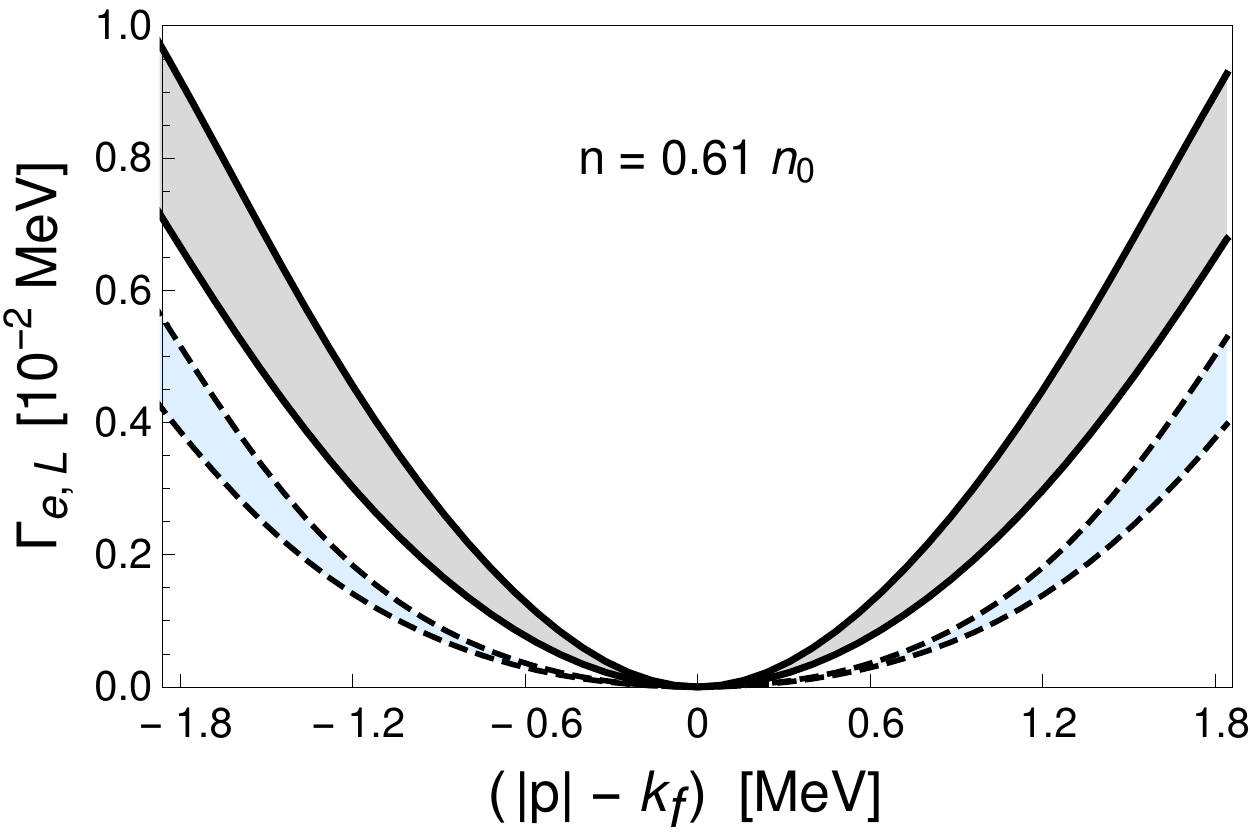}\hspace{0.85cm}\includegraphics[scale=0.56]{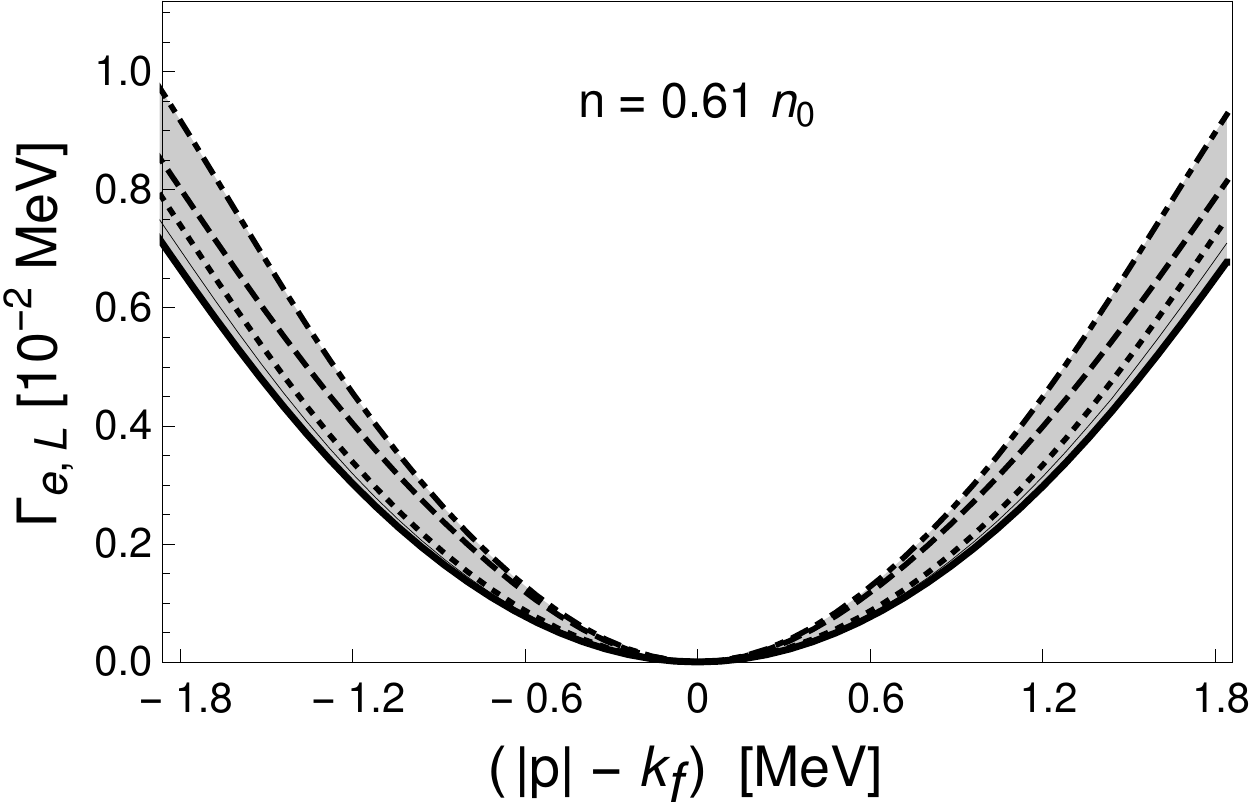}\\[2ex]\includegraphics[scale=0.56]{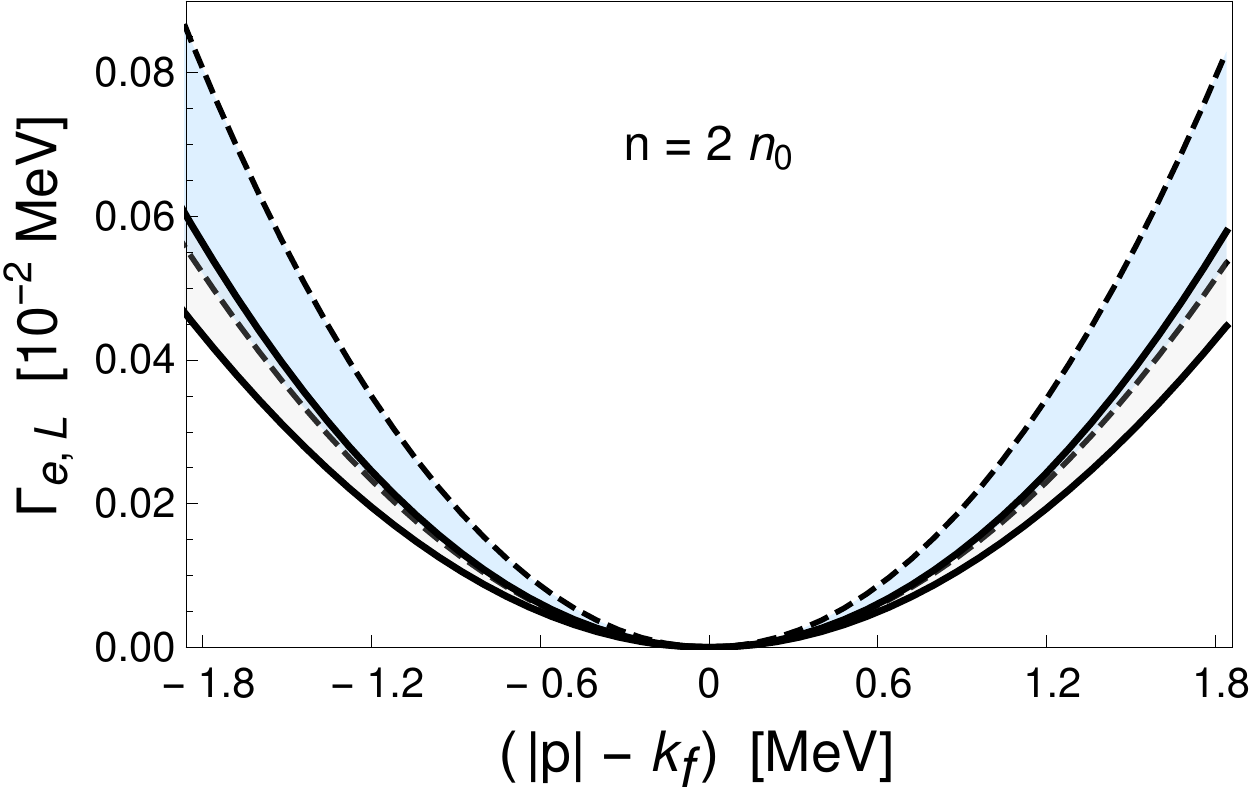}\hspace{0.75cm}\includegraphics[scale=0.56]{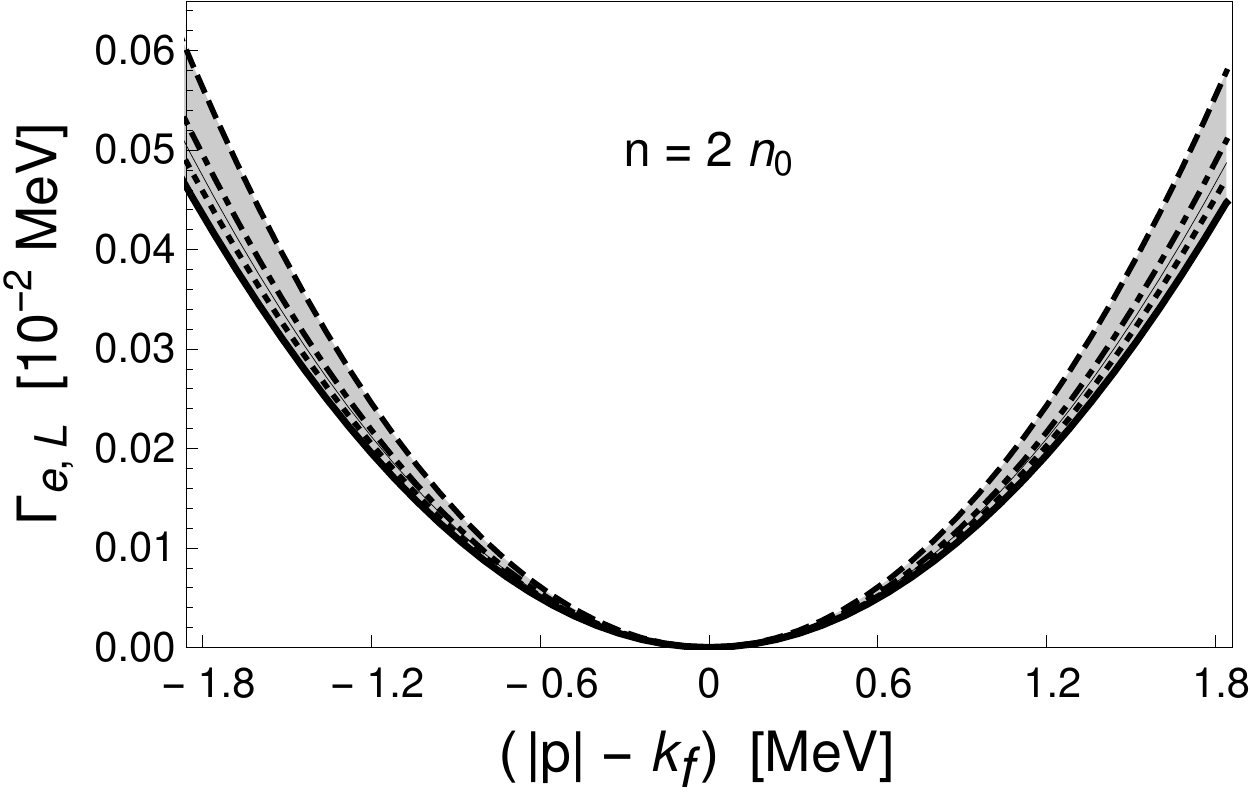}\\[2ex]
\includegraphics[scale=0.56]{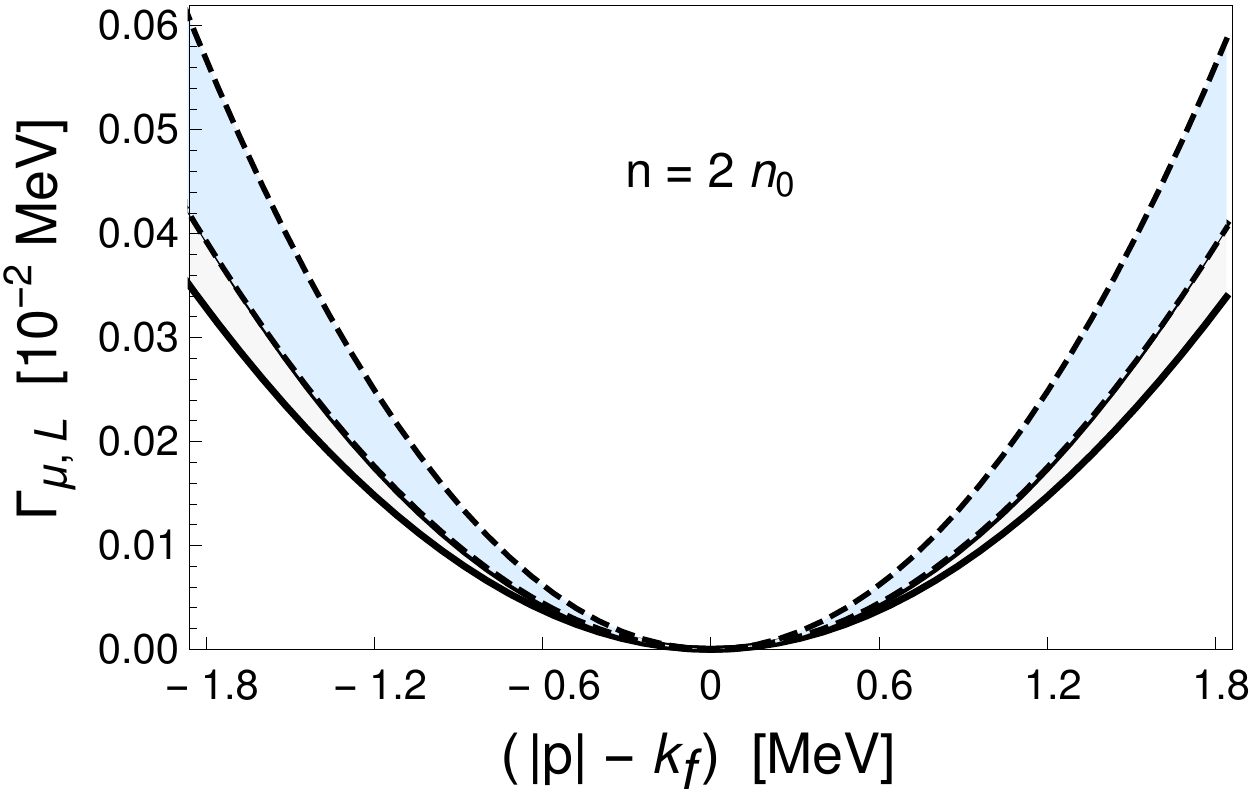}\hspace{0.85cm}\includegraphics[scale=0.56]{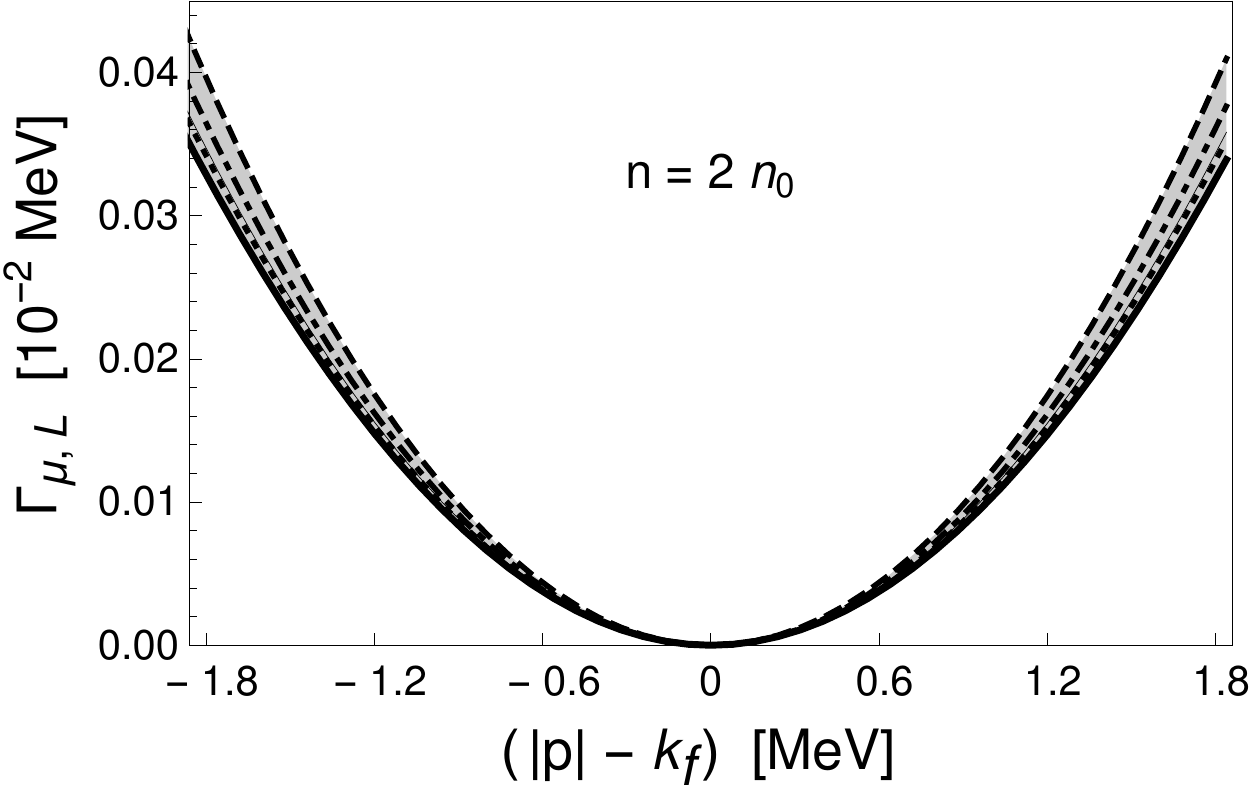}
\end{centering}
\setlength{\belowcaptionskip}{-8pt}
\caption{\label{fig:InducedComp} Scattering rates of electrons close to the crust-core boundary, and of electrons and muons deeper inside the core. Gray and blue bands represent the results for $\Gamma_L$ with and without induced interactions respectively. The panels on the right-hand side resolve the various Skyrme models that constitute the gray bands. It seems counter-intuitive at first that the width of the gray band broadens at lower densities, where Skyrme models are supposed to be well constrained. This happens mainly because the results are sensitive to the relative distance of $n$ to the critical density $n_c$, which is slightly different in each model. As expected, the impact of induced interactions at $n=0.61\,n_0$  is substantial. Even when induced interactions are ignored the scattering rate of electrons increases more than an order of magnitude going from $n=2\,n_0$ to $n=0.61\,n_0$. The muon rates increase by roughly a factor of $2$ going from $n=2\,n_0$ to $n=n_0$, see Fig. \ref{fig:inducedEN}. Deeper inside the core the scattering rates of muons become as large as those of electrons. 
}
\end{figure}
\begin{figure}
\begin{centering}
\includegraphics[scale=0.6]{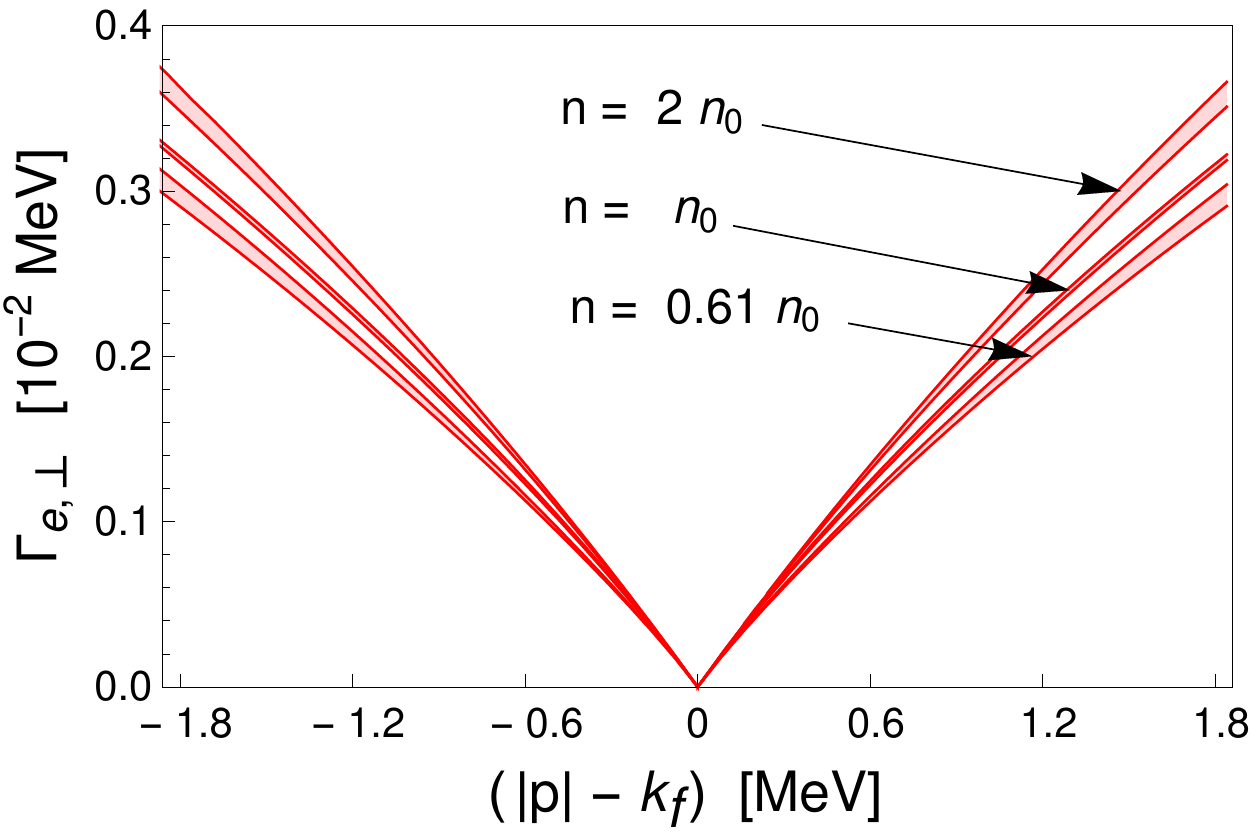}\,\hspace{0.5cm}\includegraphics[scale=0.62]{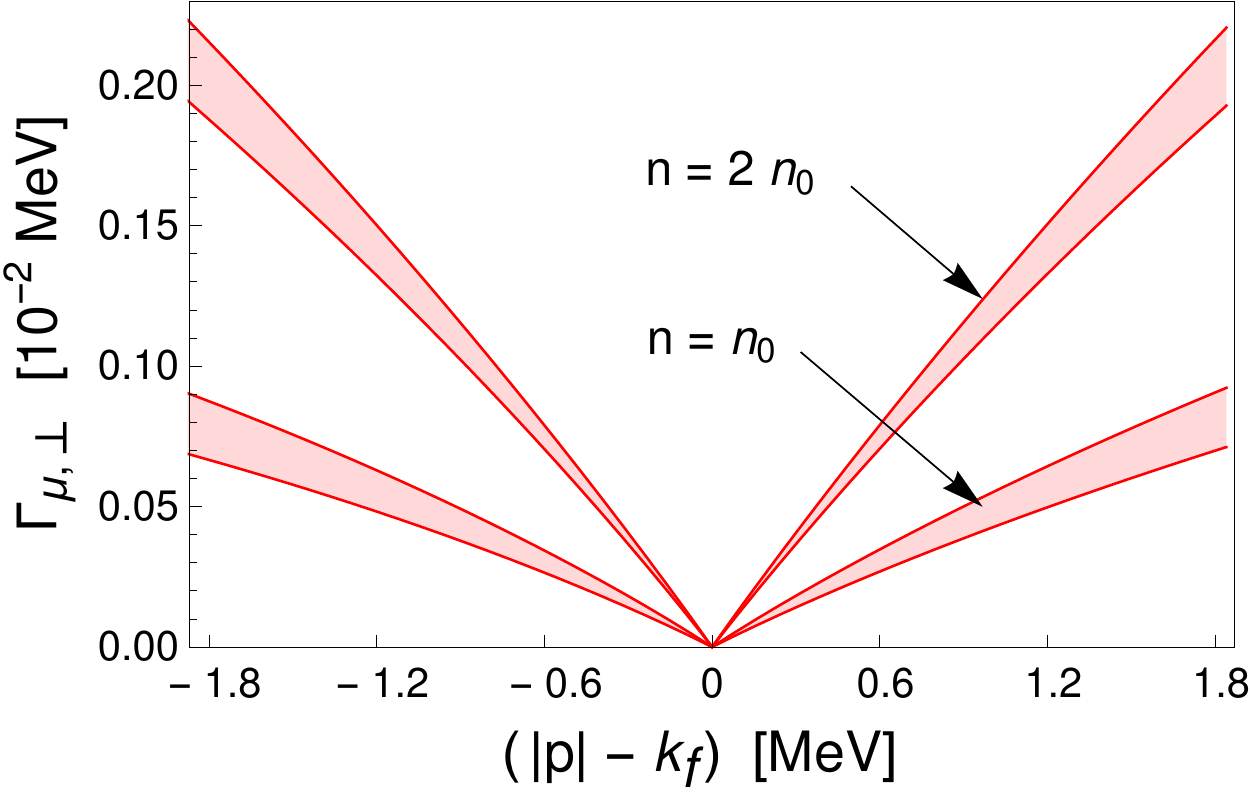}\\[2ex]
\end{centering}
\caption{\label{fig:PerpCompare} Transverse rates of electrons (left) und muons (right) at $n=0.61\,n_0$ (where muons are absent), $n=n_0$, and $n=2\,n_0$, using the Skyrme parameters listed in Tab. \ref{tab:NRAPR}. The transverse rates increase with density, in particular for muons, which become increasingly relativistic in nature. Transverse scattering of electrons hardly exhibits any model dependence. The relative model dependence for muon rates is larger, in part due to variations in the critical densities for muon onset, see Tab. \ref{tab:CritDens}.         
}
\end{figure}
\begin{figure}[!h]
\begin{centering}
\includegraphics[scale=0.5]{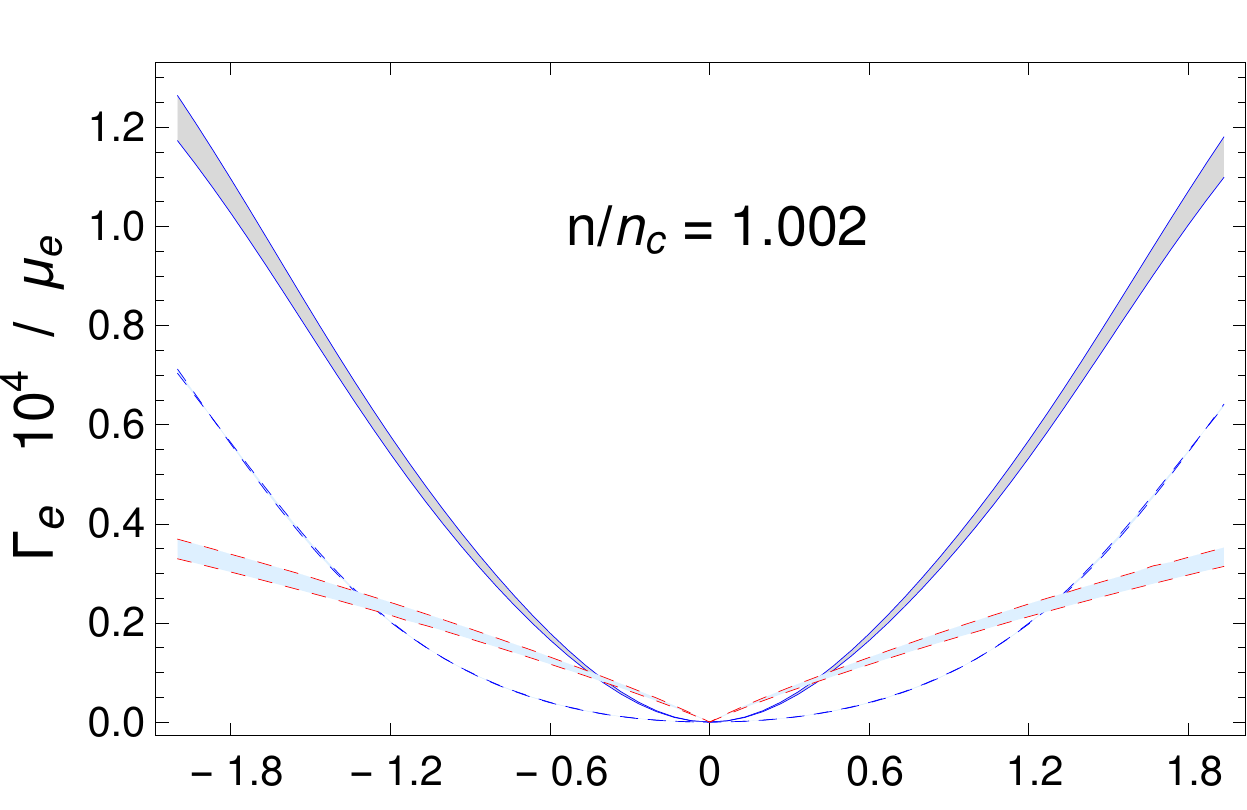}\,\includegraphics[scale=0.5]{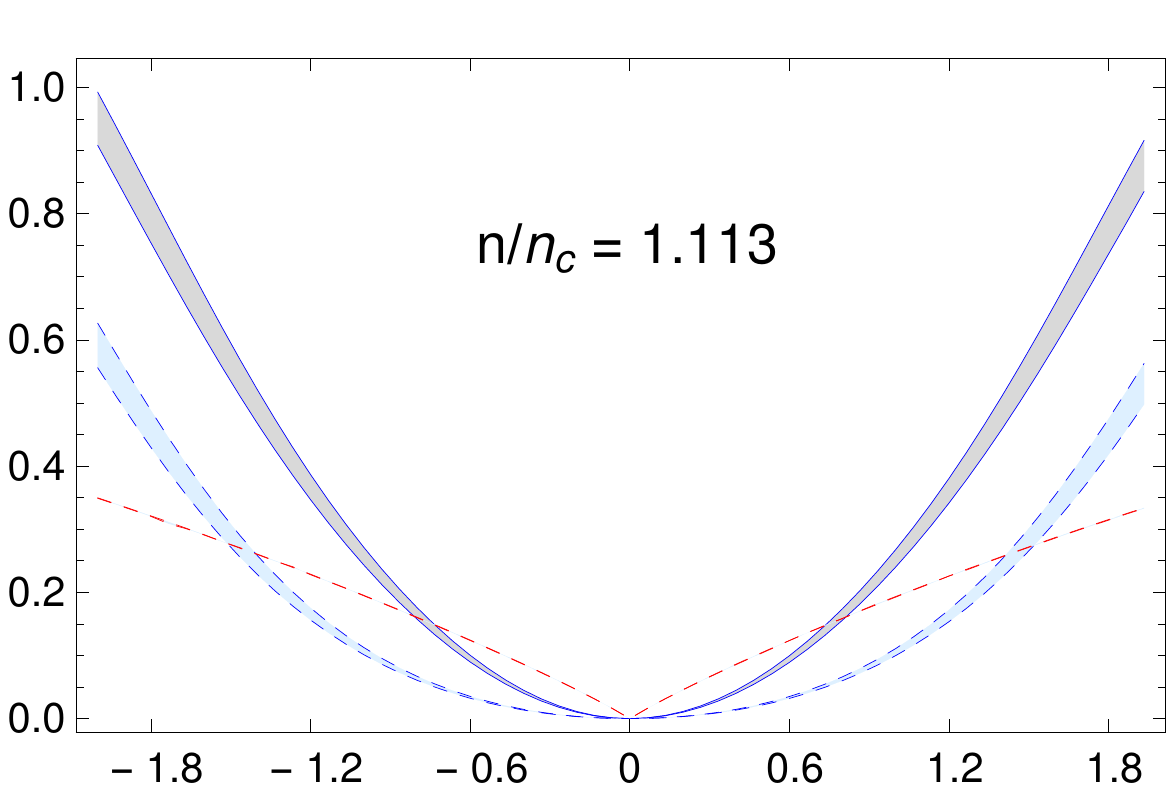}\includegraphics[scale=0.5]{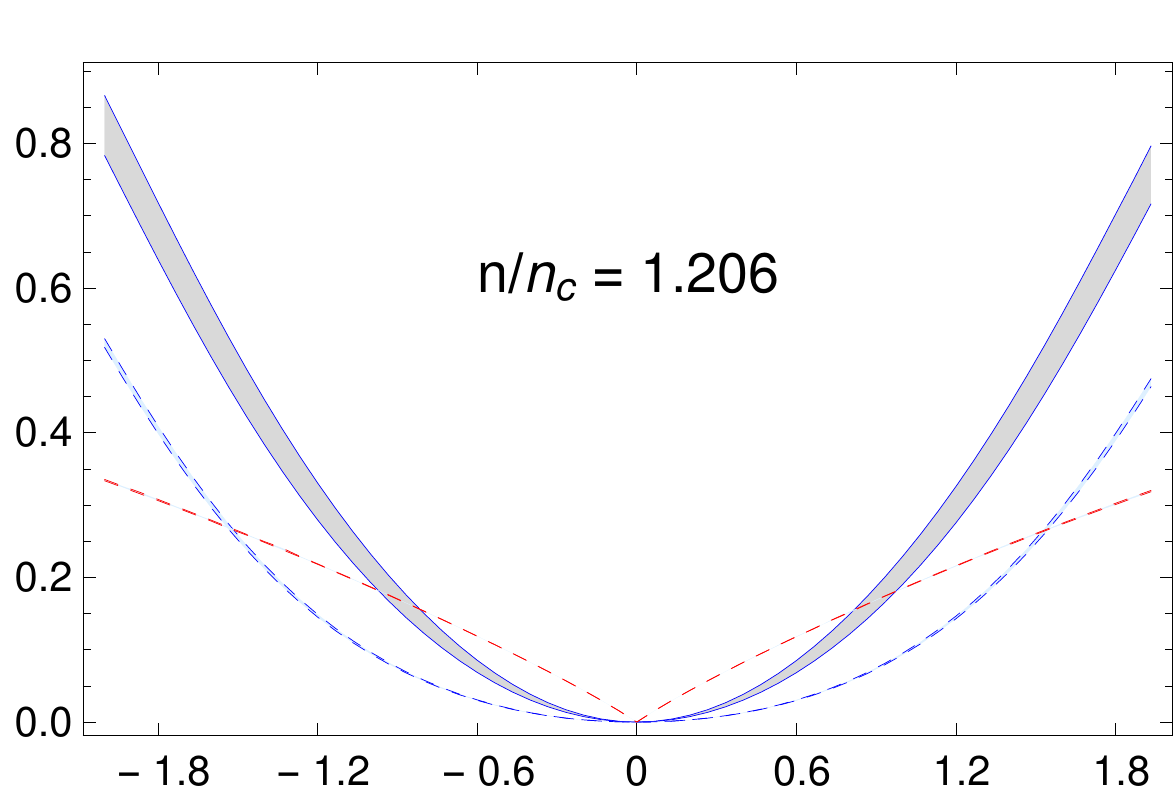}\\[2ex]
\includegraphics[scale=0.5]{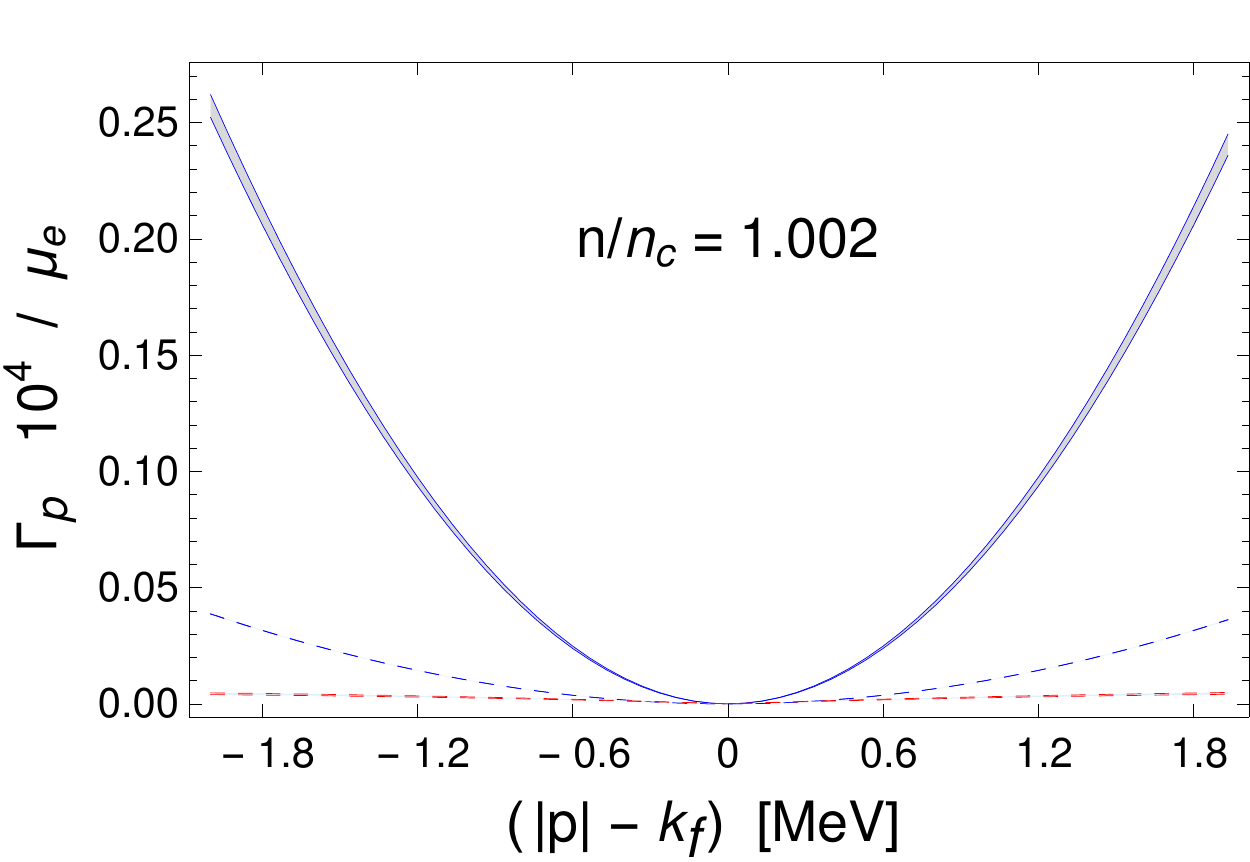}\,\includegraphics[scale=0.5]{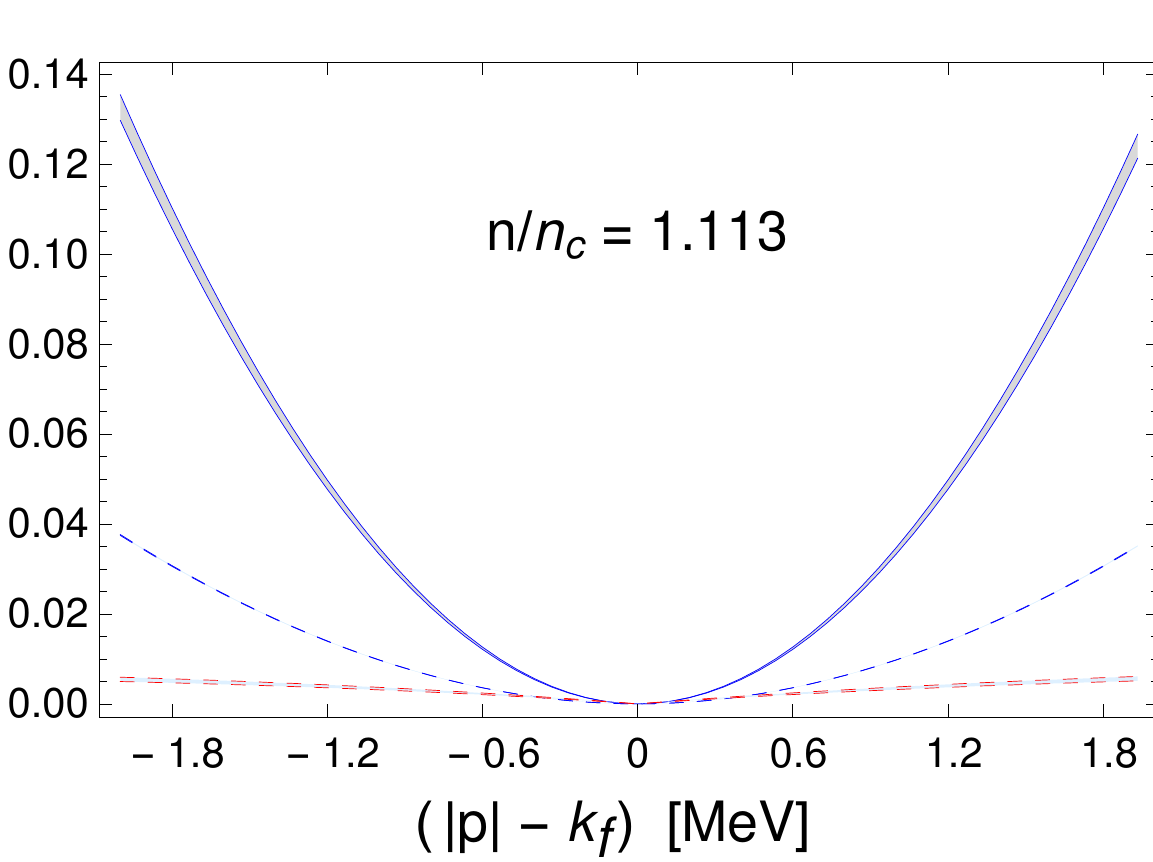}\,\includegraphics[scale=0.5]{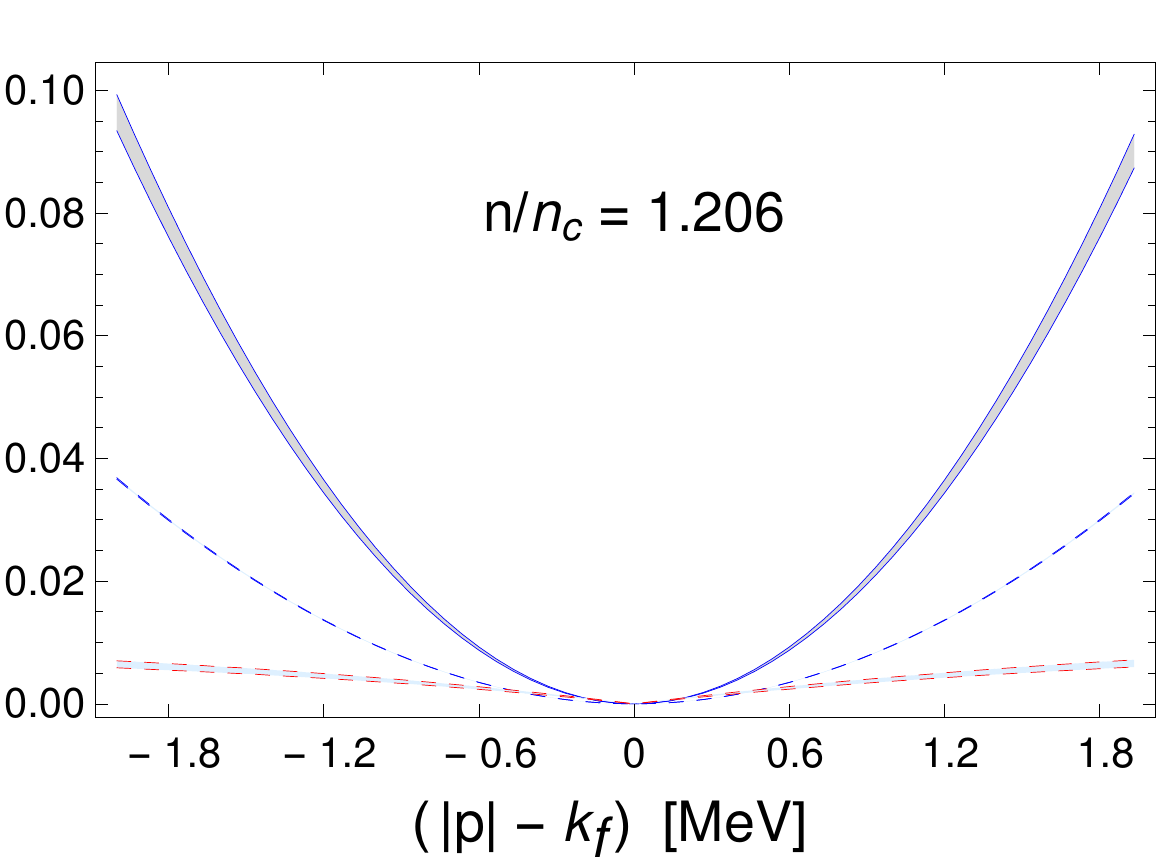}
\end{centering}
\setlength{\belowcaptionskip}{-8pt}
\caption{\label{fig:LowDens} Electromagnetic scattering of electrons and protons close to the crust-core boundary, computed using NRAPR and SQMC700 Skyrme parameters, which exhibit similar critical densities, see Tab. \ref{tab:NRAPR}. All rates are normalized over the same chemical potential $\mu_e\sim87.2$ MeV (NRAPR at $n\sim n_c$). Solid and dashed blue lines depict the longitudinal rates with and without induced interactions. Dashed red lines depict the transverse rates for comparison, for which induced interactions can safely be ignored. The three densities cover the region in which muons are absent. The transverse rates of protons are tiny. The longitudinal rates are still small compared to those of electrons, but the impact of induced interactions is striking. The inclusion of induced interactions limits the dominance of transverse rates for electrons to a tiny region around the Fermi surface, in particular at densities close to $n_c$.
}
\end{figure}
\noindent In a final step neutrons are included into the framework of electromagnetic scattering. As summarized in Sec. \ref{subsec:induced} this is a two-step process: strong interactions of protons with other protons and neutrons are resummed to obtain the dressed proton polarization function $\tilde{\Pi}_p$, which then replaces the bare polarization function $\Pi_p$ in the dressed photon propagator Eq. \ref{eq:PropEMP}. Strong interactions thus appear nested into electromagnetic ones. The properties of nuclear matter exhibit significant model dependence at higher densities, see e.g. Fig. \ref{fig:Skyrmes}, and a comparison of different Skyrme forces becomes crucial. Since the transverse channel remains mostly unaffected by induced scattering with nucleons, this section focuses on the computation of longitudinal scattering rates. The strongest impact can be expected at densities well below nuclear saturation density $n_0$, where screening effects originating from strong interactions increase rapidly, see Fig. \ref{fig:Screen}. The EMPN plasma at lower densities resembles the environment encountered in the crust-core boundary of neutron stars. Note, that Skyrme forces are well constrained at lower densities and allow for reliable predictions. Deeper inside the core, induced interactions lead to a decrease of electric screening, although by a much smaller amount. To set the stage, and to connect with the results shown in the previous subsections, it is instructive to take a look at the case $n=n_0$ first. Fig. \ref{fig:ComparisonN0} compares the longitudinal rates with and without induced interactions, employing all Skyrme models listed in subsection \ref{subsec:induced}. At fixed density each model predicts slightly different values for chemical potentials, effective masses, and residual quasiparticle interactions in $\beta$ equilibrium. The bands encompassing the rates with and without induced lepton-neutron scattering overlap almost perfectly, indicating that induced interactions are negligible at saturation density. \newline 
The impact of induced interactions at lower and higher densities is studied at $n=0.61\,n_0$ and $n=2\,n_0$. The former density is chosen such that all Skyrme models are very well within the stability region of homogeneous nuclear matter, see Tab. \ref{tab:CritDens}. The latter corresponds to a region deeper in the core, where a small muon population of about 2 \% is present, see Fig. \ref{fig:FracT}. The results are shown in Fig. \ref{fig:InducedComp}, blue and gray bands correspond to predictions for the rates without and with induced interactions respectively. Even in the absence of induced interactions, the longitudinal rate of electrons increases strongly upon approaching the crust-core boundary, compared to $n=2\,n_0$ by roughly an order of magnitude (see also Fig. \ref{fig:WeakFull}). At $n\sim0.6\,n_0$, induced interactions lead to an additional increase of more than a factor of 2, depending on the momentum of the quasi-particles. The transverse rates of electrons \textit{decrease} upon approaching the crust-core boundary, albeit by a much smaller amount, see Fig. \ref{fig:PerpCompare}. The range of momenta for which transverse scattering dominates is thus drastically reduced at lower densities. As for muons, their longitudinal rates become comparable in magnitude to those of electrons around $n=2\,n_0$ . The decrease (increase) of longitudinal (transverse) rates in between $n_0$ and $2\,n_0$ amounts in both cases to roughly a factor of $2$.\newline 
The fact that induced interactions primarily modify scattering in the longitudinal channel makes them particularly relevant for the damping rates of \textit{heavy} fermions. The absence of muons in the crust-core transition region leaves protons as the only massive particles that may engage directly in electromagnetic scattering. The proton fraction is small at lower densities, and in their interaction rates are certainly dominated by collisions with other protons and neutrons mediated by strong interactions. It is nevertheless interesting to investigate electromagnetic scattering of heavy fermions at densities close to $n_c$ on the example of protons. To perform the calculation and to compare the results with corresponding rates of electrons two Skyrme models, NRAPR and SQMC700, with almost identical critical densities are employed, see Table \ref{tab:CritDens}. The results are displayed in Fig. \ref{fig:LowDens}. The proton rates indeed benefit several times more from the impact of induced interactions than the electron rates, and reach up to $20$ \% of the latter. Given this drastic increase it seems reasonable to expect a similar effect in scattering processes mediated by strong interactions - after all the observed effect originates from a sharp increase of \textit{strong} Debye screening shown in Fig. \ref{fig:Screen}. This point will be discussed further in the outlook.
\section{Finite Temperature}\label{sec:PartDegen}
\noindent Having conducted a thorough study of scattering in (fully) degenerate nuclear matter, one may ask how the results evolve with increasing temperature. Without a sharp Fermi surface quasiparticles exhibit a finite lifetime at any given momentum. Hole states with $\epsilon_{\boldsymbol{p}}>\mu$ as well as particle states with $\epsilon_{\boldsymbol{p}}<\mu$ become available. A major obstacle is the infrared divergence of the transverse rate $\Gamma_\perp$ in the absence of strict Pauli blocking. To see this, consider the HDL expression Eq. \ref{eq:GammaHDL}, ignore the term proportional to $1-n_f$ (as it is infrared finite), and approximate the Bose distribution by $n_b\sim T/q_0$, where as usual $q_0=\epsilon_{\boldsymbol{p}}-\epsilon_{\boldsymbol{p}^\prime}\simeq \boldsymbol{v}_p\cdot\boldsymbol{q}$. The singular region can by isolated by considering collisions involving quasistatic ($q_0\rightarrow0$), ultra-soft ($\boldsymbol{q}\rightarrow0$) photons. The resulting integrand differs to that of Eq. \ref{eq:GPerpA} only by the additional factor $T/q_0$, 
\be\label{eq:diverge}
\Gamma_{\perp}\simeq\frac{e^{2}}{(2\pi)}m_{D}^{2}\,v_f^2\,T\,\int_{0}^{\infty}d\left|\boldsymbol{q}\right|\int_{0}^{v_f\,\left|\boldsymbol{q}\right|}du\,\frac{4}{16\,\boldsymbol{q}^{4}+\pi^{2}\,m_{D}^{4}\,v_f^{2}\,u^{2}/\boldsymbol{q}^{2}}=\frac{e^{2}}{2\pi^2}\,v_f\,T\int_0^{\infty}d|\boldsymbol{q}|\,\frac{1}{|\boldsymbol{q}|}\textrm{arctan}\left(\pi\frac{m_D^2\,v_f}{4\,\boldsymbol{q}^2}\right).\\[1ex]
\ee
In the limit $\boldsymbol{q}\rightarrow0$ the arctangent approaches $\pi/2$ and the transverse rate is logarithmically divergent. The consistent calculation of the lifetime of fermions in gauge theory plasmas at finite temperature is a long standing problem, see e.g. Refs.  \cite{Blaizot:1996az}, \cite{Pisarski:1988vd,Lebedev:1990kt,Rebhan:1992ak,Pisarski:1993rf,Flechsig:1995sk,Takashiba:1995qa}. In QCD, where self interactions of gluons lead to a magnetic screening mass, the introduction of an infrared cut-off seems natural. In QED the resolution of the infrared problem has to lie elsewhere, and techniques to resumm the leading order divergences have been developed for hot matter at $\mu=0$, see Refs. \cite{Blaizot:1996az}, \cite{Takashiba:1995qa,Blaizot:1997kw}. An application of these methods in the present context is certainly beyond the scope of this article. Instead, I shall compute the energy loss per distance traveled, $-dE/dx$, which is easily derivable from $\Gamma$, and which is infrared finite. From a formal point of view the quantity $-dE/dx$ is less fundamental, as it does not represent a direct link between the fermion self energy and scattering theory. In addition, the physical interpretation as the (inverse) quasiparticle lifetime is lost. From a practical point of view it is worth studying $-dE/dx$, because it is the quantity which ultimately enters the transport integral. Following Braaten and Thoma \cite{Braaten:1991jj}, $-dE/dx$ is  obtainable from $\Gamma_{L}$ and $\Gamma_{\perp}$ upon inserting $(\epsilon_{\boldsymbol{p}}-\epsilon_{\boldsymbol{p}^\prime})/v_{\boldsymbol{p}}$ into their integrands, which precisely compensates the divergence stemming from the Bose distribution in the transverse channel. Note, that the logarithmic divergence in Eq. \ref{eq:diverge} is obtained despite the inclusion of dynamical screening. In the absence of Landau damping Eq.\ref{eq:diverge} is quadratically divergent, while $-dE/dx$ is logarithmically divergent. \newline 
To connect with the zero temperate results it is worth taking a look at $-dE/dx$ in the (fully) degenerate limit. One may again ask for the behaviour in immediate proximity to the Fermi surface, which amounts to adding the factor of $u/v_f$ in the integrands of Eqs. \ref{eq:GLongA} and \ref{eq:GPerpA}. Using the same notation as in Eq. \ref{eq:LongEMPAn} the results for a given lepton species $l=\{e,\,\mu\}$ read 
\be \label{eq:dEdxT0}
-\left.\frac{dE}{dx}\right|_{L,\,l}\sim\frac{e^2}{48}\,\frac{1}{v_{f,\,l}^2}\,\left|\epsilon_{\boldsymbol{p}}-\mu_l\right|^3\,\frac{1}{M^3}\,\left(\sum_i \frac{m_{D,\,i}^2}{v_{f,\,i}}\right)\,,\hspace{1cm}-\left.\frac{dE}{dx}\right|_{\perp,\,l}\sim\frac{e^2}{18\pi}\left|\epsilon_{\boldsymbol{p}}-\mu_l\right|^2\,.\\[1ex]
\ee
The additional power of $|\epsilon_{\boldsymbol{p}}-\mu|$ in both channels results in a smaller curvature (i.e., a slower increase) of $-dE/dx$ around the Fermi surface. Note, that transport coefficients are determined by the integral of $-dE/dx$ over the momenta $|\boldsymbol{p}|$ of the incoming fermions, and thus depend on the area under $-dE/dx$ rather than the energy loss itself. The relative increase of this area due to induced interactions is particularly large if the region of interest around $k_f$ is small, and the temperatures are very small compared to $\mu$, see Fig. \ref{fig:DeDxLowT} below. In the following the scattering rates are investigated for a wide range of temperatures.
\subsection{Degenerate matter: neutron stars}
\begin{figure}[t]
\begin{centering}
\includegraphics[scale=0.55]{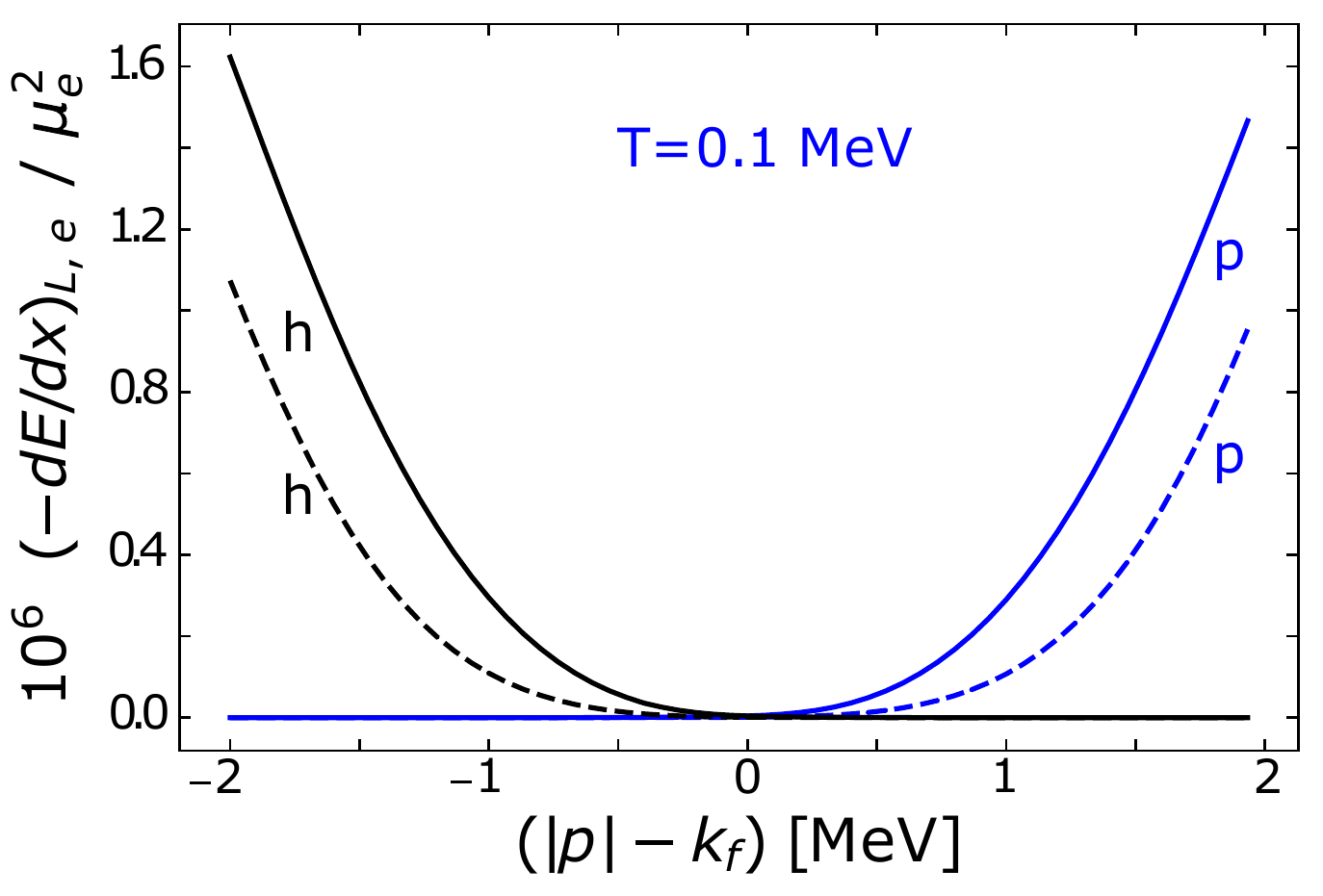}\hspace{1cm}\includegraphics[scale=0.55]{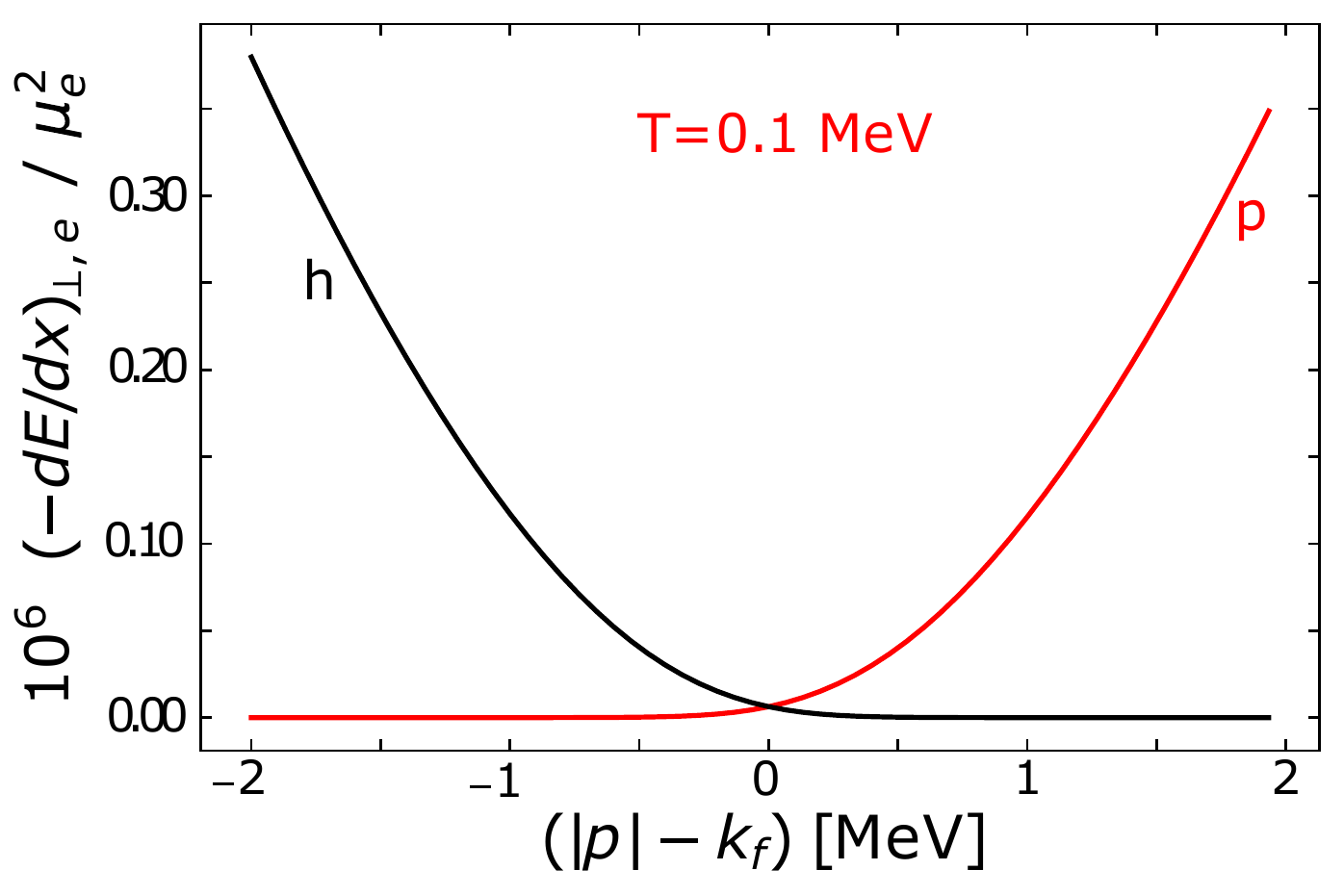}\\[1ex]
\includegraphics[scale=0.55]{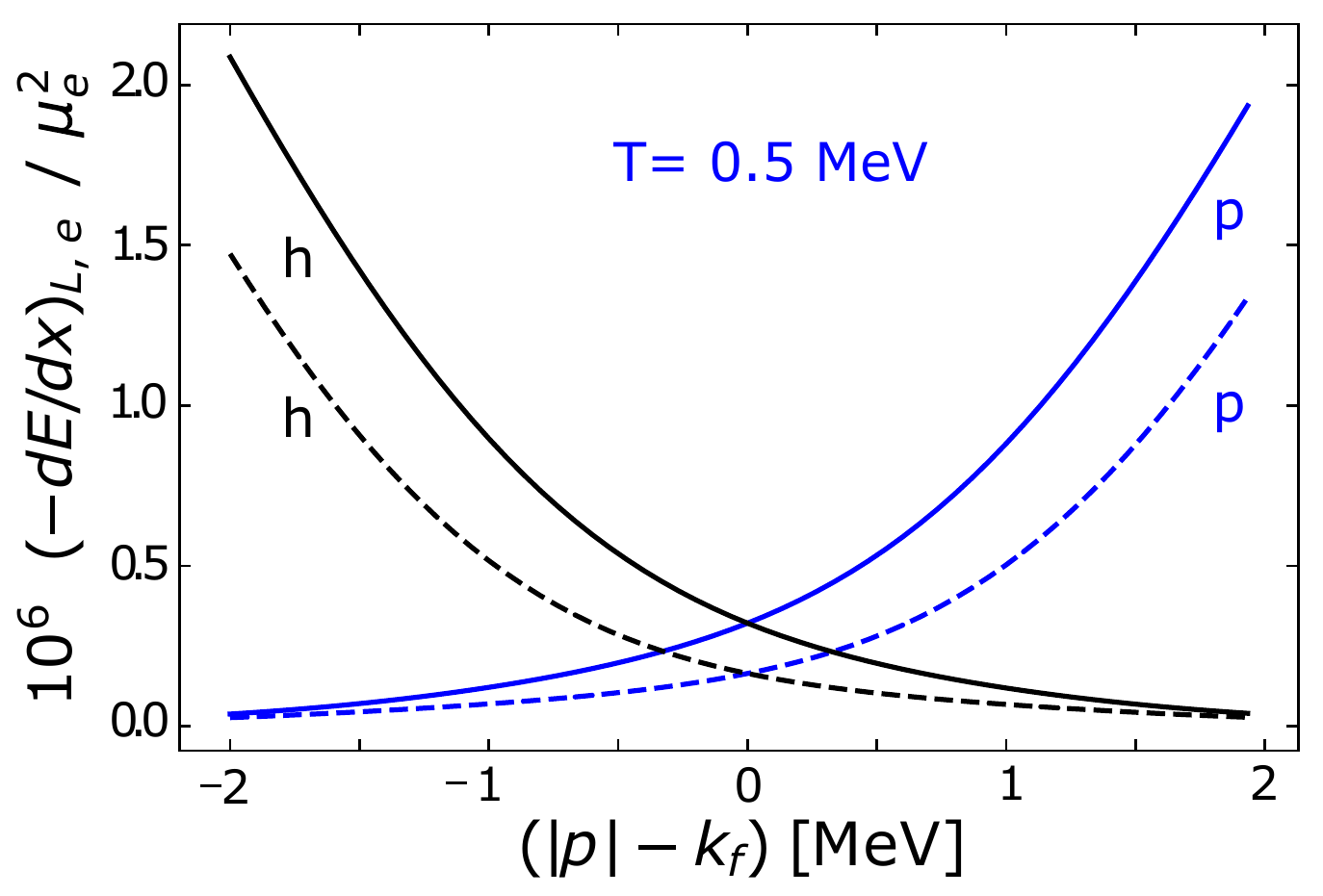}\hspace{1cm}\includegraphics[scale=0.55]{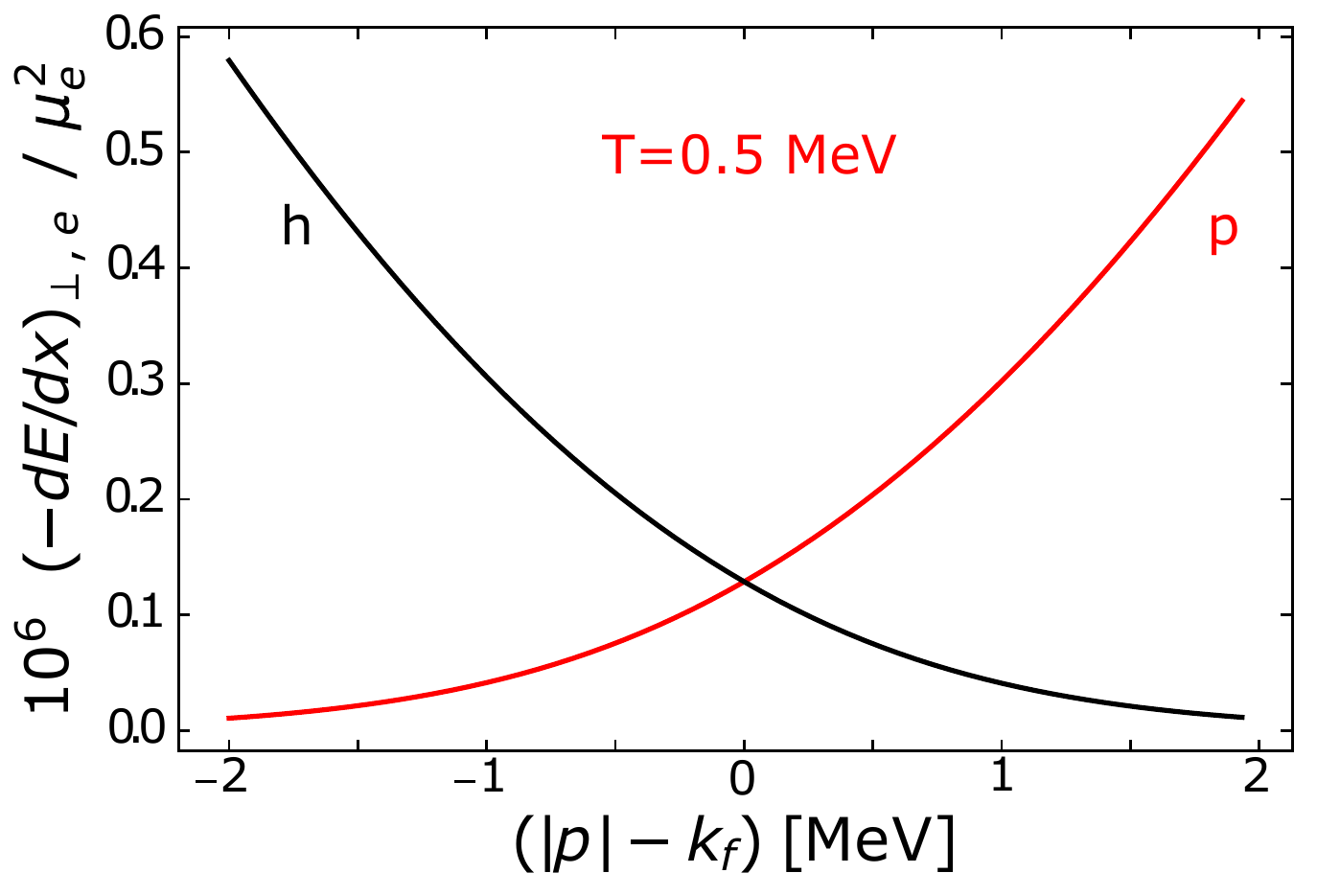}\\[1ex]
\includegraphics[scale=0.55]{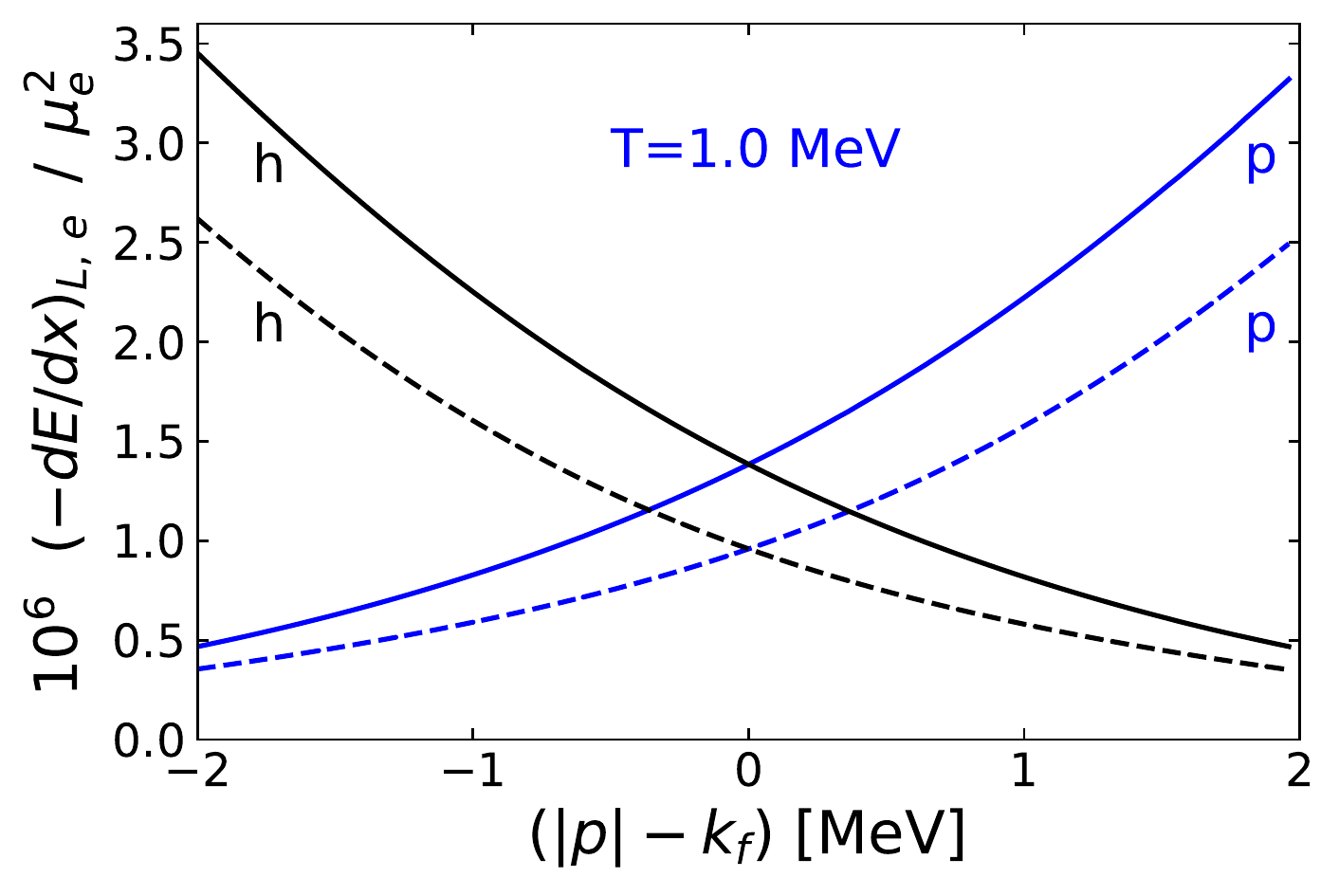}\hspace{1cm}\includegraphics[scale=0.55]{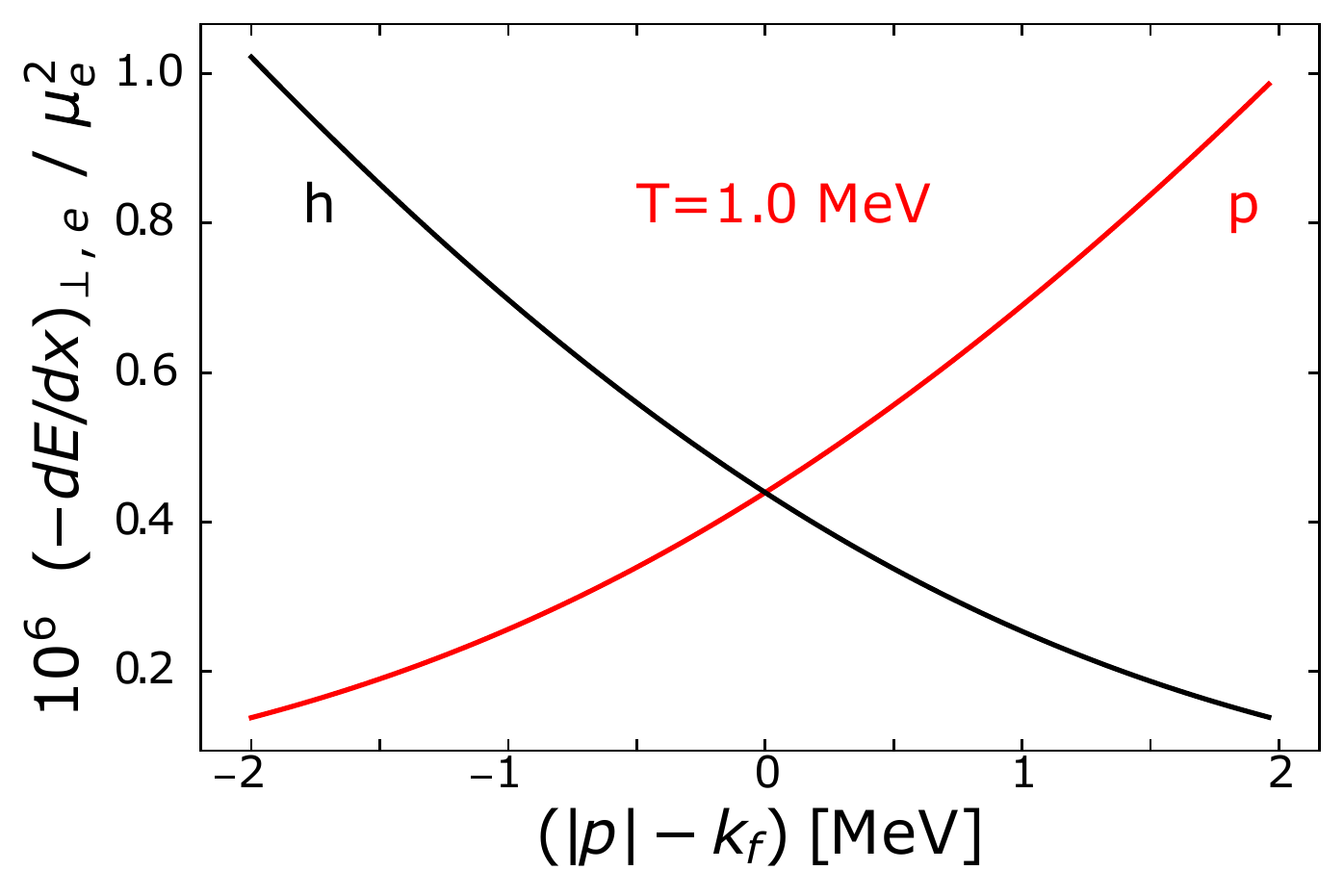}
\end{centering}
\setlength{\belowcaptionskip}{-8pt}
\caption{\label{fig:DeDxLowT} Energy loss of electrons (p) and holes (h) in nuclear matter (EP plasma) at $n=0.55\,n_0$ and three different temperatures, calculated using NRAPR Skyrme forces, see Tab. \ref{tab:NRAPR}. The conversion factor to obtain the energy loss in units of $\textrm{MeV}^2$ reads $\mu_e^2\,10^{-6}\sim7.8\cdot10^{-3}$. Blue and red lines correspond to longitudinal and transverse rates for particles, hole contributions are shown in black. Dashed lines in the longitudinal channel display the corresponding results in the absence of induced interactions. The rates at very low temperatures of $T=0.1$ MeV closely resemble the fully degenerate case. The energy loss increases rapidly with temperature, and sizeable contributions to the scattering rate of particles with $|\boldsymbol{p}|\leq k_f$ (or holes with $|\boldsymbol{p}|\geq k_f$) emerge. The impact of induced interactions on the energy loss remains large, but in relative terms it is much smaller than at $T=0.1$ MeV. Screening effects are less important, and the longitudinal rates are larger than their transverse counterpart for any given momentum $|\boldsymbol{p}|$ at $T=0.5$ MeV and $T=1$ MeV.}
\end{figure}
\noindent Temperatures in the range of $T=0.1$  MeV to $T=1$ MeV are relevant for young and middle-aged neutron stars. While tiny compared to the chemical potentials of leptons and nucleons, finite temperature effects become important for the calculation of $-dE/dx$ close to $|\boldsymbol{p}|=k_f$, where the rates otherwise go to zero. Fig. \ref{fig:DeDxLowT} displays the energy loss for electrons and electron-holes, computed using the full one-loop expression for $-dE/dx$. As before the density is set to $n=0.55\,n_0$, and temperatures of $T=0.1$ MeV, $T=0.5$ MeV, and $T=1.0$ MeV are studied. The tiny ratio of $T/\mu$ renders the numerical integration over Fermi distributions particularly challenging, and the comparison with strictly zero temperature results (which are comparatively easy to obtain) serves as valuable crosscheck. \newline 
At finite temperature $-dE/dx$ assumes a finite value for any given momentum $|\boldsymbol{p}|$. The intersection of the results for electrons and electron-holes marks the location of the Fermi surface at zero temperature, where the quasiparticles are no longer stable. The most interesting aspect to study is the interplay of finite temperature and induced interactions. At $T=0.1$ MeV the rates are almost completely identical to the results obtained at zero temperature: the imaginary part at $|\boldsymbol{p}|=k_f$ is tiny, contributions of particles (holes) with $\epsilon_{\boldsymbol{p}}<\mu$ ($\epsilon_{\boldsymbol{p}}>\mu$) play no role, and induced interactions boost the longitudinal rates by a large amount, particularly in close proximity to the (zero temperature) Fermi surface. At $T=0.5$ and $T=1.0$ MeV the available phase space for scattering increases, and longitudinal and transverse contributions to $-dE/dx$ grow accordingly. The distinct characteristics of both channels at very low temperatures, arising mainly due to screening effects, are much less noticeable. The dominance of transverse scattering has disappeared completely, energy loss due to the exchange of longitudinal plasmons is larger at any given momentum. The fact that all three temperatures satisfy the condition $T\ll\mu_e$ highlights the importance of finite temperature effects, even in highly degenerate systems. Induced interactions still lead to a sizeable increase of the longitudinal rates. However, in comparison with the corresponding energy loss due to pure electromagnetic scattering the impact is much less pronounced. In general induced interactions are important whenever screening effects are the dominant feature, and therefore play a particularly important role for old neutron stars. 
\subsection{Partially degenerate matter: neutron star mergers}
\begin{figure}[t]
\includegraphics[scale=0.55]{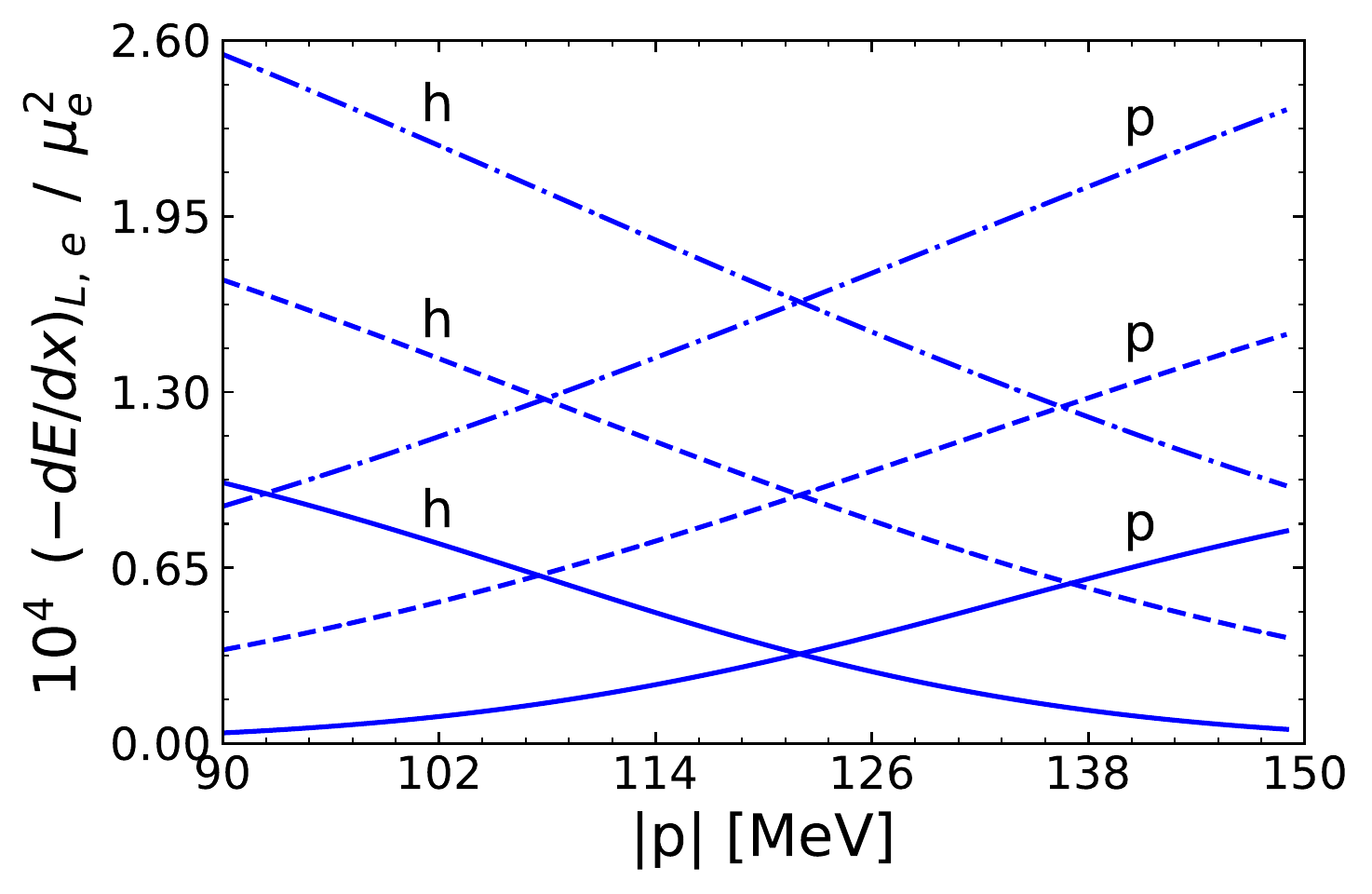}\hspace{1cm}\includegraphics[scale=0.55]{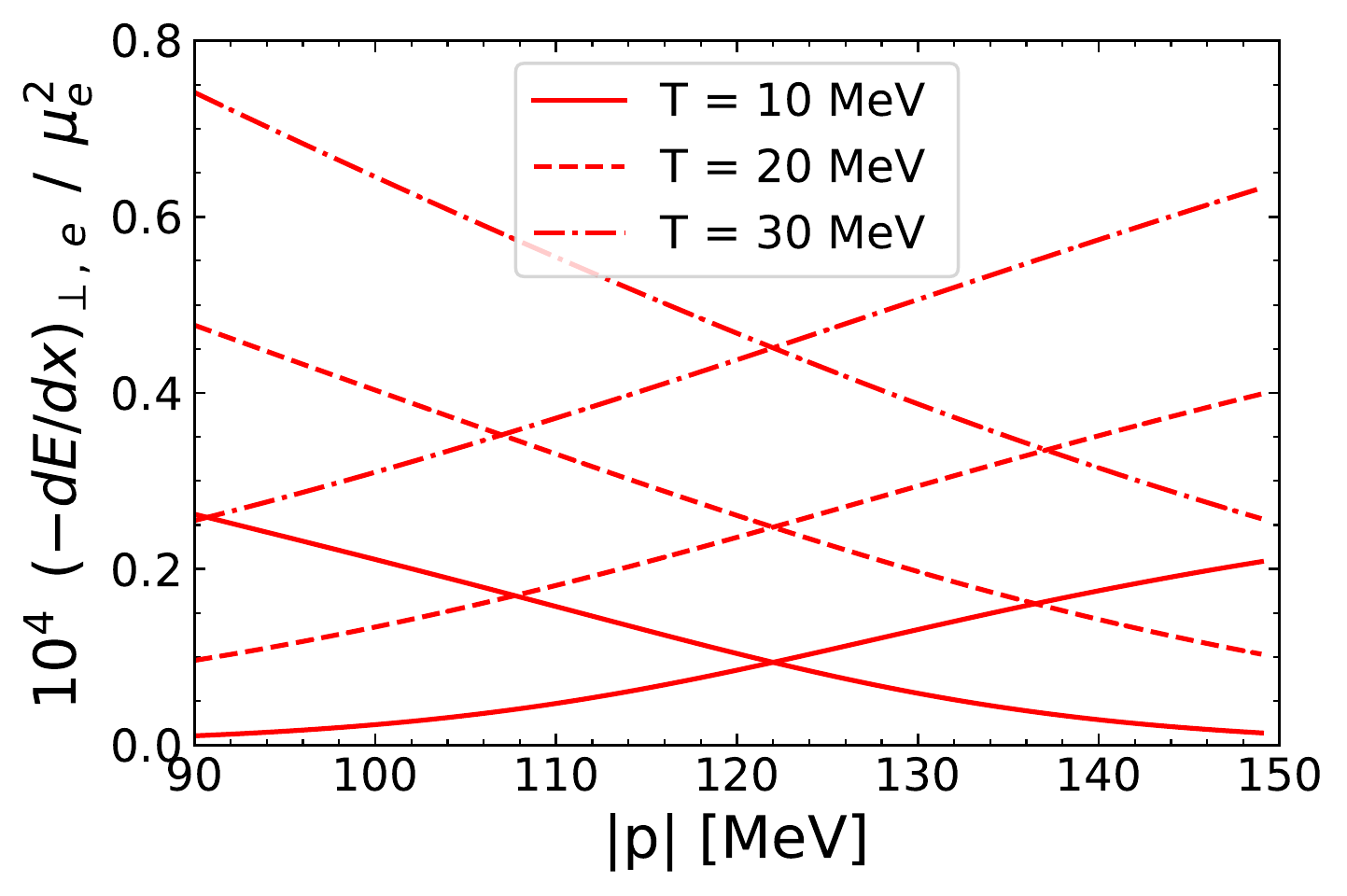}
\setlength{\belowcaptionskip}{-8pt}
\caption{\label{fig:DeDxHighTEMP}Energy loss of electrons (p) and electron-holes (h) at saturation density at $T=10$ MeV (solid), $T=20$ MeV (dashed), and $T=30$ MeV (dot-dashed), calculated using NRAPR Skyrme forces. The results differ substantially from those calculated for (fully) degenerate matter. Particles (holes) with $|\boldsymbol{p}|<k_f$ ($|\boldsymbol{p}|>k_f$) add sizeable contributions to the total rate. At $T=30$ MeV the rates increase (decrease) almost linearly with increasing momentum $|\boldsymbol{p}|$.}
\end{figure}
\noindent 

\begin{figure}[t]
\includegraphics[scale=0.55]{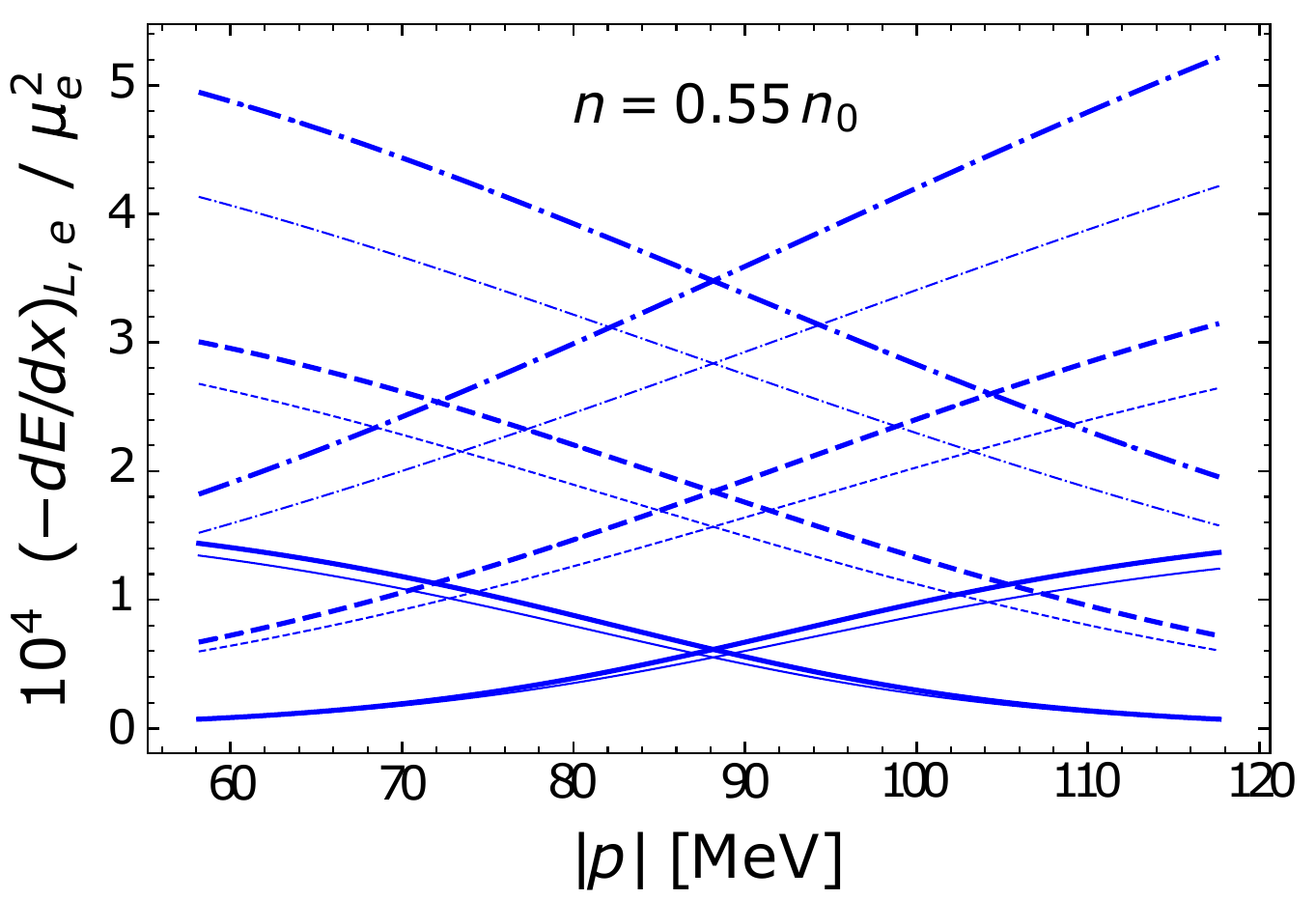}\hspace{1cm}\includegraphics[scale=0.55]{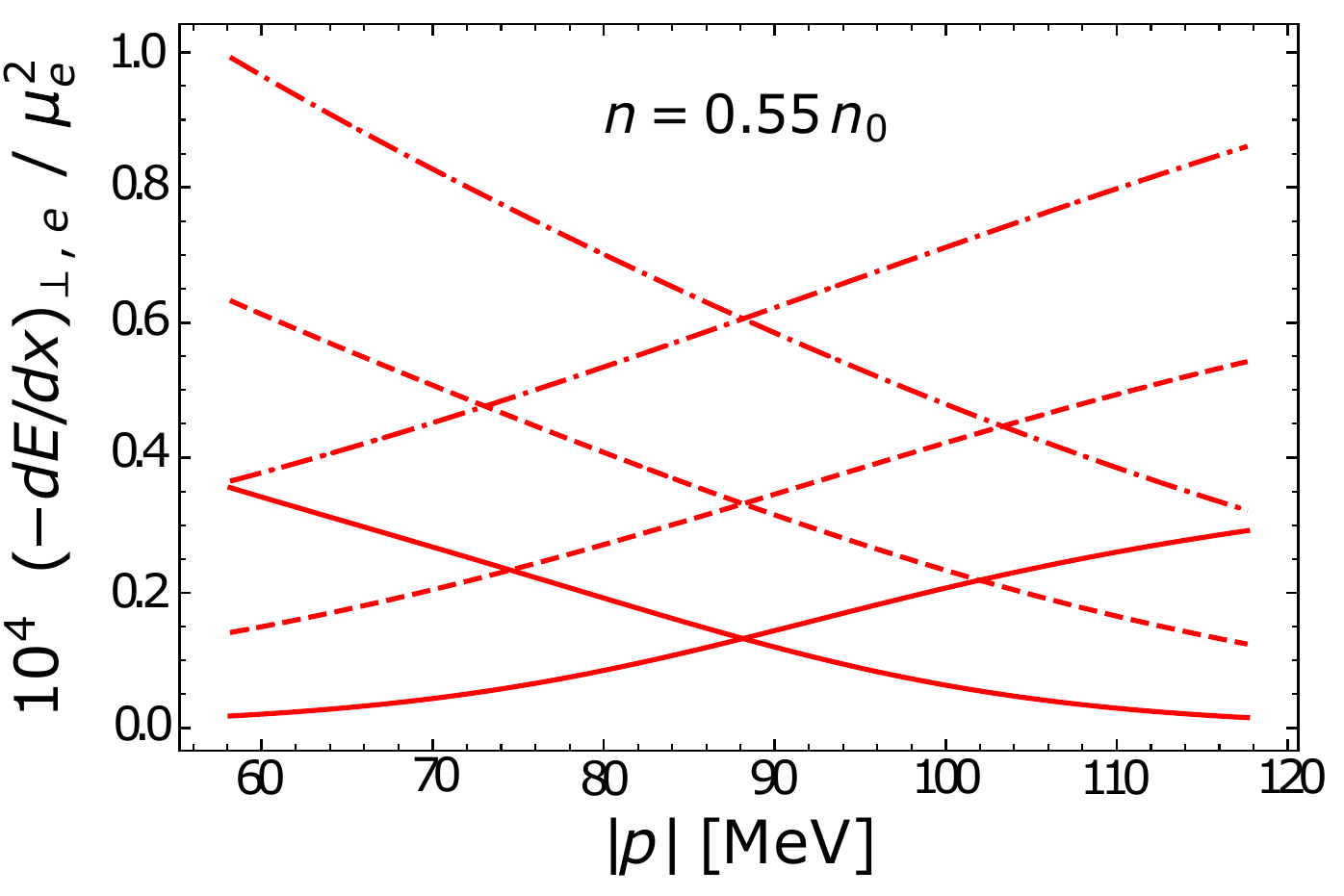}\\[2ex]
\includegraphics[scale=0.55]{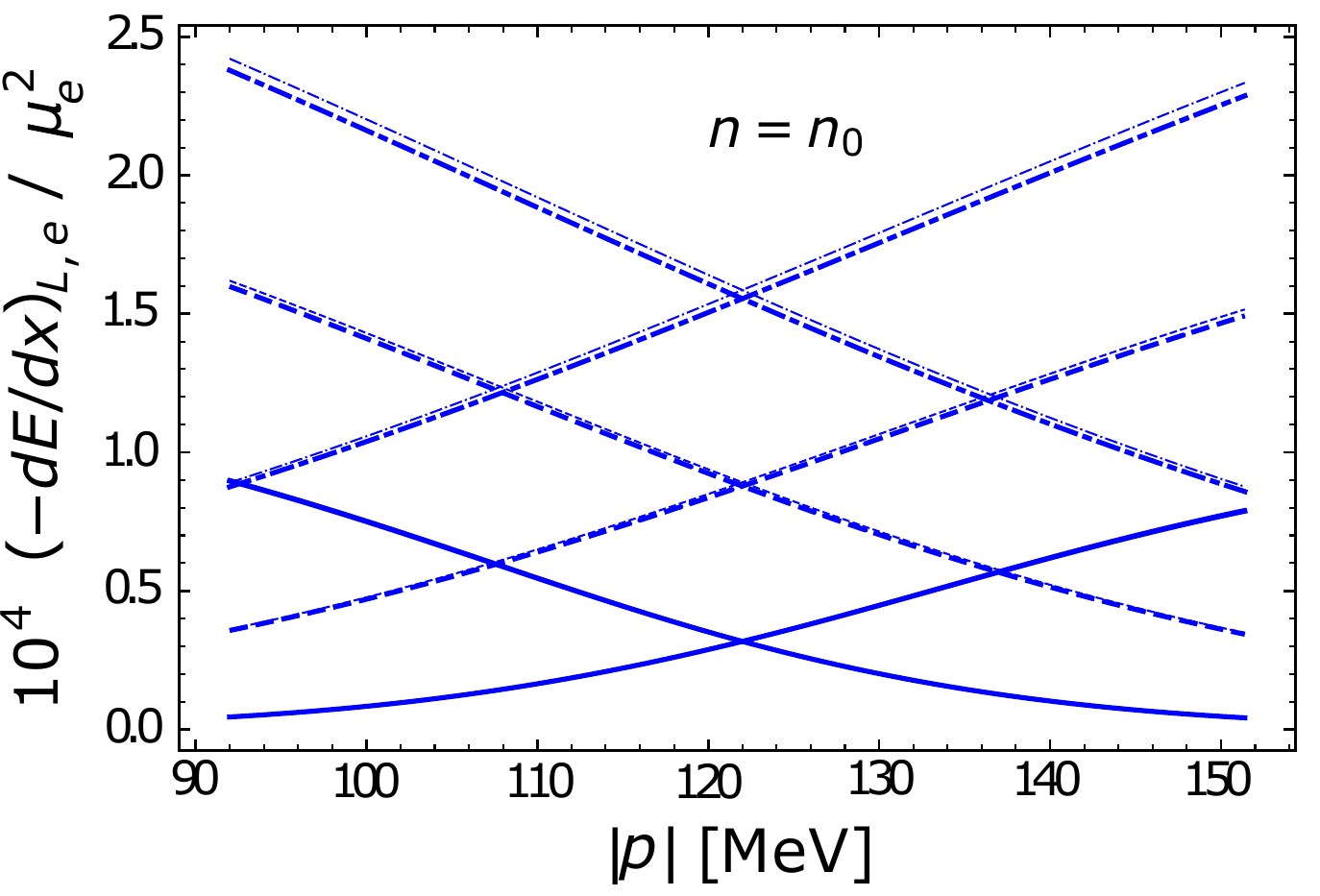}\hspace{1cm}\includegraphics[scale=0.55]{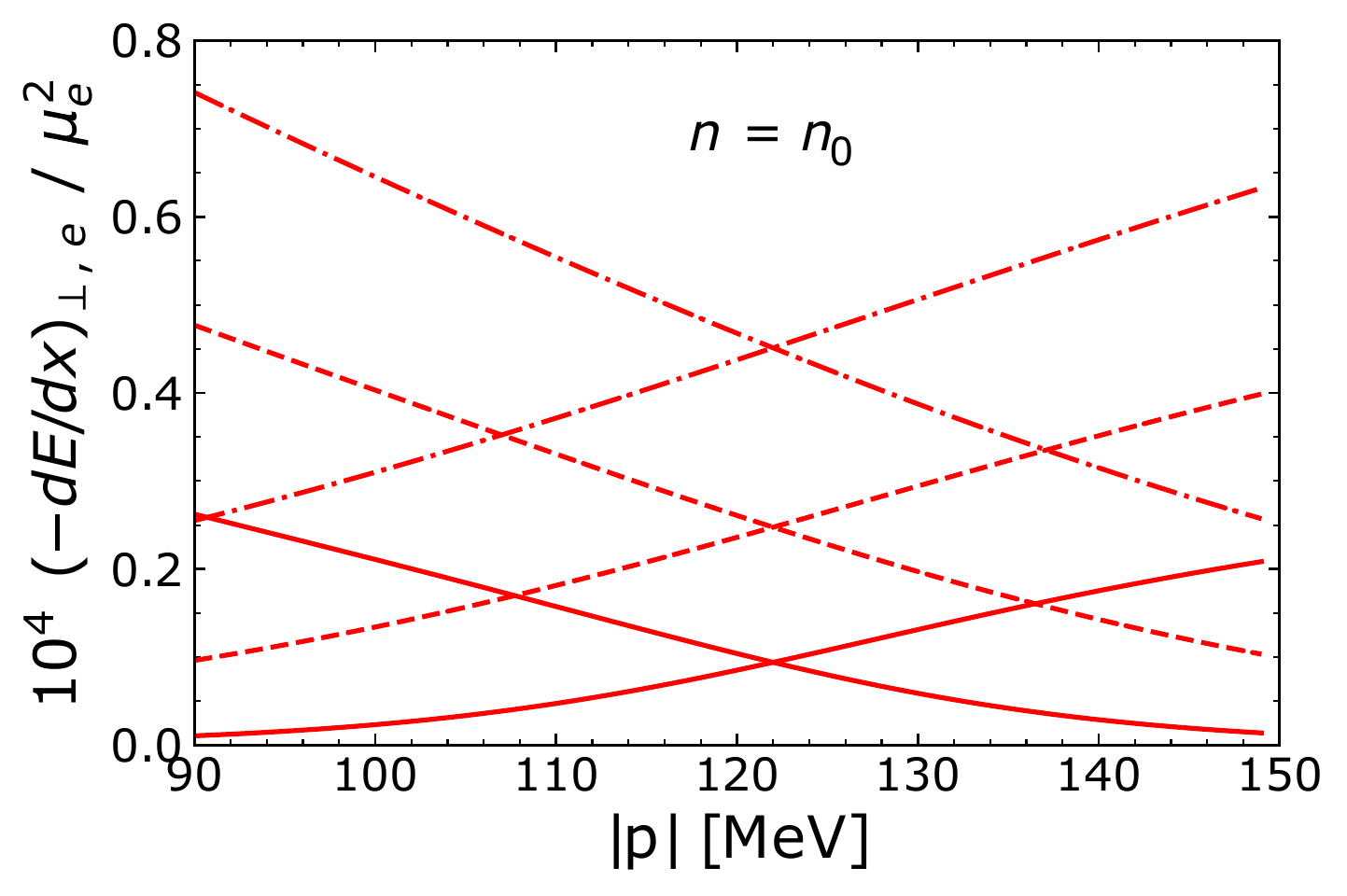}\\[2ex]
\includegraphics[scale=0.55]{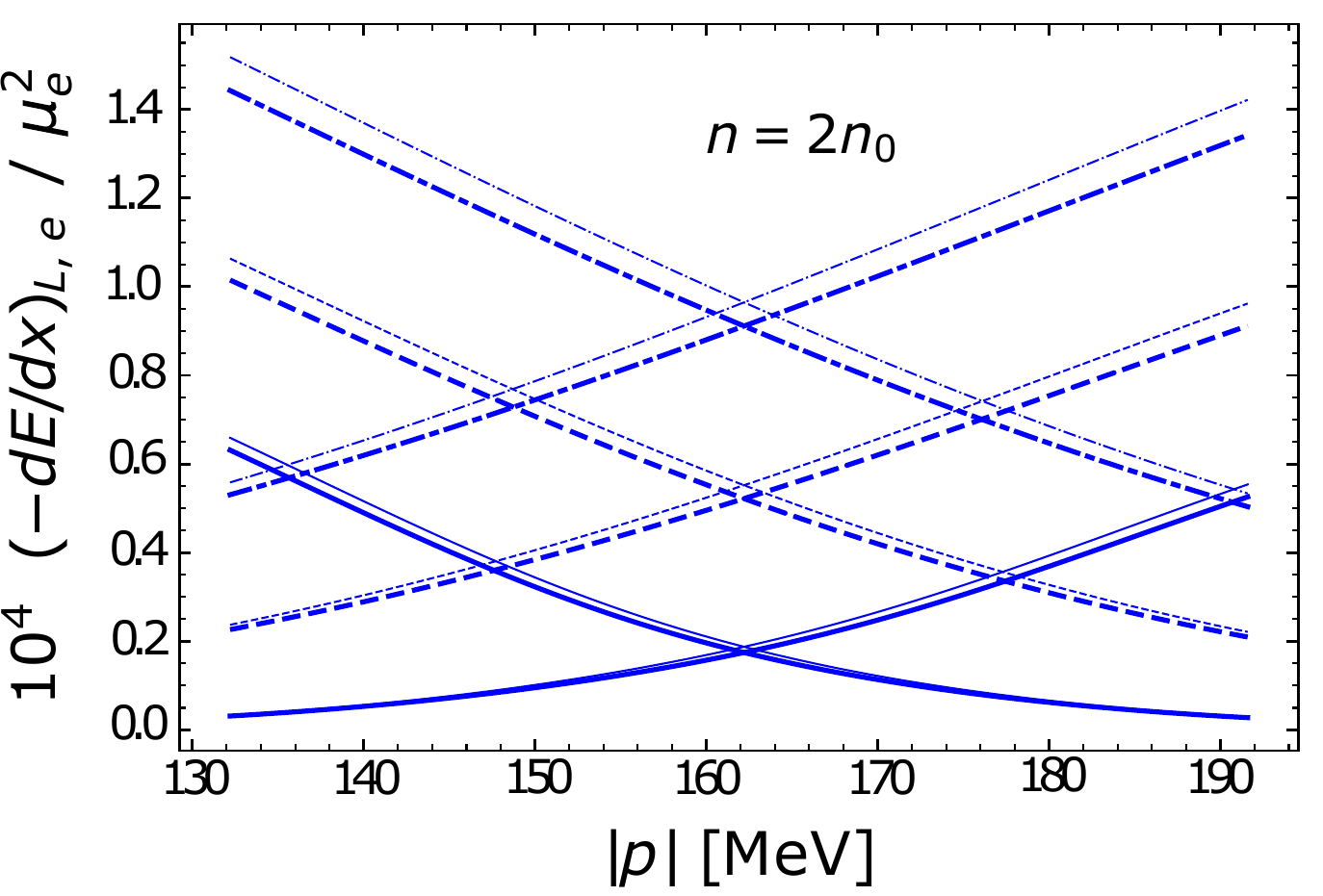}\hspace{1cm}\includegraphics[scale=0.55]{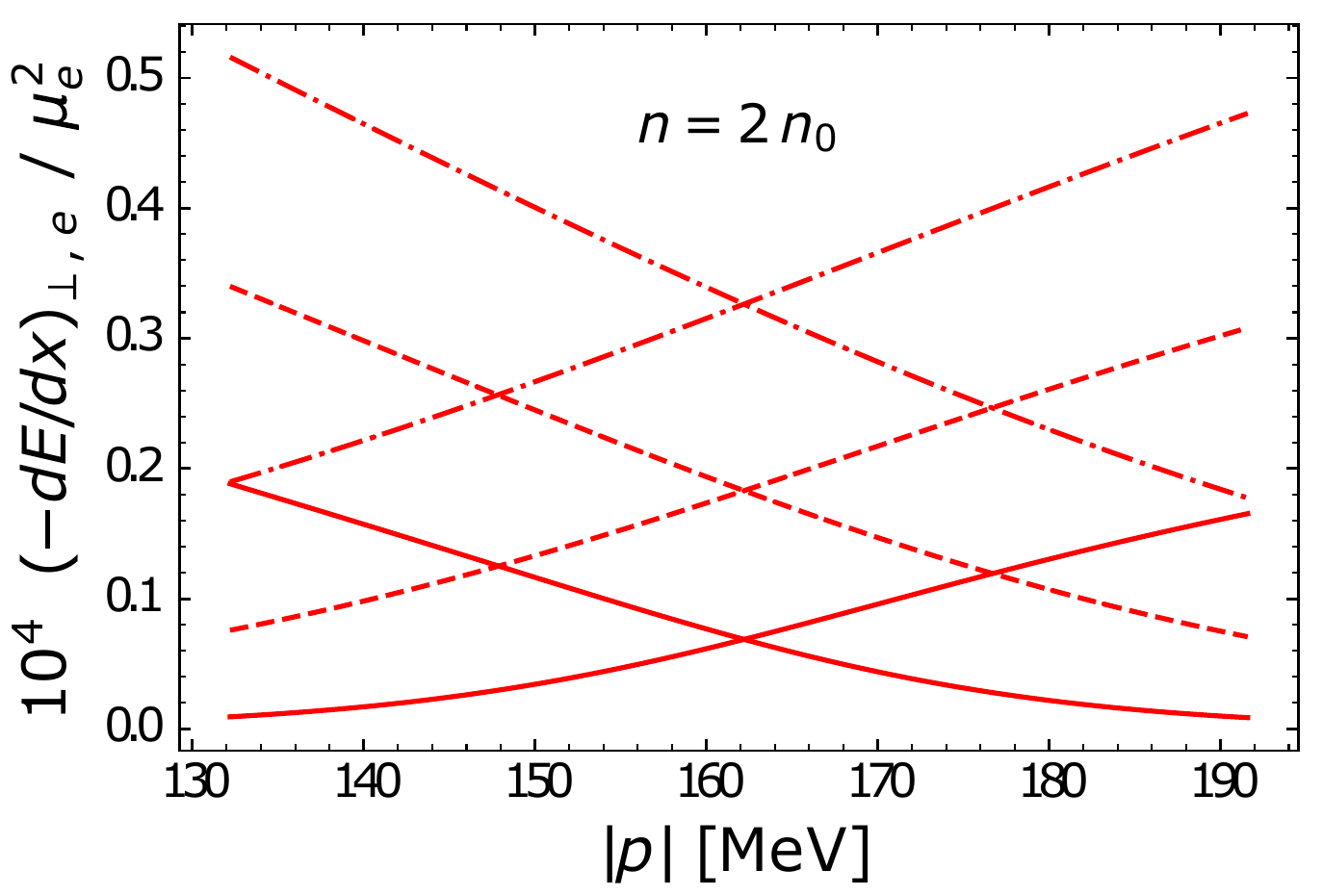}\\[2ex]
\setlength{\belowcaptionskip}{-8pt}
\caption{\label{fig:DeDxHighTINDUCED} Energy loss of electrons in an EPN plasma ($n=0.55\,n_0$), and an EMPN plasma ($n=n_0$ and $n=2\,n_0$), at $T=10$ MeV (solid), $T=20$ MeV (dashed), and $T=30$ MeV (dot-dashed). Calculations are based on NRAPR Skyrme forces, and account for induced lepton-neutron scattering. Thin lines correspond to the results neglecting induced interactions. As usual, considerable modifications occur predominantly at lower densities, but are far less relevant than in the degenerate regime. In contrast to the fully degenerate limit longitudinal and transverse rates now both decrease with density. In the longitudinal channel this decrease is more pronounced, albeit much less pronounced than in the degenerate limit.}
\end{figure}

\begin{figure}[t]
\begin{centering}
\includegraphics[scale=0.55]{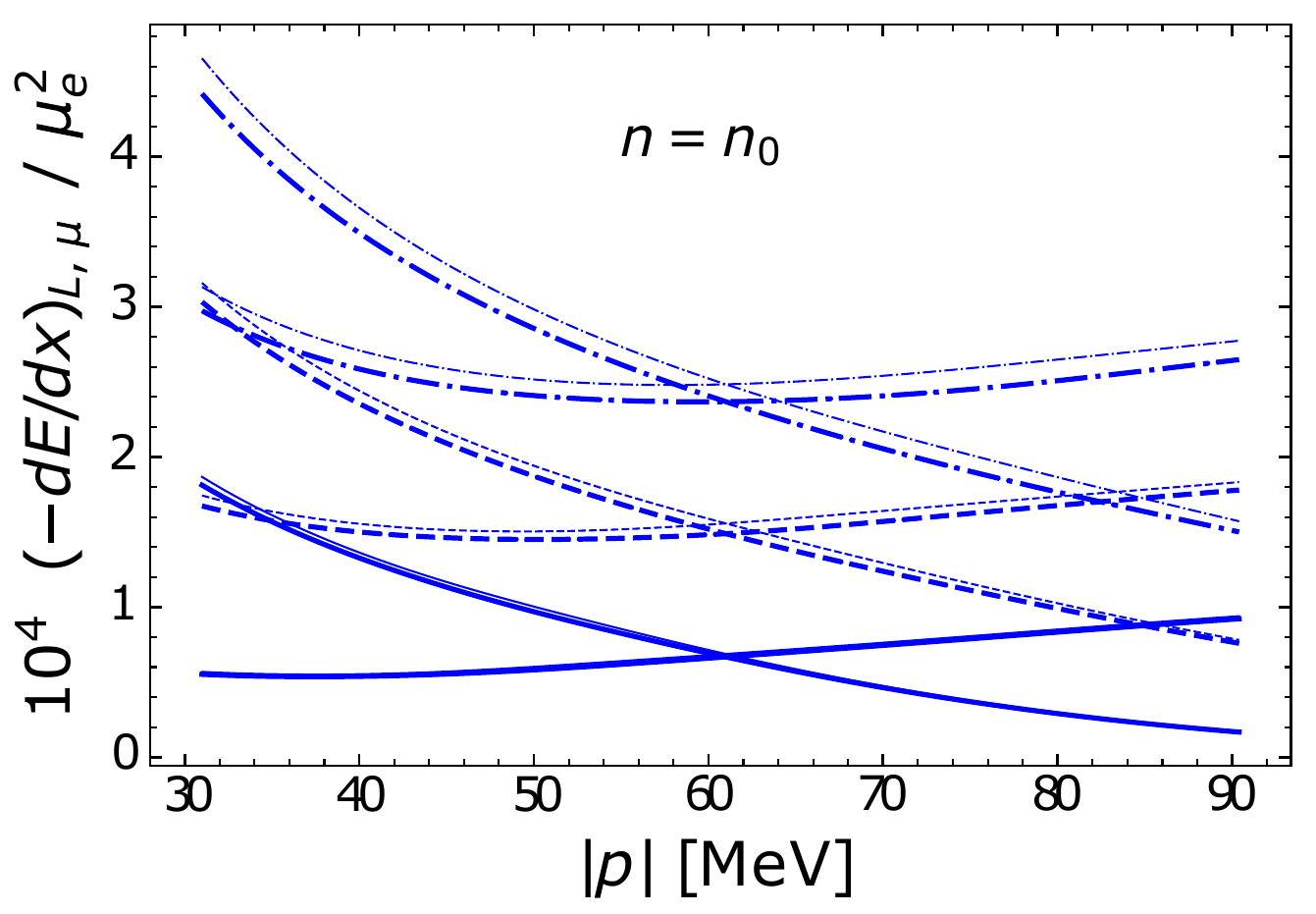}\hspace{1cm}\includegraphics[scale=0.55]{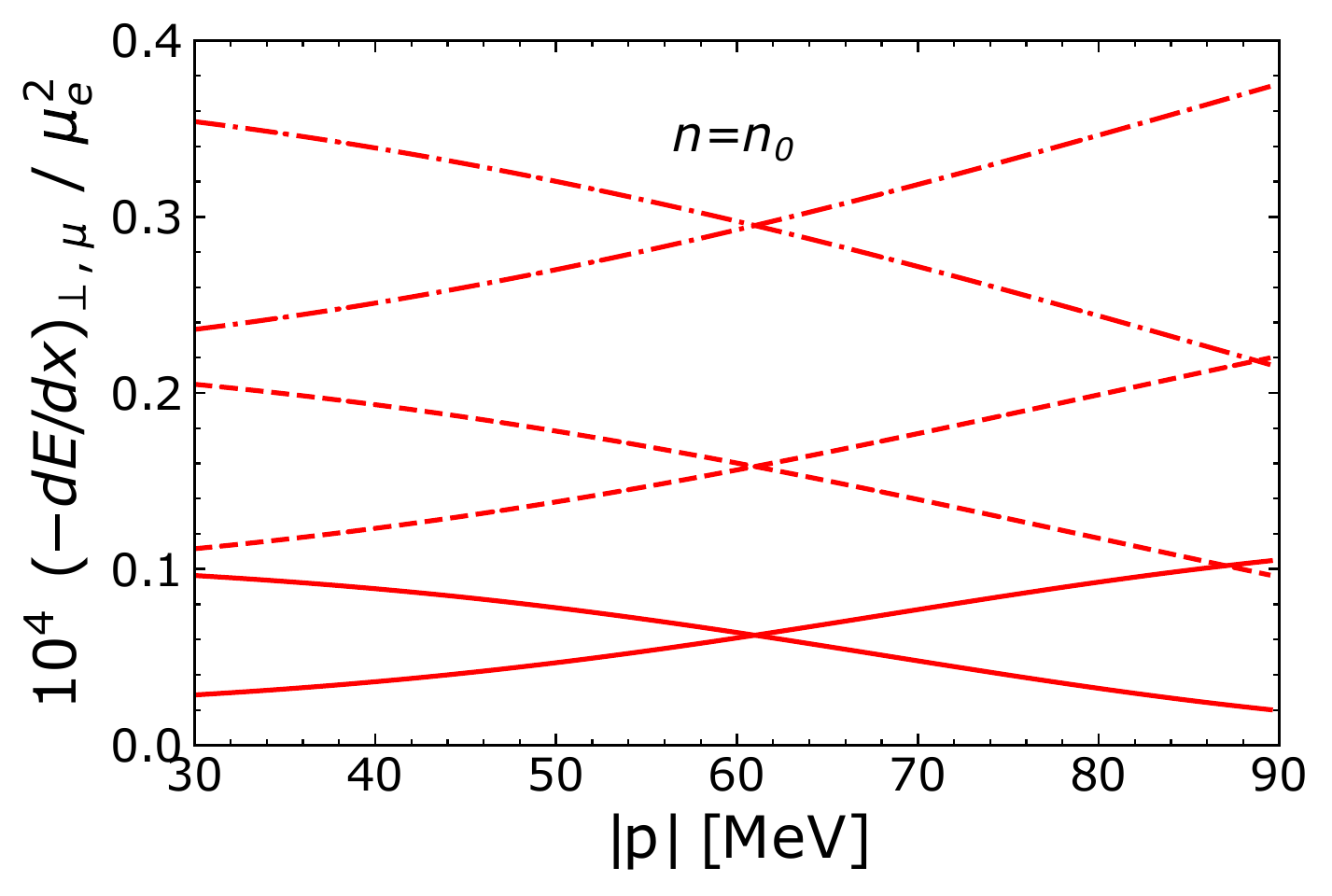}\\[2ex]
\includegraphics[scale=0.55]{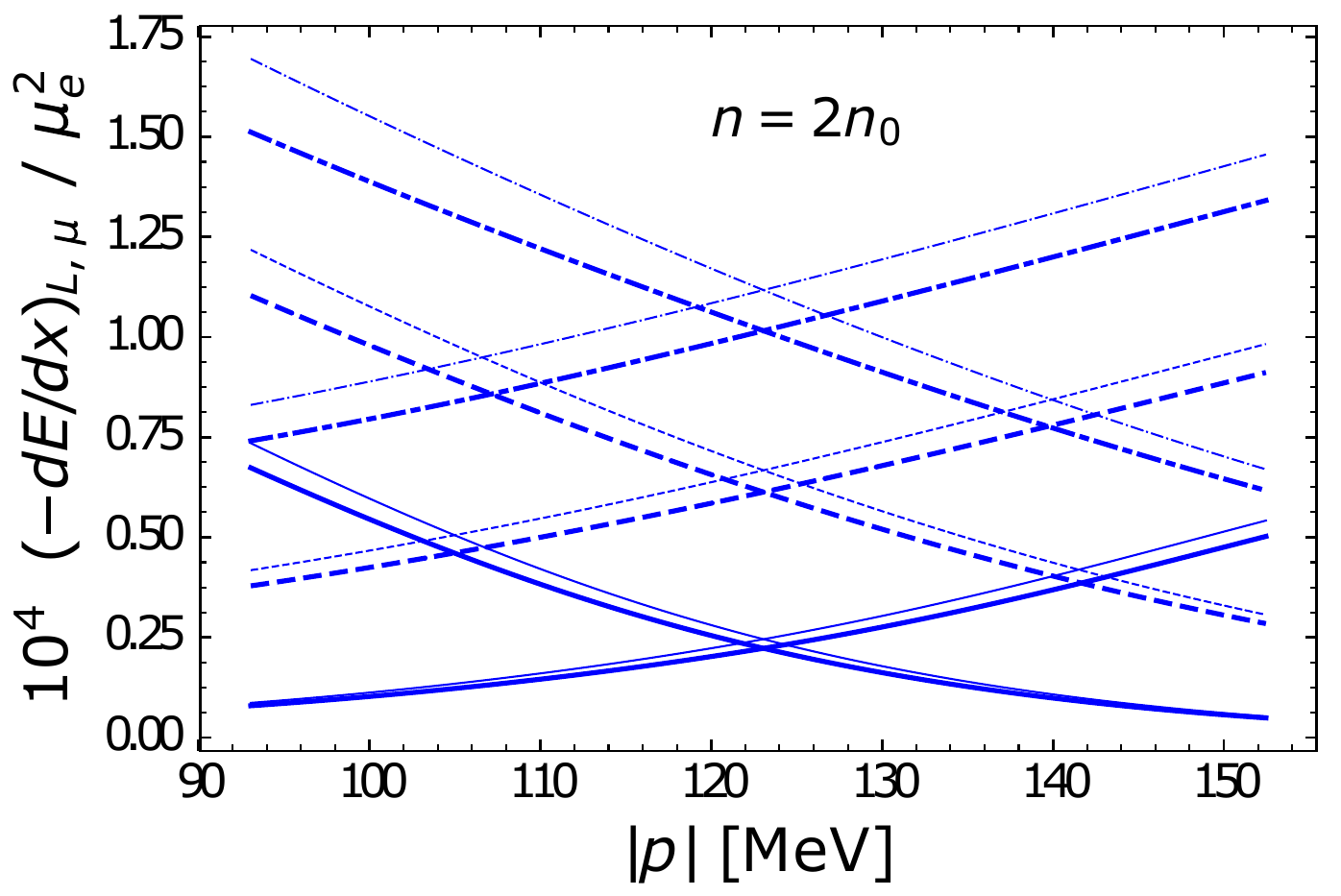}\hspace{1cm}\includegraphics[scale=0.55]{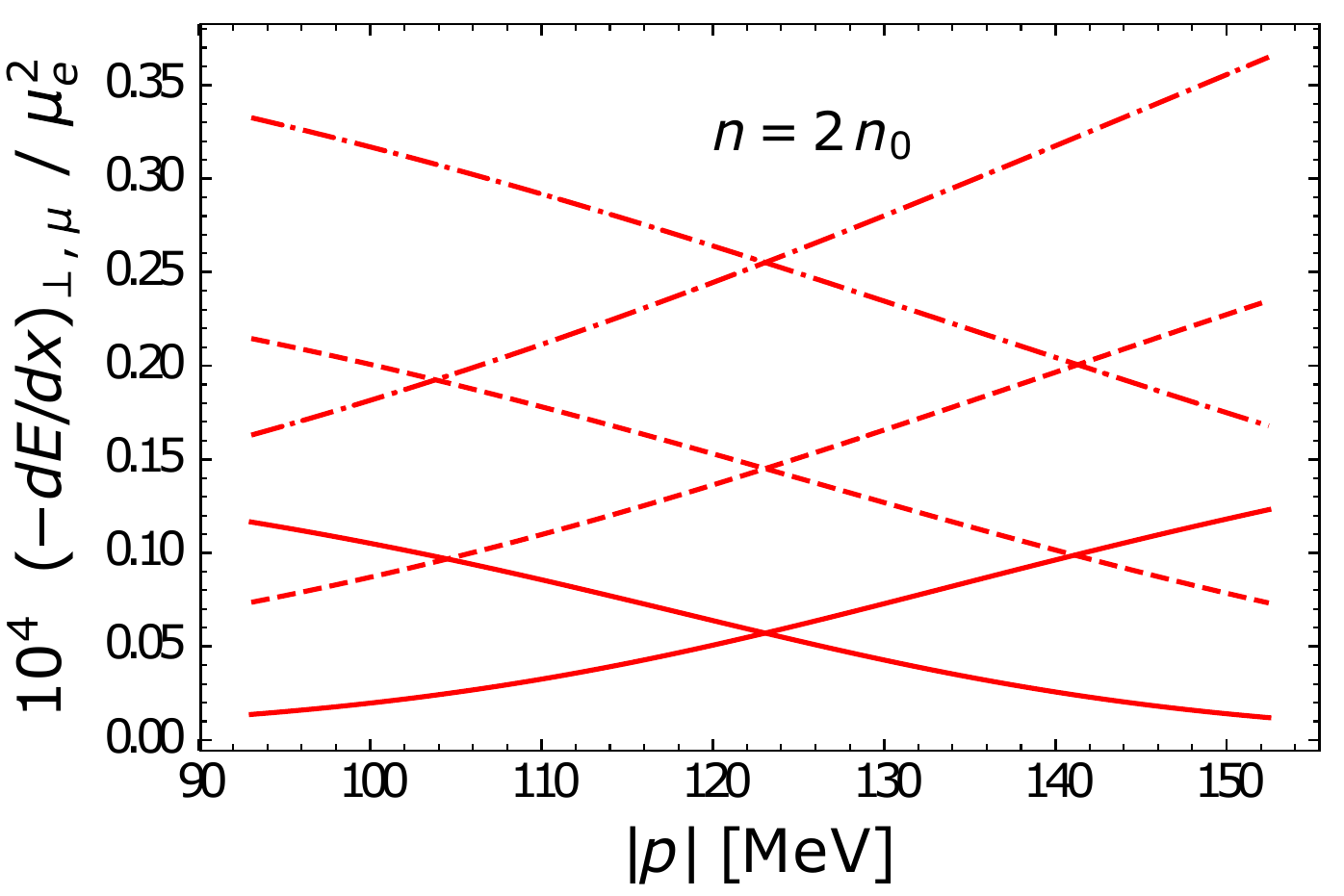}
\end{centering}
\caption{\label{fig:DeDxHighTINDUCEDMuons} Energy loss of muons in an EMPN plasma at $n=n_0$ and $n=2n_0$, taking into account induced interactions. Same parameters as in Fig. \ref{fig:DeDxHighTINDUCED}. The reduction of the energy loss at $n=2\,n_0$ is more pronounced than for electrons, see Fig. \ref{fig:DeDxHighTINDUCED}.}
\end{figure}
\noindent

\noindent Finally I take a first glimpse at scattering rates at higher temperatures, ranging from $T=10$ MeV to $T=30$ MeV. While far too high to be realized in neutron stars at any stage of their evolution, temperatures of tens of MeV are very relevant for the hot regions in neutron star mergers. The results presented in this section are subject to two major sources of uncertainty: First partially degenerate conditions in principle require for a rigorous iterative determination of effective masses, chemical potentials, and residual quasiparticle interactions for each given temperature. The simple estimates carried out in subsection \ref{subsec:multi} suggest that temperature variations of these quantities are small, but a thorough check is certainly desirable. Second the increasing thermal energy leads to larger transfer of energy and momentum during collisions. The momentum dependence of the residual quasiparticle interactions $f_{ab}$, not included in the present approach, may thus become important for the correct description of dynamical screening effects in the strong sector of the RPA resummation. Previous results indicate, that the importance of induced interactions is anyway reduced at higher temperatures, and the associated uncertainties should not change the outcome of this study on a qualitative level. In any case the results of this section have to be regarded as a first step towards the more ambitious goal of obtaining a quantitatively accurate description of electromagnetic scattering in partially degenerate nuclear matter.\newline 
Extrapolating the results of Fig. \ref{fig:DeDxLowT}, one would expect that increasing temperatures favor the longitudinal scattering rates. Fig. \ref{fig:DeDxHighTEMP} shows that the longitudinal rates are indeed several times larger than transverse ones for temperatures above $10$ MeV. The local suppression of the total rate around $|\boldsymbol{p}|=k_f$ has completely disappeared; at $T=30$ MeV the scattering rates of electrons (electron holes) increase (decrease) almost linearly with the momentum. Hole states (particle states) with $|\boldsymbol{p}|>k_f$ ($|\boldsymbol{p}|<k_f$) are increasingly populated, and their energy loss becomes an important contribution to transport. A realistic calculation of transport coefficients in partially degenerate matter should take these features into account.\newline 
It remains to take a look at the impact of induced interactions for the temperatures considered in Fig. \ref{fig:DeDxHighTINDUCED}. Induced interactions inherit their temperature dependence from the neutron and proton polarization functions $\Pi_{n,p}$, incorporated in the resummed proton polarization function $\tilde{\Pi}_p$. Increasing temperatures enlarge the low energy tail of the longitudinal photon spectrum at $q_0<v_{f\,,p}\,|\boldsymbol{q}|$, which is of primary importance for Coulomb scattering, see Fig. \ref{fig:RhoLongT}. A numerical survey at $n=0.55\,n_0$ (EPN plasma) and $n=n_0$, $n=2\,n_0$  (EMPN plasma) shows again the emergence of a familiar pattern: at lower densities the impact of induced interactions is clearly noticeable, increasing longitudinal rates by up to $20$ \%. Around saturation density the impact is negligible, and above saturation density it leads to slightly reduced rates in the longitudinal channel. The reduction at $n=2\,n_0$ is much more pronounced for muons than for electrons, see Fig. \ref{fig:DeDxHighTINDUCEDMuons}. In all cases the relative impact of induced interactions is indeed much smaller than in the degenerate regime discussed in Fig. \ref{fig:DeDxLowT}. It is noteworthy that now both, longitudinal and transverse rates, increase with decreasing density. The increase is more pronounced in the longitudinal channel, but not as large as in the degenerate limit, see Fig. \ref{fig:InducedComp}.
\section{Summary and Outlook}
\noindent I have presented an extensive study of electromagnetic scattering in the environment of dense homogeneous nuclear matter. Effective masses, chemical potentials, and residual interactions of the nucleon quasiparticles in beta equilibrium have been extracted from an energy functional based on Skyrme type interactions. The relationship between the scattering rate $\Gamma$ and the loop expansion of the fermion self energy has been discussed in detail. Debye screening and Landau damping of electromagnetic interactions due to electrons, muons, protons, and neutrons have been incorporated employing the random phase approximation (RPA). Hard dense loop (HDL) and weak-screening approximations of the full RPA result have been discussed and compared. At finite temperature the energy loss per distance travelled, $ -dE/dx$, rather than the scattering rate has been calculated, as the latter suffers from an infrared divergence within the RPA.
A central aspect of this article have been correlations of strong and electromagnetic interactions, occurring as a result of collisions with protons, which carry both charges. The induced interactions with neutrons affect electromagnetic scattering in an implicit manner, i.e., via screening effects embodied the photon propagator. While the calculation of lepton scattering rates has been the natural focus of this article, electromagnetic scattering of protons has been examined as well. In the following the main results are summarized. 
\subsection{Summary of results}
\noindent The results for the scattering rates at zero temperature highlight the importance of screening effects in (fully) degenerate matter. 
A comparison of \textit{weak-screening} approximation and full RPA results reveals, that the former delivers reliable results for the scattering rates of light plasma constituents (i.e., electrons and electron-holes) at very high densities. At lower densities of about $n\sim0.6\,n_0$ dynamical screening in the multi-component plasma beyond leading order in the photon energy $q_0$ becomes important, see Fig. \ref{fig:WeakFull}.  
The \textit{hard dense loop} resummation captures these effects correctly, and represent an excellent approximation at any given density, as long as one is interested in the energy loss of fermions with momenta in a range of $(|\boldsymbol{p}|-k_f)\lesssim \pm5$ MeV. Given that HDL results are fairly easy to handle they represent the best compromise of accuracy and computational effort, in particular if the energy transfer in collisions is not too large (which is practically always the case in the core of neutron stars). The scattering of relativistic fermions in highly degenerate plasmas has been studied in detail by Heiselberg and Pethick, who conclude that the exchange of (transverse) photons represent the dominant channel for interaction in such an environment \cite{Heiselberg:1993cr}. To which extent this remains true for leptons in nuclear matter is again a density dependent question: the longitudinal rates of electrons \textit{increase} by more than an order of magnitude going from $n=2\,n_0$ to $n=0.55\,n_0$, see Fig. \ref{fig:WeakFull}, while the transverse rates \textit{decrease} slightly, see Fig. \ref{fig:PerpCompare}. The dominance of transverse scattering is thus much less pronounced at lower densities. A similar pattern emerges for muons, which are mildly relativistic at $n=n_0$. At saturarion density their rates are much smaller than those of electrons, in particular in the transverse channel, see Fig. \ref{fig:PerpCompare}. At $n=2\,n_0$ the rates of electrons and muons are comparable in both channels, and transverse scattering clearly dominates. In general the scattering rate of a given fermion species increases considerably when subjected to a multi-component plasma. The only exception is the transverse rate of electrons, which is dominated by electron-electron scattering, and, as a result of screening effects decreases slightly in the presence of muons and protons. \newline            
\noindent When correlations with nuclear interactions are taken into account, the photon propagator becomes sensitive to ground state properties of nuclear matter. This is reflected in the proton polarization function $\tilde{\Pi}_p$, resummed in the subspace of protons and neutrons interacting via the residual quasiparticle potentials $f_{pp}$, $f_{pn}$, and $f_{nn}$. In the static limit $\tilde{\Pi}_p$ yields the (strong) Debye mass, which diverges upon approaching the critical density $n_c$ for the stability of homogeneous nuclear matter, see Fig. \ref{fig:Screen}. The quantitative repercussions on the scattering rates are to some extent model dependent, and the robustness of the results has been tested using five different modern Skyrme forces. The sharp increase of static screening close to $n_c$ enlarges the low energy tail of the photon spectrum, see Fig. \ref{fig:RhoLongT}, which in turn increases the longitudinal scattering rates by more than a factor of $2$. At densities above $n_0$ induced interactions \textit{reduce} the Debye screening of protons (and consequently $\Gamma_L$), although by a much smaller amount. Muons are absent at densities below $n\sim0.7\,n_0$, where the impact of induced interactions is most pronounced, and thus only experience the decreasing effects at higher densities. The lack of static screening in the \textit{transverse} channel entails, that induced interactions are of minor importance for $\Gamma_\perp$. The additional boost of longitudinal scattering further represses the dominance of transverse rates at densities close to $n_c$, see Fig. \ref{fig:LowDens}. To study the impact of induced interactions on \textit{heavy} fermions at the edge of stability, the scattering rates of protons have been computed, see Fig. \ref{fig:LowDens}. The impact on the longitudinal rates of protons is indeed striking, resulting in a boost of roughly a factor of $5$. All of the above results are robust, and to a large extent independent of the chosen Skyrme model. At lower densities variations in the numerical results occur at fixed density $n$, because the impact of induced interactions is very sensitive to the relative distance to the critical density $n_c$, which computes to a slightly different value in each model. Using Skyrme parameter sets with similar critical densities results again in very narrow bands, see Fig.\ref{fig:LowDens}. \newline
\noindent At finite temperature the energy loss per distance travelled, $-dE/dx$, has been computed under degenerate conditions, with $T\leq 1$ MeV, and under partially degenerate conditions, with $10\,\textrm{MeV}
\leq T \leq 30$ MeV. 
At temperatures of $T=0.1$ MeV the scattering rates are basically indistinguishable from those calculated at zero temperature. As the phase space available for scattering increases with the temperature, the characteristics of the scattering rates change in several ways: temperatures of $T=(0.5 -1)$ MeV, though tiny compared to the electron chemical potential, lead to a substantial increase of the energy loss of fermions with $|\boldsymbol{p}|\sim k_f$. At a distance of about $|\boldsymbol{p}|\sim(k_f\pm1)$ MeV from the Fermi surface the rates at $T=1$ MeV are still more than twice as large as those at $T=0$, see Fig. \ref{fig:DeDxLowT}. The dominance of the transverse channel has completely disappeared, as longitudinal contributions are larger for fermions at any given momentum. Induced interactions lose much of their importance: in the absence of a (sharp) Fermi surface, there is no longer a region around $k_f$, where induced lepton-neutron scattering represent the dominant contribution to the longitudinal rate.  
This trend continues in partially degenerate matter at temperatures of $T=(10-30)$ MeV. Longitudinal rates easily outgrow transverse ones, in the case of muons by about 5-10 times, in the case of electrons still by about 3 times, see Fig. \ref{fig:DeDxHighTEMP}. At $T=30$ MeV the rates of electrons (holes) increase (decrease) almost linearly with their momenta, sharing little in common with the typical characteristics of scattering in fully degenerate matter. Longitudinal and transverse rates both decrease with increasing density, in contrast to degenerate matter where transverse rates increase, see Fig. \ref{fig:PerpCompare}. In most cases induced interactions are of minor importance, with the exception of electrons at lower densities, where the increase be can as large as $20$\%, see Fig. \ref{fig:DeDxHighTINDUCED}.         
\subsection{Relevance for neutron star phenomenology and outlook}
\noindent A particular focus of this article has been the computation of electron and muon scattering rates under conditions similar to those realized in the crust-core boundary of neutron stars. The transport properties of this regions play an important role for the spin evolution of neutron stars. Among the biggest mysteries in nuclear astrophysics is the observation of high-frequency pulsars. As a neutron star spins up by accretion of matter various oscillation modes, among them r-modes, are excited. R-modes are known to be generically unstable with respect to the emission of gravitational waves, to which they transmit angular momentum. As a result, the spin-up of the star should be limited by a certain critical frequency. The r-mode amplitudes grow provided that potential damping mechanisms operate on a longer timescale than the emission gravitational waves. One of them is viscosity in the crust-core interface, which, however, would have to be several times larger than previously calculated \cite{Ho:2011tt} to stabilize fast spinning stars. Several effects, including suerpconductivity/superfluidity \cite{Gusakov:2013jwa} or meson condensation \cite{Kolomeitsev:2014gfa} have  have been considered in order to explain the appearant discrepancy. Because electrons are weakly interacting and ultra-relativistic, they are considered a dominant contributor to transport. Induced interactions strongly increase the scattering rates of electrically charged particles at densities which are just about high enough to support stable homogeneous nuclear matter, i.e., at the boundary to the crust. In addition, they are particularly pronounced at the very small temperatures typical of accreting neutron stars. The energy loss $-dE/dx$ of electrons integrated in a small range of momenta around $k_f$ increases by more than a factor of $3$. The increase of the scattering rate (or in other words the reduction of the mean free path) indicates that the capability of electrons to transport heat, charge, or momentum has previously been overestimated. How large the friction between crust and core can become depends, among other things, on the existence of a ``nuclear pasta phase" \cite{Pethick:1995di}, which potentially smears out the transition, and reduces the damping effect of the viscous boundary layer \cite{Haskell:2012vg}. The rapid increase of Debye screening particularly impacts heavy fermions, and motivates a refined study of the transport properties of nucleons in the crust-core interface. It is certainly interesting to ask whether or not nuclear matter in its simplest manifestation is sufficient to explain the observation of fast spinning stars.\newline If gravitational wave asteroseismology of isolated neutron stars becomes available it would  represent an outstanding tool to probe the interior of neutron stars. Neutron star mergers are well within reach of current gravitational-wave detector sensitivities, and the relevance of transport phenomena for their simulation is the subject of several ongoing studies. The calculations carried out under partially degenerate conditions reveal substantial differences to the scattering rates in highly degenerate matter. These should be taken into account in the calculation of transport coefficients, allowing for a better assessment of their importance. Induced interactions most likely play a minor role at higher temperatures. 
\newline
\newline
\noindent Several approximations employed in this article warant further studies. Vertex corrections, briefly discussed in section \ref{sec:OT}, might become important in the calculation of scattering rates at higher temperatures. Without vertex corrections interference contributions to Moeller scattering cannot be extracted from the fermion self energy. An example is the diagram depicted in Fig. \ref{fig:scatter} (c), which may be expressed as  $\Sigma(p)\propto\int d^4 q\,\Gamma(q)^\mu\,D_{\mu\nu}(q)\,\gamma^\nu\,S(p-q)$, where $\Gamma^\mu$ is the one-loop vertex. It is certainly not unusual to include vertex corrections in the Braaten and Pisarski resummation programm. However, how resummed vertices relate to screening effects in the various scattering channels has not been explored in detail yet. In addition, the momentum dependence of residual particle-hole interactions should be extracted from 
microscopic approaches, see e.g. Ref. \cite{Benhar:2017oli}. To do so would allow for a rigorous study of dynamical screening effects in the strong section of the RPA resummation. Finally, a fully iterative determination of the mean-field energies and interaction potentials of nucleon quasiparticles at finite temperature would be desirable. The refined results for effective masses and chemical potentials improve the accuracy of the RPA calculation of scattering rates in partially degenerate nuclear matter. \newline 
In addition to the technical aspects mentioned above it would be important to asses the impact of superconductivity and magnetic fields on the calculations presented in this article. It is commonly, though not unanimously, projected that protons in the outer core of neutron stars are superconducting. Scattering at temperatures below the corresponding critical temperature is subject to the Meissner effect, which introduces static screening to the transverse channel. The resulting interplay of the Meissner effect and induced interactions should be studied in detail. Finally, to calculate the scattering rates in the presence of a magnetic field requires for a careful reevaluation of the fermion self energy (i.e., the scattering matrix element). 
\section*{Acknowledgements}
\noindent I would like to thank Sanjay Reddy and Ermal Rrapaj for many useful discussions during the execution of this project, Ingo Tews and Ermal Rrapaj for a critical reading of the manuscript, and Alessandro Roggero for providing valuable help in the development of the numerical code. I have been supported by a Schroedinger Fellowship of the Austrian Science Fund FWF, project no. J3639.  

\appendix
\section{Fermion Self Energy}
\label{sec: RTFcalc}
\begin{figure}[t]
\includegraphics[scale=1.0]{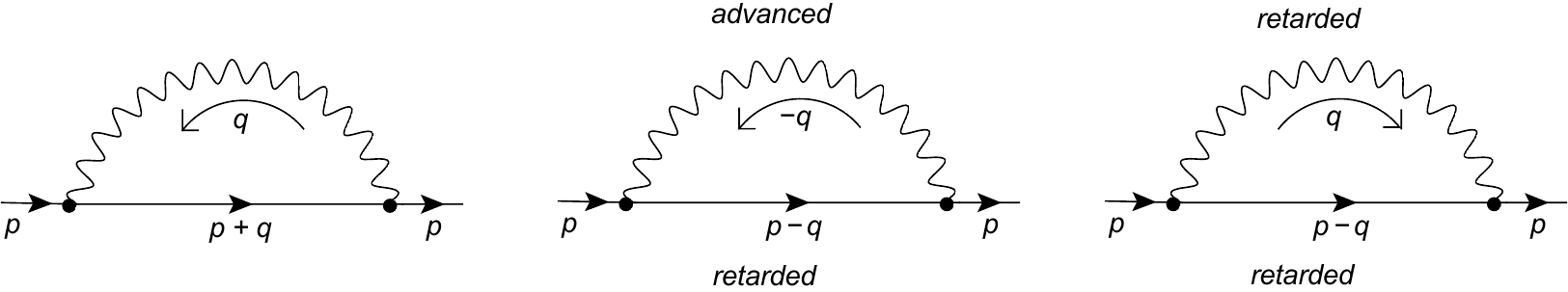}
\caption{\label{fig:SelfEnergyElectron}Assigned momenta for the electron self energy. The formation a closed loop with  photon momentum $q$ requires the electron to carry the momentum $p+q$. The electron line is retarded, the photon line is advanced. Reversing the loop momentum
leads to an advanced photon propagator with momentum $-q$ which corresponds to a retarded photon propagator with momentum $+q$. }
\end{figure}
\noindent This appendix reviews the calculation of the retarded
fermion self energy using the real time formalism (RTF) in Keldysh representation. Retarded, advanced and symmetric propagators of fermions and photons are given by
\begin{eqnarray}
S_{R/A}(p) & = & \frac{\cancel{p}+m}{p^{2}-m^{2}\pm i\textrm{sgn}(p_{0})\epsilon}\,,\,\,\,\,\,\,\, D_{R/A}^{\mu\nu}(q)=\frac{1}{q^{2}\pm i\textrm{sgn}(q_{0})\epsilon}G^{\mu\nu}(q)\,,\label{eq:PropRA}\\[3ex]
S_{S}(p) & = & -2\pi i\left(\cancel{p}+m\right)\,(1-2N_{f}(p_{0}))\delta(p^{2}-m^{2})\,,\label{eq:PropSFermi}\\[3ex]
D_{S}^{\mu\nu}(q) & = & -2\pi i(1+2n_{b}(q_{0}))\delta(q^{2})G^{\mu\nu}(q)\,.\label{eq:PropSBose}
\end{eqnarray}
\newline
\noindent The Fermi distribution $N_f$ covers particles and antiparticles, see Eq. \ref{eq:FermiDistr}. I work in Coulomb gauge, where the gauge fixing dependent factor $G^{\mu\nu}$ of the photon propagator
reads
\begin{equation}
G^{\mu\nu}(q)=\frac{q^{2}}{\boldsymbol{q}^{2}}\, g^{\mu0}g^{\nu0}+\delta^{\mu i}\delta^{\mu j}(\delta_{ij}-\hat{q}_{i}\hat{q}_{j})\,.\label{eq:PhotonGauge2}
\end{equation}
\noindent Following the momentum flow of an external leg into a retarded
amplitude and tracing through the diagram one assigns a retarded propagator
to each momentum aligned with the tracing direction, and an advanced
propagator to each momentum aligned opposite to it. Reading off the
momentum assignment in Fig. \ref{fig:SelfEnergyElectron} one consequently
finds
\noindent 
\begin{equation}
\Sigma_{R}(p)=\frac{ie^{2}}{2}\int\frac{d^{4}q}{(2\pi)^{4}}\left[\gamma_{\mu}S_{R}(p+q)\gamma_{\nu}D_{S}^{\mu\nu}(q)+\gamma_{\mu}S_{S}(p+q)\gamma_{\nu}D_{A}^{\mu\nu}(q)\right]\,.\label{eq:SigmaR}
\end{equation}
\newline
\noindent Additional pairings of two retarded or advanced propagators
which in general occur, but equate to zero under $q_{0}$ integration,
have been ignored. In relating the kinematics of the self energy to
those of the scattering process discussed in Sec. \ref{sec:OT} it would be advantageous if the fermions carried momentum $p^{\prime}=p-q$
rather than $p^{\prime}=p+q$. This can be achieved by replacing
the loop momentum $q$ with $-q$. While this doesn't affect the symmetric
propagators, it does affect the retarded and advanced ones which according
to \ref{eq:PropRA} satisfy $D_{R/A}^{\mu\nu}(-q)=D_{A/R}^{\mu\nu}(q)$.
As a result, the retarded self energy is written in terms of pairings
of symmetric and retarded propagators only (see right panel of Fig.
\ref{fig:SelfEnergyElectron})
\noindent 
\begin{equation}
\Sigma_{R}(p)=\frac{ie^{2}}{2}\int\frac{d^{4}q}{(2\pi)^{4}}\left[\gamma_{\mu}S_{R}(p-q)\gamma_{\nu}D_{S}^{\mu\nu}(q)+\gamma_{\mu}S_{S}(p-q)\gamma_{\nu}D_{R}^{\mu\nu}(q)\right]\,.\label{eq:SigmaR2}
\end{equation}
\noindent Symmetric propagators are purely imaginary, the real part of $\Sigma_{R}$ is obtained by pairing them with the principal value piece of the retarded propagators \ref{eq:PropRA} (mind the global factor $ie^2$), which results in
\newline
\begin{equation}
\text{Re}\,\Sigma_{R}(p)=\frac{1}{2}\frac{e^{2}}{(2\pi)^{3}}\int d^{4}q\,\gamma_{\mu}\,(\slashed{p}^{\prime}+m)\,\gamma_{\nu}\,G^{\mu\nu}(q)\,\left[(1+2n_{b}(q_{0}))\frac{1}{p^{\prime2}-m^{2}}\delta(q^{2})+(1-2N_{f}(p_{0}^{\prime}))\frac{1}{q^{2}}\delta(p^{\prime2}-m^{2})\right]\,.\label{eq:ReSigma}
\end{equation}
The imaginary part is easily obtained via 
\begin{equation}
\text{Im}[\Sigma_{R}(p)]=\frac{1}{2i}\left[\Sigma_{R}(p)-\Sigma_{A}(p)\right]\,.
\end{equation}
Since $S_{R}(p)-S_{A}(p)=(-2\pi i)\,\text{sign}(p_{0})\,\delta(p^{2}-m^{2})\,(\cancel{p}+m)$
(and similarily for the photon propagator) one immediately finds 
\begin{equation}
\text{Im}\Sigma(p)=-\frac{e^{2}}{4}\int\frac{d^{4}q}{(2\pi)^{2}}\,\gamma_{\mu}\left(\cancel{p}^{\prime}+m\right)\gamma_{\nu}G^{\mu\nu}(q)\,I_{DB}(q_0,\,p_0)\,\delta(p^{\prime2}-m^{2})\delta(q^{2}).\label{eq:ImSigma-1}
\end{equation}
with the definition of the detailed balance factor
\begin{equation}
\text{I}_{DB}(q_0,\,p_0)=\text{sign}(p_{0}^{\prime})\left[1+2n_{b}(q_{0})\right]+\text{sgn}(q_{0})\left[1-2N_{f}(p_{0}^{\prime})\right]\,.
\end{equation}
 
\section{Imaginary part of the photon polarization tensor}\label{sec:ImPi}
\noindent The calculation of the retarded one-loop photon polarization tensor runs mostly analogous to the calculation of the fermion self-energy. The generic expression reads
\be \label{eq:PIformal}
\Pi_{R}^{\mu\nu}(q)  =  -ie^{2}\int\frac{d^{4}k}{(2\pi)^{4}}\left\{ \text{Tr}\left[\gamma^{\mu}\,S_{S}(k^{\prime})\,\gamma^{\nu}S_{R}(k)\right]+\text{Tr}\left[\gamma^{\mu}\,S_{A}(k^{\prime})\,\gamma^{\nu}S_{S}(k)\right]\right\} \,,
\ee
with the propagators \ref{eq:PropRA} - \ref{eq:PropSBose}. The evaluation of this object is discussed in great detail in Ref. \cite{Stetina:2017ozh}, here we take a quick look at calculation of the imaginary part, which is needed in the calculation of the scattering rate below. The imaginary part can be obtained directly from Eq. \ref{eq:PIformal} via
\begin{equation}
\text{Im}\,\Pi^{\mu\nu}(q)=\frac{1}{2i}\left[\Pi_{R}^{\mu\nu}(q)-\Pi_{A}^{\mu\nu}(q)\right]\,.\label{eq:IMPhoton}
\end{equation}
Using $S_{R}(k^{\prime})-S_{A}(k^{\prime})=-2\pi i\,\text{sign}(k_{0}^{\prime})\,\delta(k^{\text{\ensuremath{\prime}}2}-m^{2})\left(\cancel{k}^{\prime}+m\right)$ one immediately finds the generic expression for $\text{Im}\,\Pi^{\mu\nu}$, 
\begin{equation}
\text{Im}\,\Pi^{\mu\nu}(q)=-\frac{e^{2}}{4}\int\frac{d^{4}k}{(2\pi)^{2}}\,\tilde{I}_{DB}\,\,\delta(k^{\prime2}-m^{2})\,\delta(k^{2}-m^{2})\,T^{\mu\nu}(k)\,,\label{eq:ImPhotonBare}
\end{equation}
where the detailed balance factor (labeled $\tilde{I}_{DB}$ to make it distinct from $I_{DB}$ included in the fermion self energy) and the trace are
\bea 
\tilde{I}_{DB} &=&\text{sgn}(k_{0})(1-2N_{f}(k_{0}^{\prime}))-\text{sgn}(k_{0}^{\prime})\,(1-2N_{f}(k_{0}))\,,\\[3ex]
T^{\mu\nu}&=&\text{Tr}\left[\gamma^{\mu}\left(\cancel{k}+m\right)\gamma^{\nu}\left(\cancel{k}^{\prime}+m\right)\right]\,.
\eea

\section{Scattering rates to order $\alpha_f^2$} \label{sec:scatrate}
\noindent This appendix demonstrates the calculation of the scattering rate of fermions to leading order in $\alpha_f=e^2$. The calculation is carried out using two different approaches: once by applying Fermi's golden rule and once via the optical theorem (i.e., using the imaginary part of the fermion self energy). The results are shown to be equivalent. To calculate the damping rate from the fermion self energy, we start from Eq. \ref{eq:SigmaPlus}. The imaginary part of $\Sigma$ is given by Eq. \ref{eq:ImSigma}, where the bare on-shell photon has to be replace with a corresponding spectral function. For the sake of the current discussion the $e^2\propto\alpha_f$ dependence of $\Pi$ is made explicit. An expansion $\rho_\perp$ to leading order yields
\newline
\be
\rho_{\perp}(q) = - \frac{1}{\pi}\frac{\alpha_f\,\text{Im}\,\Pi_\perp}{(\alpha_f\,\text{Re}\,\Pi_\perp-q^{2})^{2}+(\alpha_f\,\text{Im}\,\Pi_\perp)^{2}}\sim-\frac{1}{\pi}\frac{1}{q^{4}}\alpha_f\text{\, Im}\,\Pi_\perp+\mathcal{O}(\alpha_f^2)\,.
\ee
Note that we are specifically interested in the transverse photon polarization, as it survives if the exchanged momentum $q$ is hard. To continue working in a covariant notation we shall express $\rho$ as 
\bea
\delta(q^2)\,G^{\mu\nu} \rightarrow \rho^{\mu\,\nu}\sim-\frac{\alpha_f}{\pi}\,G^{\mu\alpha}\, G^{\nu\beta}\,\,\text{Im}\,\Pi_{\alpha\,\beta}\,,
\eea
where $G$ is given by Eq. \ref{eq:PhotonGauge}, and in principle contains the longitudinal projector as well. We may still pick out the transverse polarization in the final result. Putting everything together results in 
\be \label{eq:Im1}
\text{Im}\,\Sigma(p)=\frac{e^4}{4\,\pi}\int\frac{d^4 q}{(2\,\pi)^2}\,I_{DB}\,\,\gamma_{\mu}\left(\cancel{p}^{\prime}+m\right)\gamma_{\nu}\,G^{\mu\alpha}\, G^{\nu\beta}\,\,\text{Im}\,\Pi_{\alpha\,\beta}\,\,\delta(p^{\prime\,2}-m^2)\,, 
\ee
where the detailed balance factor as defined in Eq. \ref{eq:IDB}, and the imaginary part of the one-loop photon polarization tensor is given by Eq. \ref{eq:ImPhotonBare}. The scattering rate Eq. \ref{eq:OTheorem} for a single particle excitation now becomes
\be \label{eq:Gamma1}
\Gamma(\epsilon_{\boldsymbol{p}})=-\frac{1}{2\,\epsilon_{\boldsymbol{p}}}\left[ 1 - \,n_f^-(p_0)\right]\text{Tr}\left[\left(\slashed{p}+m\right)\,\text{Im}\,\Sigma(p_0=\epsilon_{\boldsymbol{p}},\,\boldsymbol{p})\right]\,.
\ee
If the exciation under consideration is a hole, the factor $1-n_f$ has to be replaced by $n_f$. To simplify the above expression, we may exploit the kinematics of the scattering process: The external momenta $k$, $k^{\prime}$, $p$, and $p^\prime$ are all on-shell,i.e., $k^{2}=m^{2}$
and $k^{\prime2}=m^{2}$, and the momentum transfer is thus
\noindent 
\begin{equation}
q^{2}=(k^{\prime}-k)^{2}=2\left(m^{2}-k_{0}\,k_{0}^{\prime}+\boldsymbol{k}\cdot\boldsymbol{k}^{\prime}\right)\,.
\end{equation}
\noindent Scattering processes imply $q^{2}<0$, and therefore, by virtue of the
Schwartz inequality, $k_{0}\,k_{0}^{\prime}=\sqrt{\boldsymbol{k}^{2}+m^{2}}\sqrt{\boldsymbol{k}^{\prime2}+m^{2}}\geq m^{2}+\left|\boldsymbol{k}\right|\left|\boldsymbol{k}^{\prime}\right|$. Consequently $k_{0}$ and $k_{0}^{\prime}$ must both either be positive (corresponding
to incident particles), or negative (corresponding to incident anti-particles), where the latter case can be ignored. Identical considerations hold for the vertex involving the momenta $p$ and $q$. Exploiting these kinematic constraints, the detailed balance factors included the imaginary part $\textrm{Im}\,\Sigma$ Eqs. \ref{eq:Im1} and the imaginary part  $\textrm{Im}\,\Pi$ Eq. \ref{eq:ImPhotonBare} evaluate to 
\bea \label{eq:IdbP}
\left[1-\,n_f^-(p_0)\right]\,\,I_{DB}\,\tilde{I}_{DB}&=&\left[1-n_{f}^{-}(p_{0})\right]\,\left[1+n_{b}(q_{0})-n_{f}^{-}(p_{0}^{\prime})\right]\,\left[n^-_{f}(k_{0})-n^-_{f}(k_{0}^{\prime})\right]\,\nonumber \\[2ex] &=&n^-_{f}(k_{0})\,\left[1-n^-_{f}(k_{0}^{\prime})\right]\,\left[1-n^-_{f}(p_{0}^{\prime})\right]\,,
\eea
where it is easy to check the second equality by explicit calculation. As expected, the detailed balance in the second line describes the scattering of a single fermion with energy $p_0$ on another fermion with energy $k_0$, which belongs to the Fermi sea. As a result the two new particles appear with energies $k_0^\prime$ and $p_0^\prime$, which, forced by Pauli blocking are extracted from the Fermi sea. If we are interested in the scattering rate of a hole, the thermal distribution functions in Eq. \ref{eq:Gamma1} can be written as
\be \label{eq:IdbH}
n_f^-(p_0)\,I_{DB}\,\tilde{I}_{DB}=n^-_{f}(k_{0}^{\prime})\,\,n^-_{f}(p_{0}^{\prime})\,\left[1-n^-_{f}(k_{0})\right]\,.
\ee
This time, the expression on the right hand side matches with the detailed balance of the inverse process, in which a hole with energy $p_0$ is filled by a particle "falling" into it from above in the Fermi sea, leaving behind another hole with energy $p_0^\prime$. The energy difference is transferred to another particle with energy $k_0^\prime=k_0+q_0$ ($q_0<0$ for inverse processes), which is then extracted out of the Fermi sea. It remains to deal with the momentum integration. Collecting all integration variables in Eq. \ref{eq:Gamma1} yields
\begin{equation}
\int d^{4}q\,\delta\left(p^{\prime\,2}-m^{2}\right)\,\int d^{4}k\,\delta\left(k^{\prime\,2}-m^{2}\right)\delta\left(k^{2}-m^{2}\right)\,.
\end{equation}
These can be reorganized by exploiting the conservation of four-momentum at each vertex,
\begin{equation}
1=\int d^{4}k^{\prime}\,\delta^{(4)}(k^{\prime}-k-q)\,,\,\,\,\,\,\,\,\,\,1=\int d^{4}p^{\prime}\,\delta^{(4)}(p^{\prime}-p+q)\,,
\end{equation}
such that after integrating out $q$ and dropping anti-particle contributions one obtains
\bea
& &\int d^{4}k\,\int d^{4}k^{\prime}\,\int d^{4}p^{\prime}\,\delta(k^{2}-m^{2})\delta(k^{\prime2}-m^{2})\,\delta(p^{\prime2}-m^{2})\,\delta^{(4)}(p+k-p^{\prime}-k^{\prime})\,\nonumber \\[2ex]
&=& \frac{1}{2\,\epsilon_{\boldsymbol{k}}}\int d^{3}\boldsymbol{k}\,\frac{1}{2\,\epsilon_{\boldsymbol{k}^\prime}}\int d^{3}\boldsymbol{k}^{\prime}\,\frac{1}{2\,\epsilon_{\boldsymbol{p}^\prime}}\int d^{3}\boldsymbol{p}^{\prime}\,\delta^{(4)}\,(p+k-p^{\prime}-k^{\prime})\,,\label{eq:DeltaDist}
\eea 
where all energies in $\delta^{(4)}$ are on-shell. Collecting all signs and factors, the final result for particles reads
\newline
\bea 
\Gamma(\epsilon_{\boldsymbol{p}})&=&\frac{e^{4}}{(2\pi)^{5}}\,\frac{1}{2\,\epsilon_{\boldsymbol{p}}}\,\frac{1}{2\,\epsilon_{\boldsymbol{k}}}\int d^{3}\boldsymbol{k}\,\,n_{f}^{-}(\epsilon_{\boldsymbol{k}})\,\frac{1}{2\,\epsilon_{\boldsymbol{k}^\prime}}\int d^{3}\boldsymbol{k}^{\prime}\,\,\left[1-n_{f}^{-}(\epsilon_{\boldsymbol{k}^{\prime}})\right]\frac{1}{2\,\epsilon_{\boldsymbol{p}^\prime}}\int d^{3}\boldsymbol{p}^{\prime}\,\,\left[1-n_{f}^{-}(\epsilon_{\boldsymbol{p}^{\prime}})\right]\cdot\delta^{(4)}(p+k-p^{\prime}-k^{\prime})\,\nonumber\\[2ex]&&\cdot\frac{1}{4}\,\frac{1}{q^{4}}\,\text{Tr}\left[\left(\cancel{p}+m\right)\gamma_{\mu}\left(\cancel{p}^{\prime}+m\right)\gamma_{\nu}\right]\cdot\text{Tr}\left[\left(\gamma\cdot k^{\prime}+m\right)\gamma_{\alpha}\left(\gamma\cdot k+m\right)\gamma_{\beta}\right]\,G^{\mu\alpha}\, G^{\nu\beta}\,. \label{eq:RateCompare}
\eea 
\newline
This is precisely the result one obtains from a calculation of the interaction rate according to Fermi's golden rule,
\begin{equation}
\Gamma(\epsilon_{\boldsymbol{p}})=\frac{1}{2\epsilon_{\boldsymbol{p}}}\,\int_{k}\,n_{f}^{-}(\epsilon_{\boldsymbol{k}})\,\int_{k^{\prime}}\,\left[1-n_{f}^{-}(\epsilon_{\boldsymbol{k}^{\prime}})\right]\,\int_{p^{\prime}}\,\left[1-n_{f}^{-}(\epsilon_{\boldsymbol{k}^{\prime}})\right]\,\left(2\pi\right)^{4}\delta(p+k-p^{\prime}-k^{\prime})\,\left|M\right|^{2}\,,\label{eq:FermiRate}
\end{equation}
with the short-hand notation for the integration measure,
\be 
\int_{p}=\int\frac{d^{3}\boldsymbol{p}}{(2\pi)^{3}}\frac{1}{2\epsilon_{\boldsymbol{p}}}\,,\label{eq:measure}
\ee
and the squared t-channel matrix element $|M_t|^2$, averaged over the initial spin states, which reads (in Coulomb gauge)
\be
\left|M_t\right|^2 = \frac{1}{4}\frac{e^4}{q^4}\sum_{\textrm{spins}}\left[\bar{u}_k^\prime\,\gamma_\mu\,u_k\right]\,\left[\bar{u}_p^\prime\,\gamma_\alpha\,u_p\right]\,\left[\bar{u}_k^\prime\,\gamma_\nu\,u_k\right]^\dagger\,\left[\bar{u}_p^\prime\,\gamma_\beta\,u_p\right]^\dagger\,G^{\mu\alpha}\,G^{\nu\beta}, 
\ee
and, after a after summing over the final spin states results in the second line of Eq. \ref{eq:RateCompare}. The u-channel matrix element can be obtained by a simple relabelling of the momenta. The matrix element of the interference term, however, reads e.g.,
\bea 
M_t\,M^*_u &=& \frac{1}{4}\frac{e^4}{q^2\,(p-k-q)^2}\sum_{\textrm{spins}}\left[\bar{u}_k^\prime\,\gamma_\mu\,u_k\right]\,\left[\bar{u}_p^\prime\,\gamma_\alpha\,u_p\right]\,\left[\bar{u}_p^\prime\,\gamma_\nu\,u_k\right]^\dagger\,\left[\bar{u}_k^\prime\,\gamma_\beta\,u_p\right]^\dagger\,G^{\mu\alpha}\,G^{\nu\beta}\,\\[2ex]
 &=&\frac{1}{4}\frac{e^4}{q^2\,(p-k-q)^2}\textrm{Tr}\left[\gamma_\mu\left(\cancel{k}+m\right)\,\gamma_\nu\left(\cancel{p}^\prime+m\right)\,\gamma_\alpha\left(\cancel{p}+m\right)\,\gamma_\beta\left(\cancel{k}^\prime+m\right)\,G^{\mu\alpha}\,G^{\nu\beta}\right]\,.
\eea 
Note, that the Dirac trace does not factorize. This contributions has to be extracted from the two-loop diagram in Fig. \ref{fig:2loop} (c), which cannot be constructed by inserting one self-energy into another.

\newpage

\bibliographystyle{apsrev}
\bibliography{fermions.bib}
\end{document}